\documentclass[aps,prb,amsmath,amssymb,
twocolumn,tightenlines]{revtex4-2}
\usepackage{graphicx}% Include figure files
\usepackage{dcolumn}% Align table columns on decimal point
\usepackage{bm}% bold math
\usepackage[mathscr]{euscript}
\begin{document}
%\preprint{APS/123-QED}
\title{Low-energy peak in the one-particle spectral function of the electron gas
\\ at metallic densities}

%%%%%%%%%%%%%%%%%%%%%%%%%%%%%%%%%%%%%%%%%%%%%%%%%%%%%%%%%%%%%%%%%%%%%%%%%%%%%%%%%%%%%
\author{Yasutami Takada}
\affiliation{Institute for Solid State Physics, The University of Tokyo, 
Kashiwa, Chiba 277-8581, Japan}

\date{\today}    % \today is  today, but any date may be explicitly specified.      %
%%%%%%%%%%%%%%%%%%%%%%%%%%%%%%%%%%%%%%%%%%%%%%%%%%%%%%%%%%%%%%%%%%%%%%%%%%%%%%%%%%%%%
\begin{abstract}
Based on a nonperturbative scheme to determine the self-energy 
$\Sigma({\bm k},i\omega_n)$ with automatically satisfying the Ward identity and 
the total-momentum conservation law, a fully self-consistent calculation is done 
in the electron gas at various temperatures $T$ to obtain $G({\bm k},i\omega_n)$ 
the one-particle Green's function with fulfilling all known conservation laws, 
sum rules, and correct asymptotic behaviors; here, $T$ is taken unprecedentedly 
low, namely, $k_{\rm B}T/\varepsilon_{\rm F}$ down to $10^{-4}$ with 
$\varepsilon_{\rm F}$ the Fermi energy, and tiny mesh $\Delta k$ as small as 
$10^{-4}k_{\rm F}$ is chosen near the Fermi surface in ${\bm k}$ space with 
$k_{\rm F}$ the Fermi momentum. By analytically continuing $G({\bm k},i\omega_n)$ 
to the retarded function $G^R({\bm k},\omega)$, we find a novel low-energy peak, 
in addition to the quasiparticle (QP) peak and one- and two-plasmon high-energy 
satellites, in the spectral function $A({\bm k},\omega)
[=\!-{\rm Im}G^R({\bm k},\omega)/\pi]$ for $k_{\rm B} T \! \lesssim \! 
10^{-3}\varepsilon_{\rm F}$ in the simple-metal density region 
($2\!<\!r_s\!<\!6$ with $r_s$ the dimensionless density parameter). 
This new peak is attributed to the effect of excitonic attraction on 
$\Sigma({\bm k},i\omega_n)$ arising from multiple excitations of tightly-bound 
electron-hole pairs in the polarization function $\Pi({\bm q},i\omega_q)$ for 
$|{\bm q}|\! \approx \! 2k_{\rm F}$ and $|\omega_q| \! \ll \! \varepsilon_{\rm F}$ 
and thus it is dubbed ``excitron''. Although this excitron peak height is only 
about a one-hundredth of that of QP, its excitation energy is about 
a half of that of QP for $|{\bm k}| \! \approx \! k_{\rm F}$, seemingly 
in contradiction to the Landau's hypothesis as to the one-to-one 
correspondence of low-energy excitations between a free Fermi gas and an interacting 
normal Fermi liquid. As for the QP properties, our results of both the effective mass 
$m^*$ and the renormalization factor $z^*$ are in good agreement with 
those provided by recent quantum Monte Carlo simulations and available experiments.
\end{abstract}
% APS now has a new classification scheme for physics, 
% PhySH - Physics Subject Headings, which replaced PACS. 
% With the transition to PhySH, authors will no longer need to provide PACS. 
%\pacs{71.10.Ca, 71.45.Gm, 05.30.Fk, 71.10.Hf}
%05.30.Fk Fermion systems and electron gas (in Statistical Physics)
%         (see also 71.10.-w Theories and models of many-electron systems; 
%          see also 67.10.Db Fermion degeneracy in quantum fluids) 
%31.15.E- Density-functional theory (in Atomic and Molecular Physics)
%31.15.eg Exchange-correlation functionals (in current density functional theory) 
%67.85.Lm Degenerate Fermi gases
%71.10.Ca Electron gas, Fermi gas 
%71.10.Hf Non-Fermi-liquid ground states, electron phase diagrams 
%         and phase transitions in model systems
%71.15.Mb Density functional theory, local density approximation, 
%         gradient and other corrections
%71.15.-m Methods of electronic structure calculations
%71.10.-w Theories and models of many-electron systems
%71.45.Gm Exchange, correlation, dielectric and 
%         magnetic response functions, plasmons 
%%%%%%%%%%%%%%%%%%%%%%%%%%%%%%%%%%%%%%%%%%%%%%%%%%%%%%%%%%%%%%%%%%%%%%%%%%%%%%%%%%%%%
\maketitle
%%%%%%%%%%%%%%%%%%%%%%%%%%%%%%%%%%%%%%%%%%%%%%%%%%%%%%%%%%%%%%%%%%%%%%%%%%%%%%%%%%%%%
%\tableofcontents
%%%%%%%%%%%%%%%%%%%%%%%%%%%%%%%%%%%%%%%%%%%%%%%%%%%%%%%%%%%%%%%%%%%%%%%%%%%%%%%%%%%%%
%%%%%%%%%%%%%%%%%%%%%%%%< Section 1 >%%%%%%%%%%%%%%%%%%%%%%%%%%%%%%%%%%%%%%%%%%%%%%%%
\section{Introduction}
\label{sec:1}

%%%%%%%%%%%%%%%%%%%%%%%%%%< Paragraph 1: Landau theory >%%%%%%%%%%%%%%%%%%%%%%%%%%%%
The Landau Fermi-liquid theory (FLT)~\cite{Landau_1,Landau_2} is very useful 
in describing low-temperature physics in ordinary metals through the concept 
of quasiparticle (QP). This key concept is verified to infinite order in perturbation 
expansion in the quantum field theory~\cite{GM_1958,Landau_3,AGD_1963,Nozieres_1964}. 
It is also confirmed in both the renormalization-group approach~\cite{Shankar_1994} 
and the multidimensional bosonization~\cite{Houghton_1994,Kwon_1995,Houghton_2000}.

%%%%%%%%%%%%%%%%%< Paragraph 2: Breakdown of the Landau theory >%%%%%%%%%%%%%%%%%%%
In these several decades, FLT is found to break down in a number of exotic metals 
broadly referred to as non-Fermi liquids (NFLs), including the one-dimensional 
(1D) Luttinger liquids~\cite{Mattis_1965,Dzyaloshinskii_1974,Everts_1974,Haldane_1981,
Voit_1994}, the strange-metal phase in high-$T_c$ cuprates~\cite{Anderson_2004,
Lee_2006,Anderson_2009,Keimer_2015,Phillips_2016,Varma_2020}, and the fluctuation 
regime around a quantum critical point (QCP)~\cite{Gan_1993,Sachdev_1999,Abanov_2003,
Lohneysen_2007,Mross_2010,Fitzpatrick_2013,Sur_2014,Torroba_2014,Schlief_2018,Trott_2018,
Chowdhury_2018a,Xu_2020,Driskell_2021,Shi_2022}. 

%%%%%%%%%%%%%%%%%< Paragraph 3: Route along Gauge theory >%%%%%%%%%%%%%%%%%%%
It has been quite a challenge to fully understand the routes to NFLs from normal 
Fermi liquids~\cite{Metzner_1998,Castellani_1999,Varma_2002,Wolfle_2018,Su_2018,
Lee_2018,Else_2021}. In 3D simple metals, we can envisage a couple of routes to NFLs. 
The first one is related to the long-range nature of the Coulomb interaction 
$V({\bm q})$($=\!4\pi e^2/{\bm q}^2$). In fact, this has been investigated in detail 
in the past~\cite{Bares_1993,Nayak_1994,Bartosch_1999} and it is concluded that 
$V({\bm q})$ at $|{\bm q}|\! \to \!0$ is not singular enough to break FLT in 3D metals 
due to the screening effect. Incidentally, transverse electromagnetic fields give 
rise to unscreened long-range interactions, leading to the breakdown of FLT even 
in alkali metals~\cite{Holstein_1973,Polchinski_1994,Mandal_2020}, but this 
occurs only at unrealistically low temperatures because of its very weak coupling 
controlled by the small factor $(k_{\rm F}/mc)^2$, where $k_{\rm F}$ is the Fermi 
momentum, $m$ is the mass of a free electron, and $c$ is the velocity of light. 
Hereafter we employ units in which $\hbar\!=\!k_{\rm B}\!=\!1$.  

%%%%%%%%%%%%%%%%%%%%%%%%%< Paragraph 4: 2k_F anomaly >%%%%%%%%%%%%%%%%%%%%%%%%%%%%%%
The second route is concerned with the response function at $|{\bm q}|\! \approx \!
2k_{\rm F}$ and possible $2k_{\rm F}$ singularities in it~\cite{Khveshchenko_1994,
Altshuler_1994,Kwon_1995,Perali_1996,Bartosch_1999,Houghton_2000,Chubukov_2003,
Mross_2010,Chubukov_2012}. This is a problem which has not been examined in detail, 
especially in the context of 3D simple metals. The density response at low energies 
and short wavelengths will be an essential ingredient in the present study, 
as we now explain.

%%%%%%%%%%%%%< Paragraph 5: excitonic collective mode vs CDW >%%%%%%%%%%%%%%%%%%%%%
The low-lying excited states in simple metals are well described by those of 
a 3D homogeneous electron gas, an assembly of $N$ electrons embedded in a uniform 
positive rigid background. In a recent paper~\cite{Takada_2016}, it is shown 
in the low-density electron gas that an excitonic collective mode $\omega_{\rm ex}(q)$ 
made of correlated electron-hole pair excitations appears as a soft mode for 
$q\! \approx \!2k_{\rm F}$. If this $\omega_{\rm ex}(q)$ vanishes at $q\!=\!q_c\! 
\approx \! 2k_{\rm F}$, then a macroscopic number of electron-hole pairs are produced 
spontaneously to form a CDW state with the wave number $q_c$ which is 
exactly the state predicted by Overhauser~\cite{Overhauser_1978}. 
In actual simple metals for which the density parameter $r_s$ defined by 
\begin{align}
\label{eq:02}
r_s=\frac{1}{\alpha k_{\rm F} a_{\rm B}} \ {\rm with}\ 
\alpha=\left (\frac{4}{9\pi} \right )
^{1/3} \approx 0.5211
\end{align}
and $a_{\rm B}$ the Bohr radius is in the range from 2 to 6, 
this kind of CDW does not seem to exist, but from the perspective of QCP physics, 
the correlated electron-hole pair excitations around $2k_{\rm F}$ may be regarded 
as quantum fluctuations around the quantum critical CDW transition point. In this 
regard, those bosonic excitations in simple metals can be considered 
as marginally relevant processes in the sense of renormalization group to break FLT.

%%%%%%%%%%%%%< Paragraph 6: excitonic collective mode vs CDW >%%%%%%%%%%%%%%%%%%%%%
In fact, the above-mentioned $\omega_{\rm ex}(q)$ mode is confirmed as an incipient 
excitonic mode in the dynamic structure factor $S({\bf q},\omega)$ at $|{\bm q}|\!  
\approx \! 2k_{\rm F}$ by recent {\it ab initio} path integral Monte Carlo simulations 
performed in the electron gas for $r_s\!=\!2$-$10$~\cite{Dornheim_2018,Dornheim_2022,
Dornheim_2023} and $r_s\!=\!2$-$36$~\cite{Filinov_2023}, where the mode is referred to 
as a ``roton''. 

%%%%%%%%%%%%%%%%%%%%< Paragraph 7: choice of theoretical framework >%%%%%%%%%%%%%% 
In this paper, by exploiting the precise polarization function in the charge channel 
derived from Monte Carlo data for the density response, we will accurately calculate 
the self-energy $\Sigma(K)$ with $K \! \equiv \! \{{\bm k},i\omega_n\}$, 
a combined notation of momentum ${\bm k}$ and fermion Matsubara frequency $\omega_n$, 
in the 3D homogeneous electron gas for $r_s\!<\!6$ with the motivation 
to inspect the validity of FLT in simple metals. For this inspection,  
we are not allowed to adopt a theory based on any kinds of perturbation expansion, 
because NFL cannot be described by perturbation expansion starting 
from the noninteracting Green's function $G_0(K)$; actually, the so-called $GW$ 
approximation to the Hedin's closed set of equations, derived by using the screened 
interaction $W$ as an expansion parameter~\cite{Hedin_1965,Onida_2002,Kutepov_2009,
Houcke_2017,Reining_2018,Vacondio_2022}, and its refinements~\cite{Ren_2015,
Kutepov_2016,Wang_2021,Lorin_2023} have not shown any indication of the breakdown 
of FLT. The same is true both in the many-body perturbation theory with using 
some appropriate local-field factor~\cite{Rice_1965,Bueche_1990,YT_1993,
Richardson_1996,Simion_2008,Kutepov_2017,Cai_2022} and in the effective-potential 
expansion (EPX) method~\cite{YT_1991a,YT_1991b}. 

%%%%%%%%%%%%%%%%%%%%< Paragraph 8: GW\Gamma >%%%%%%%%%%%%%%%%%%%%%%%%%%%
Considering this situation, we shall employ the GW$\Gamma$ scheme~\cite{Maebashi_2011} 
which is improved on the original one~\cite{YT_2001} and satisfies all the 
known conservation laws and sum rules, including the Ward identity (WI)~\cite{Ward_1950}, 
total-energy and total-momentum conservation laws, three sum rules concerning 
the momentum distribution function $n({\bm k})$~\cite{Takada_2016}, and correct 
asymptotics. This scheme is an intrinsically nonperturbative approach 
applicable to both Fermi and Luttinger liquids in a unified manner~\cite{Maebashi_2014}, 
but so far, it was too complicated to be implemented in 3D systems. In particular, 
the total-momentum conservation law (and thus the backflow effect) was not 
well respected, so that the QP effective mass $m^*$ could not 
be reliably determined. Therefore, one of the aims in this paper is to develop a 
feasible code to implement this advanced scheme with keeping the 
total-momentum conservation law.

%%%%%%%%%%%%%%%%%%%%< Paragraph 9: implementation >%%%%%%%%%%%%%%%%%%%%%%%%%%%
By applying this newly developed code to the electron gas at $r_s\!=\!2,07$, $3.25$, 
$3.93$, $4.86$, $5.20$, and $5.62$ relevant to Al, Li, Na, K, Rb, and Cs, respectively, 
we successfully obtain a convergent result of $\Sigma(K)$ at each $r_s$ and various 
values of $T$ down to $10^{-4} \varepsilon_{\rm F}$, where $\varepsilon_{\rm F}$ 
is the Fermi energy. For $T \lesssim 10^{-3}\varepsilon_{\rm F}$, 
the obtained $\Sigma(K)$ is not smooth enough at $|{\bm k}|\!\approx\! k_{\rm F}$ 
and $\omega_{k}\! \approx \!0$ to safely confirm FLT; by a heuristic analysis, 
this anomalous behavior is found to be well described by a branch-cut singularity, 
exhibiting a symptom of possible breakdown of FLT. 

%%%%%%%%%%%%%%%%%%%%%%%%%%%%%%%%%%%%< Figure 02 >%%%%%%%%%%%%%%%%%%%%%%%%%%%%%%%%%%%%
\begin{figure}[t]
\begin{center}
\includegraphics[scale=0.41
,keepaspectratio]{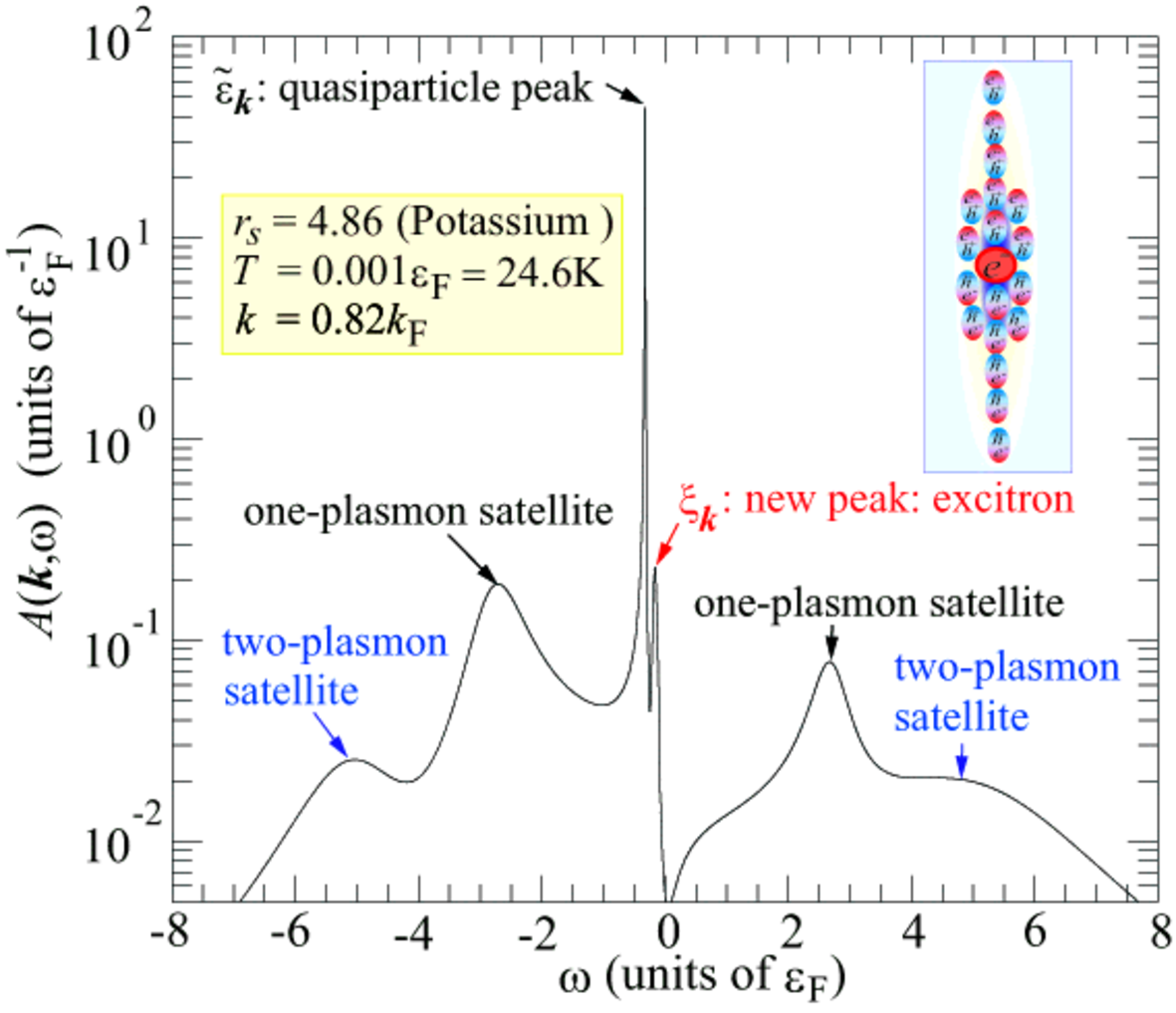}
\end{center}
\caption[Fig.02] {One-particle spectral function $A({\bm k},\omega)$ for 
the 3D homogeneous electron gas at $r_s\!=\!4.86$ corresponding to potassium, 
$T\!=\!0.001\varepsilon_{\rm F}$, and $|{\bm k}|\!=\!0.82k_{\rm F}$.}
\label{fig:02}
\end{figure}
%%----------------------------------------------------------------------------------%

%%%%%%%%%%%%%%%%%%%%< Paragraph 10: excitron >%%%%%%%%%%%%%%%%%%%%%%%%%%%
This situation is more clearly seen in the one-particle spectral function 
$A({\bm k},\omega)\! \equiv \!-{\rm Im} G^R({\bm k},\omega\!+\!i\gamma)/\pi$ with 
$\gamma\!=\!\pi T$ and $G^R({\bm k},\omega)$ the retarded Green's function obtained 
by an analytic continuation of $G(K)$ in complex $\omega$ plane through Pad\`{e} 
approximants. A typical example of $A({\bm k},\omega)$, a quantity to be observed 
by angle-resolved photoemission spectroscopy (ARPES), is shown in Fig.~\ref{fig:02} 
in which a new peak indicated by ``excitron'' appears in addition to the predominant 
QP peak and one- and two-plasmon satellites. It is evident from this figure that 
an energy resolution in ARPES should be of the order of 1 meV or less to 
experimentally distinguish this weak excitron peak from the very strong QP peak. 

%%%%%%%%%%%%%%%%%%%%< Paragraph 11: m^* & z^* >%%%%%%%%%%%%%%%%%%%%%%%%%%%
Since the excitron spectral strength is quite weak and $G(K)$ is dominated 
by the QP pole singularity, bulk properties of the electron gas at metallic 
densities will be well explained by FLT in which the quantitative evaluation of 
$m^*$ is a long-standing yet hot issue. We can determine $m^*$ from the QP peak 
position at $|{\bm k}|$ in the vicinity of $k_{\rm F}$; our values for $m^*$ are 
in good agreement with those given in recent quantum Monte Carlo (QMC) 
simulations~\cite{Haule_2022,Holzmann_2023}. The same is true for the QP 
renormalization factor $z^*$; our values for $z^*$ determined by the jump of 
$n({\bm k})$ at $|{\bm k}|\!=\!k_{\rm F}$ agree well with those in recent 
QMC simulations~\cite{Holzmann_2011,Hiraoka_2020} 
and available experiments~\cite{Hiraoka_2020,Suortti_2000,Huotari_2010}.

%%%%%%%%%%%%%%%%%%%%< Paragraph 12: Contents of this manuscript  >%%%%%%%%%%%%%%%%%%%
This paper is organized as follows: In Sec.~\ref{sec:2} we explain our framework 
to calculate $\Sigma(K)$ in the electron gas through a nonperturbative 
iteration loop with rigorously satisfying the Ward identity and the total-momentum 
conservation law. A proposed functional form for the vertex function 
$\Gamma(K,K\!+\!Q)$ in Sec.~\ref{sec:2E} is one of our main results. 
In Sec.~\ref{sec:3} we give the calculated results on various properties in the 
electron gas, including a new low-energy peak dubbed excitron and 
two-plasmon satellites in $A({\bm k},\omega)$. Although we treat the electron 
densities corresponding to all simple metals, the main stress is laid on Na, 
because Na is known to be an almost perfect realization of the electron gas 
with $r_s=3.93$. In Sec.~\ref{sec:4} we make a detailed account of the excitron, 
ascribing it to the effect of excitonic attraction arising from multiple excitations 
of tightly bound electron-hole pairs. Finally in Sec.~\ref{sec:5} we give a summary 
of this paper and discuss related and future issues. Appendices 
A$-$E provide some details of our numerical calculation.

%%%%%%%%%%%%%%%%%%%%%%%%%%%%%%%%%%%%%%%%%%%%%%%%%%%%%%%%%%%%%%%%%%%%%%%%%%%%%%%%%%%%%
%%%%%%%%%%%%%%%%%%%%%%%%%%< Section 2 >%%%%%%%%%%%%%%%%%%%%%%%%%%%%%%%%%%%%%%%%%%%%%%
\section{Self-Energy Calculation Scheme}
\label{sec:2}
%%%%%%%%%%%%%%%%%%%%%%%%%%< Section 2-1 >%%%%%%%%%%%%%%%%%%%%%%%%%%%%%%%%%%%%%%%%%%%%
\subsection{Hamiltonian}
\label{sec:2A}
%%%%%%%%%%%%%%%%%%%%%%%%%%%%< Paragraph 13: Hamiltonian  >%%%%%%%%%%%%%%%%%%%%%%%%%%%
The Hamiltonian $H$ for a 3D homogeneous electron gas is written 
in second quantization as
\begin{align}
H=&\sum_{{\bm k}\sigma}\varepsilon_{{\bm k}}c_{{\bm k}\sigma}^{+}c_{{\bm k}\sigma}
\nonumber \\
&+\frac{1}{2}\sum_{{\bm q}\neq {\bm 0}}\sum_{{\bm k}\sigma}\sum_{{\bm k'}\sigma'}
V({\bm q})c_{{\bm k}+{\bm q}\sigma}^{+}
c_{{\bm k'}-{\bm q}\sigma'}^{+}c_{{\bm k'}\sigma'}c_{{\bm k}\sigma}
\label{eq:05},
\end{align}
where $c_{{\bm k}\sigma}$ is the annihilation operator of an electron with 
momentum ${\bm k}$ and spin $\sigma$ whose single-particle energy is given by 
$\varepsilon_{{\bm k}}\! \equiv \!{\bm k}^2/(2m)\!-\!\varepsilon_{\rm F}$ with 
$\varepsilon_{\rm F}\!=\!k_{\rm F}^2/(2m)$. We will consider the system in unit 
volume and the total number of electrons $N$ is nothing but the electron density 
$n$, given in terms of the Fermi momentum $k_{\rm F}$ as $n\!=\!k_{\rm F}^3/(3\pi^2)$. 

 %%%%%%%%%%%%%%%%%%%< Paragraph 14: Thermodynamic relations  >%%%%%%%%%%%%%%%%%%%%%%
In this system, the correlation energy per electron $\varepsilon_c$ is already 
known accurately as a function of $r_s$ by the 
Green's-Function Monte Carlo (GFMC) method~\cite{CA1980} and the interpolation 
formulas to reproduce the GFMC data~\cite{VWN1980,PW1992}. With the use of 
$\varepsilon_c(r_s)$, $\mu_c$ the correlation contribution to the chemical potential 
$\mu$ and the compressibility $\kappa$ are, respectively, obtained as
\begin{align}
\frac{\mu_c}{\varepsilon_{\rm F}}&=\alpha^2 r_s^2 \left ( \varepsilon_c-\frac{r_s}{3}
\varepsilon' \right ),
\label{eq:06} \\
\frac{\kappa_{\rm F}}{\kappa}&=\frac{d\,\mu}{d\,\varepsilon_{\rm F}}=
1-\frac{\alpha r_s}{\pi}
+\frac{\alpha^2 r_s^3}{6}\left(r_s \varepsilon_c''-2\varepsilon_c' \right ),
\label{eq:07}
\end{align}
where $\kappa$ is given through the thermodynamic relation of $\kappa\!=\!
(d\,n/d\,\mu)/n^2$ and $\kappa_{\rm F}$ is the compressibility in the noninteracting 
electron gas, given by $\kappa_{\rm F}\!=\!D_{\rm F}/n^2\!=\!3m/(k_{\rm F}^2n)$ 
with $D_{\rm F}\!=\!d\,n/d\,\varepsilon_{\rm F}\!=\! mk_{\rm F}/\pi^2$ the density 
of states at the Fermi level in the noninteracting system. 

%%%%%%%%%%%%%%%%%%%%%%%%%%< Section 2-2 >%%%%%%%%%%%%%%%%%%%%%%%%%%%%%%%%%%%%%%%%%%%%
\subsection{One-particle Green's function}
\label{sec:2B}
%%%%%%%%%%%%%%%%%%%%%%%%< Paragraph 15: G(K) & \Sigma(K)  >%%%%%%%%%%%%%%%%%%%%%%%%%
The Dyson equation relates the one-particle Green's function $G(K)$ 
with the self-energy $\Sigma(K)$ through 
\begin{align}
G^{-1}(K) 
\!=\! i\omega_n \!+\!\mu_x\!+\!\mu_c\!-\!\varepsilon_{{\bm k}} - \Sigma(K). 
\label{eq:08}
\end{align} 
Here, $\mu$ is divided into $\mu \!=\! \varepsilon_{\rm F}+\mu_x+\mu_c$ 
with  $\mu_x$ the exchange contribution to $\mu$, given by
\begin{align}
\frac{\mu_x}{\varepsilon_{\rm F}}=-\frac{2\alpha r_s}{\pi}.
\label{eq:09}
\end{align}
Let us divide $\Sigma(K)$ into odd and even parts in $\omega_n$ as
\begin{align}
 \Sigma(K)=\left [1-Z(K)\right ] i\omega_n\!+\!\chi(K),
\label{eq:10}
\end{align}
where both $Z(K)$ and $\chi(K)$ are not only even functions in $\omega_n$ but also 
real functions, as seen by combining Eq.~(\ref{eq:10}) with the analyticity property 
$\Sigma({\bm k},-i\omega_n)=\Sigma^*({\bm k},i\omega_n)$. Then, $G^{-1}(K)$ is rewritten as 
\begin{align}
G^{-1}(K) 
=Z(K)i\omega_n-E(K),
\label{eq:11}
\end{align} 
with the introduction of $E(K)\!\equiv \! {\bm k}^2/(2m)\!+\!\chi(K)\!-\!\mu$. 

%%%%%%%%%%%%%%%%%%%%%%%%%%< Section 2-3 >%%%%%%%%%%%%%%%%%%%%%%%%%%%%%%%%%%%%%%%%%%%%
\subsection{Momentum distribution function}
\label{sec:2C}
%%%%%%%%%%%%%%%%%%%%%%%%%%%%< Paragraph 16: n(k)  >%%%%%%%%%%%%%%%%%%%%%%%%%%%
Once $G(K)$ is known, the momentum distribution function $n({\bm k})\ (= \langle 
c_{{\bm k}\sigma}^{+}c_{{\bm k}\sigma} \rangle )$ is calculated as 
\begin{align}
n({\bm k})=T\sum_{\omega_n}G(K) e^{-i\omega_n 0^+},
\label{eq:12}
\end{align}
where the Matsubara sum is taken by the procedure explained in Appendix A. 
Accuracy of $n({\bm k})$ may be checked by the three sum rules related to 
total electron number, total kinetic energy, and total kinetic-energy 
fluctuation, as derived in Ref.~\cite{Takada_2016}. Those sum rules are conveniently 
expressed in terms of the $n$th-power integral $I_n$, given as
\begin{align}
I_n \equiv \int_0^{\infty}dx \,n(x)x^n,
\label{eq:13}
\end{align} 
with $x\!=\!|{\bm k}|/k_{\rm F}$ and $n(x)\!=\!n({\bm k})$. The rigorous values for $I_n$ 
with $n\!=\!2$, $4$, and $6$ are: $I_2\!=\!1/3$, $I_4\!=\!1/5+\alpha^2r_s^2(-\varepsilon_c
\!-\!r_s\varepsilon')/3$, and $I_6\!=\!8/105\!+\!5I_4^2/3\!+\!5\alpha r_s[B(r_s)
\!-\!2/3\!+\!2g(0)/3]/(9\pi)$ with $g(0)$ the on-top value of the pair distribution 
function and $B(r_s)$ specified in Eq.~(51) in Ref.~\cite{Takada_2016}. 
After several tentative calculations, we find that $n({\bm k})$ as obtained from 
Eq.~(\ref{eq:12}) satisfies the $I_2$ sum rule up to five digits or more and 
$I_4$ up to three digits, but $I_6$ up to only a single digit in most cases, 
indicating that $n({\bm k})$ for $|{\bm k}|\! \gtrsim \!2k_{\rm F}$ is not accurate 
enough, which reflects the fact that the $k$-mesh (or grid) in ${\bm k}$ space 
in the iterative calculation of $G(K)$ is not dense enough for 
$|{\bm k}|\! \gtrsim \!2k_{\rm F}$. 

%%%%%%%%%%%%%%%%%%%%%%%%%%%%< Paragraph 17: n_c(k)  >%%%%%%%%%%%%%%%%%%%%%%%%%%%
As a remedy for this problem in $n({\bm k})$, we have modified $n({\bm k})\ [=\!n(x)]$ 
to $n_c(x)$ by the procedure explained in Appendix B in which the behavior in the 
region of $x\! \gtrsim \! 2$ is rectified by the introduction of $n_{\rm IGZ}(x)$ 
obtained by the parametrization scheme described in Ref.~\cite{Takada_2016}. 
In actual calculations, $n_c(x)$ always satisfies all the three sum rules up to 
at least five digits (and mostly seven or eight digits). As an example, the results 
of $n({\bm k})$, $n_c(x)$, and $n_{\rm IGZ}(x)$ are shown for $r_s=3.93$ in 
Fig.~\ref{fig:11}(a) in which we can barely see the difference among those three 
functions on the scale of the figure.
 
%%%%%%%%%%%%%%%%%%%%%%%%%%< Section 2-4 >%%%%%%%%%%%%%%%%%%%%%%%%%%%%%%%%%%%%%%%%%%%%
\subsection{Polarization function}
\label{sec:2D}
%%%%%%%%%%%%%%%%%%%%%< Paragraph 18: definition of Pi(Q)  >%%%%%%%%%%%%%%%%%%%%%%%%%
The charge response function $Q_c(Q)$ is related to the polarization function 
in the charge channel $\Pi(Q)$ through 
\begin{align}
\label{eq:14-a}
Q_c(Q)=-\frac{\Pi(Q)}{1+V({\bm q})\Pi(Q)},
\end{align}
and the formal definition of $\Pi(Q)$ is written as 
\begin{align}
\label{eq:14-b}
\Pi(Q)&=-2\sum_K G(K)G(K\!+\!Q)\Gamma(K,K\!+\!Q)
\nonumber \\
&\equiv -2\,T\sum_{\omega_n}\sum_{{\bm k}}G(K)G(K\!+\! Q)\Gamma(K,K\!+\!Q), 
\end{align}
where $\Gamma(K,K\!+\!Q)$ is the three-point vertex function in the charge channel.

%%%%%%%%%%%%%%%%%%%%%< Paragraph 19: accurate Pi(Q)  >%%%%%%%%%%%%%%%%%%%%%%%%%
In the many-body problem, it is often the case that $\Pi(Q)$ is less 
singular than $G(K)$. In fact, in 1D Luttinger liquids, $\Pi(Q)$ is easily 
obtained and does not exhibit any NFL-related singularity~\cite{Dzyaloshinskii_1974}. 
In 3D Fermi liquids, it is shown that no nonanalytic corrections are contained 
in $\Pi(Q)$, in sharp contrast to the polarization function in the spin channel 
in which a nonanalytic ${\bm q}^2\ln|{\bm q}|$-term exists~\cite{Chubukov_2003}. 
Thus we can expect that in the 3D electron gas, even if there were a branch-cut 
singularity in $G(K)$ at low $T$, it would not induce any singular effects 
on $\Pi(Q)$, implying that $\Pi(Q)$ will be reliably determined even at $T$ 
of the order of $0.1\varepsilon_{\rm F}$. On this assumption, we consider 
$Q_c({\bm q},0)$ obtained by Monte Carlo simulations~\cite{Moroni_1995,
Groth_2019,Haule_2019,Bonitz_2020,Kukkonen_2021,LeBlanc_2022,Dornheim_2023a} 
as sufficiently accurate data, based on which various parametrized forms 
for the conventional charge local-field factor $G_{+}(Q)$ have been 
proposed~\cite{Richardson_1994,Corradini_1998,Qian_2002,Ruzsinszky_2020,
Nepal_2020,Kaplan_2022,Kaplan_2023}.

%%%%%%%%%%%%%%%%%%%%%%%%< Paragraph 20: G_s(Q)  >%%%%%%%%%%%%%%%%%%%%%%%%%
In view of this situation, we regard $\Pi(Q)$ as a quantity precisely known 
from the outset in the self-consistent iteration loop. In actual calculations, 
$\Pi(Q)$ is given either with $G_+(Q)$ as
\begin{align}
\label{eq:14}
\Pi(Q)=\frac{\Pi_{0}(Q)}{1-V({\bm q})G_+(Q)\Pi_{0}(Q)},
\end{align}
or with $G_s(Q)$ due to Richardson and Ashcroft~\cite{Richardson_1994} as
\begin{align}
\label{eq:15}
\Pi(Q)=\frac{\Pi_{\rm WI}(Q)}{1-V({\bm q})G_s(Q)\Pi_{\rm WI}(Q)},
\end{align}
where the Lindhard function $\Pi_0(Q)$ is calculated as~\cite{Lindhard_1954} 
\begin{align}
\label{eq:04}
\Pi_0(Q)&=-2\sum_K G_0(K) G_0(K\!+\! Q)  
\nonumber \\
&=4\int \frac{d^3k}{(2\pi)^3} \ n_0({\bm k})
\frac{\varepsilon_{{\bm k}+{\bm q}}\!-\!\varepsilon_{\bm k}}
{\omega_q^2\!+\!(\varepsilon_{{\bm k}+{\bm q}}\!-\!\varepsilon_{\bm k})^2},
\end{align}
with $n_0({\bm k})\!=\!\theta(k_{\rm F}\!-\!|{\bm k}|)$ the step function 
and $\Pi_{\rm WI}(Q)$ is given by
\begin{align}
\label{eq:16}
\Pi_{\rm WI}(Q)=
4\int \frac{d^3k}{(2\pi)^3} \ n({\bm k})
\frac{\varepsilon_{{\bm k}+{\bm q}}\!-\!\varepsilon_{\bm k}}
{\omega_q^2\!+\!(\varepsilon_{{\bm k}+{\bm q}}\!-\!\varepsilon_{\bm k})^2}.
\end{align}
Note that $G_s(Q)$ is obtained from $G_+(Q)$ as 
\begin{align}
\label{eq:17}
G_s(Q)=G_+(Q)+\frac{1}{V({\bm q})\Pi_{0}(Q)}
\frac{\Pi_{0}(Q)\!-\!\Pi_{\rm WI}(Q)}{\Pi_{\rm WI}(Q)},
\end{align}
but in this paper, we shall adopt the function form (a slightly modified 
Richardson-Ashcroft parametrization) prescribed in Eq.~(58) 
in Ref.~\cite{Takada_2016} for $G_s(Q)$.

%%%%%%%%%%%%%%%%%%%%%%%%< Paragraph 21: Pi_WI(Q)  >%%%%%%%%%%%%%%%%%%%%%%%%%
In Eq.~(\ref{eq:16}), we employ $n_c(x)$ for $n({\bm k})$ to make 
$\Pi_{\rm WI}({\bm q},0)$ correctly behave in the limit of $|{\bm q}| \to \infty$. 
On the other hand, the behaviors of $\Pi_0(Q)$ and $\Pi_{\rm WI}(Q)$ 
in the limit of $Q \to Q_0\! \equiv \! \{{\bm 0},0\}$ can be derived directly 
from Eqs.~(\ref{eq:04}) and (\ref{eq:16}), respectively; in the $\omega$-limit 
(i.e., ${\bm q} \to {\bm 0}$ first, and then $\omega_q \to 0$), we obtain
\begin{align}
\label{eq:18}
\Pi_{0}(Q)=\Pi_{\rm WI}(Q)=\frac{D_{\rm F}}{3}
\frac{v_{\rm F}^2{\bm q}^2}{\omega_q^2},
\end{align}
with $v_{\rm F}=k_{\rm F}/m$ and in the $q$-limit (i.e., $\omega_q \to 0$ 
first, and then ${\bm q} \to {\bm 0}$), we obtain
\begin{align}
\label{eq:19}
\Pi_{0}(Q)=D_{\rm F},\ \ {\rm and} \ \ \Pi_{\rm WI}(Q)=D_{\rm F}I_0,
\end{align}
where $I_0$ is defined in Eq.~(\ref{eq:13}) with $n\!=\!0$.
Combining the above-mentioned behavior of $\Pi_{0}(Q)$ with the constraints 
imposed on $G_+(Q)$ (or that of $\Pi_{\rm WI}(Q)$ with those on $G_s(Q)$) at 
$Q \to Q_0$, we see that in the $\omega$-limit, 
\begin{align}
\label{eq:20}
\Pi(Q)=\frac{D_{\rm F}}{3}\frac{v_{\rm F}^2{\bm q}^2}{\omega_q^2}
=\frac{n{\bm q}^2}{m\omega_q^2},
\end{align}
and in the $q$-limit, 
\begin{align}
\label{eq:21}
\Pi(Q)=D_{\rm F}\frac{\kappa}{\kappa_{\rm F}}
=\frac{d\, n}{d\, \varepsilon_{\rm F}}
\frac{d\,\varepsilon_{\rm F}}{d\,\mu}
=\frac{d\, n}{d\, \mu}.
\end{align}
The relations in Eqs.~(\ref{eq:20}) and (\ref{eq:21}) are, respectively, 
known as the f-sum rule and the compressibility sum rule.

%%%%%%%%%%%%%%%%%%%%%%%%%%< Section 2-5 >%%%%%%%%%%%%%%%%%%%%%%%%%%%%%%%%%%%%%%%%%%%%
\subsection{Vertex function}
\label{sec:2E}
%%%%%%%%%%%%%%%%%%%%%%%%< Paragraph 22: \Gamma(K,K+Q)  >%%%%%%%%%%%%%%%%%%%%%%%%%
Formally, we can calculate $\Sigma(K)$ rigorously by 
\begin{align}
\label{eq:22}
\Sigma(K)\!=\! -\sum_{Q} W(Q) G (K\!+\!Q)\Gamma(K,K\!+\!Q),
\end{align}
where $W(Q)$ is the effective interaction, given by 
\begin{align}
\label{eq:23}
W(Q)\!=\! \frac{V({\bm q})}{1\!+\!V({\bm q})\Pi(Q)}.
\end{align}
Since we regard $\Pi(Q)$ as a precisely known quantity, $W(Q)$ is already known. 
As for $\Gamma(K,K\!+\!Q)$, we adopt the improved $GW\Gamma$ scheme described in 
Ref.~\cite{Maebashi_2011}. According to Eqs.~(54)-(58) in Ref.~\cite{Maebashi_2011}, 
$\Gamma (K,K\!+\!Q)$ is given in the product of two components as 
\begin{align}
\Gamma(K,K\!+\!Q) = \widetilde{\Gamma}_{\rm WI}(K,K\!+\!Q) 
\Gamma_{\Pi}(K,K\!+\!Q),
\label{eq:24}
\end{align}
with
\begin{subequations}
\label{eq:25ab}
\begin{align}
&\widetilde{\Gamma}_{\rm WI}(K,K\!+\!Q)\!=\!\frac{
G^{-1}(K\!+\!Q)\! - \! G^{-1}(K)} 
{i \omega_q  \!-\!  (\varepsilon_{{\bm k} \!+\! {\bm q}}
 \!-\! \varepsilon_{{\bm k}}) 
{\tilde \eta}_{1} (K\!+\!Q/2)} ,
\label{eq:25a}
\\
&\Gamma_{\Pi}(K,K\!+\!Q) \!=\! 
\frac{\Pi(Q)}{\widetilde{\Pi}_{\rm WI}(Q)}
\nonumber \\
& \hspace{2.1cm}  
+ \frac{\Pi(Q)}{D_{\rm F}} 
\Biggl \{ \frac{3i\omega_q}{v_{\rm F}^2{\bm q}^2} 
\biggl [ -i\omega_q \left ( 1\!-\!\frac{\Pi_1(Q)}{\widetilde{\Pi}_{\rm WI}(Q)}
\right ) 
\nonumber \\
& \hspace{2.5cm}  
- \left (\varepsilon_{{\bm k}+{\bm q}}-\varepsilon_{\bm k} \right )
\Bigl (1-{\tilde \eta}_{1} (K\!+\!Q/2)\Bigr  ) \biggr ]
\nonumber \\
& \hspace{2.5cm}  
+{\tilde \eta}_{2} (K;Q)-\frac{\Pi_2(Q)}{\widetilde{\Pi}_{\rm WI}(Q)}
\Biggr \},
\label{eq:25b}
\end{align}
\end{subequations}
where the functional ${\tilde \eta}_{1} (K)$ is introduced as
\begin{align}
{\tilde \eta}_{1} (K) &\equiv  
- \frac{\partial G^{-1} (K)}{\partial \varepsilon_{\bm k}} 
\bigg/ \frac{\partial G^{-1} (K)}{\partial (i \omega_n)},
\label{eq:26}
\end{align}
and three ``modified polarization functionals'', $\widetilde{\Pi}_{\rm WI}(Q)$, 
$\Pi_1(Q)$, and $\Pi_2(Q)$, are, respectively, defined as  
\begin{subequations}
\label{eq:27abc}
\begin{align}
\widetilde{\Pi}_{\rm WI}(Q) & \!\equiv\! 2 \! \sum_{K} 
\frac{
G(K\!+\!Q) \! - \! G(K)} 
{i \omega_q  \!-\!  (\varepsilon_{{\bm k} \!+\! {\bm q}} 
\!-\! \varepsilon_{{\bm k}} ) {\tilde \eta}_{1} (K\!+\!Q/2)},
\label{eq:27a}
\\
\Pi_1(Q) & \!\equiv\! 2 \! \sum_{K} 
\frac{
[G(K\!+\!Q) \! - \! G(K)] (\varepsilon_{{\bm k} \!+\! {\bm q}} 
\!-\! \varepsilon_{{\bm k}})
}{[i \omega_q\!-\!(\varepsilon_{{\bm k} \!+\! {\bm q}} \!-\! \varepsilon_{{\bm k}} ) 
{\tilde \eta}_{1} (K\!+\!Q/2)] i \omega_q},
\label{eq:27b}
\\
\Pi_2(Q) & \!\equiv\! 2 \! \sum_{K} 
\frac{
[G(K\!+\!Q) \! - \! G(K)] {\tilde \eta}_{2} (K;Q)}
{i \omega_q \!-\!(\varepsilon_{{\bm k} \!+\! {\bm q}} \!-\! \varepsilon_{{\bm k}}) 
{\tilde \eta}_{1} (K\!+\!Q/2)}.
\label{eq:27c}
\end{align}
\end{subequations}
The functional ${\tilde \eta}_{2} (K;Q)$ will be specified later. 

%%%%%%%%%%%%%%%%< Paragraph 23: background of this derivation  >%%%%%%%%%%%%%%%%%%%%%%
In Ref.~\cite{Maebashi_2011}, this functional form for $\Gamma (K,K\!+\!Q)$ was 
derived from the perspective of FLT, i.e., with the assumption that $G(K)$ 
is given in such a form as
\begin{align}
\label{eq:01}
G(K)\!=\!\frac{z_{\bm k}}{i\omega_n\!-\!\tilde{\varepsilon}_{\bm k}
\!+\!i\ {\rm sgn}(\omega_n)(2\tau_{\bm k})^{-1}}\!+\!G_{\rm incoh}(K),
\end{align}
for $|{\bm k}|\! \approx \!k_{\rm F}$ and $|\omega_n| \!\ll \! \varepsilon_{\rm F}$,
where $z_{\bm k}$, $\tilde{\varepsilon}_{\bm k}$, and $\tau_{\bm k}$ are, 
respectively, the QP renormalization factor, the QP dispersion, and the QP lifetime. 
Here, $|\tilde{\varepsilon}_{\bm k}|\tau_{\bm k}\!\gg \! 1$ is assumed and 
$G_{\rm incoh}(K)$ is a smooth function, corresponding to the incoherent 
smooth background in $A({\bm k},\omega)$.
Note that QP appears as a pole in $G(K)$ as long as $z_{\bm k}\!\neq \!0$ and 
this pole-singularity is intimately connected with the condition that $\Sigma(K)$ 
is smooth enough to be analytically expanded around the Fermi point 
$K_{\rm F}\! \equiv \! \{{\bm k}_{\rm F},0 \}$.
 
%%%%%%%%%%%%%%%%%%%%%%%%< Paragraph 24: Justification  >%%%%%%%%%%%%%%%%%%%%%%%%%
The final result in Eq.~(\ref{eq:24}), together with 
Eqs.~(\ref{eq:25ab})-(\ref{eq:27abc}), however, 
does not explicitly contain any FLT-specific parameters such as the Landau 
interaction and $m^*$, implying that this functional form itself can be applied to NFL 
as well. In fact, ${\tilde \eta}_{1} (K)$, a key quantity leading to $m/m^*$ at 
$K \! \to \!K_{\rm F}$ in FLT, appears in this formalism as a direct consequence of 
the total-current (or total-momentum) conservation law that should be satisfied even 
in NFL, indicating that ${\tilde \eta}_{1} (K)$ must also be a key quantity in NFL.

%%%%%%%%%%%%%%%%%%%%%%%%< Paragraph 25: {\tilde \eta}_1(K,K+Q)  >%%%%%%%%%%%%%%%%%%%
It is the most important advantage in this formalism that the Ward identity 
(WI)~\cite{Ward_1950} is automatically satisfied, whatever choice one may make 
for ${\tilde \eta}_{1}(K)$ and ${\tilde \eta}_{2}(K;Q)$; if we choose 
${\tilde \eta}_{1}(K)\!=\!{\tilde \eta}_{2}(K;Q)\!=\!1$, the present vertex 
function is nothing but the one in the original $GW\Gamma$ scheme~\cite{YT_2001} 
and $\widetilde{\Gamma}_{\rm WI}(K,K\!+\!Q)$ is reduced to the canonical form 
appearing in connection to WI as
\begin{align}
\Gamma_{\rm WI}(K,K\!+\!Q)\!=\!\frac{
G^{-1}(K\!+\!Q)\! - \! G^{-1}(K)} 
{i \omega_q  \!-\!  \varepsilon_{{\bm k} \!+\! {\bm q}}
 \!+\! \varepsilon_{{\bm k}}}.
\label{eq:28}
\end{align}
Thus, we note that by just changing $\Gamma_{\rm WI}(K,K\!+\!Q)$ into 
$\widetilde{\Gamma}_{\rm WI}(K,K\!+\!Q)$, we can enter into a more advanced 
stage in which both WI and the total-momentum conservation law are 
simultaneously fulfilled. 

%%%%%%%%%%%%%%%%%%%%%%%%< Paragraph 26: \eta_1(Q)\Gamma(K,K+Q)  >%%%%%%%%%%%%%%%%%%%%%%%%%
In actual numerical calculations, however, it turns out that it is not easy 
to proceed at each iteration step with ${\tilde \eta}_{1}(K)$ evaluated in 
accordance with Eq.~(\ref{eq:26}). Because ${\tilde \eta}_{1}(K)$ reflects the 
total-momentum conservation law only in its value at $K\!=\!K_{\rm F}$, we can 
approximate ${\tilde \eta}_{1}(K\!+\!Q/2)$ as $\eta_1(Q)$, a function depending 
only on $Q$ with the condition that $\eta_1(Q_0)\!=\!{\tilde \eta}_{1}(K_{\rm F})$ 
in the $q$-limit. Under this approximation, we can reduce $\Gamma_{\Pi}(K,K\!+\!Q)$ 
into 
\begin{align}
&\Gamma_{\Pi}(K,K\!+\!Q) \!=\! 
\frac{\Pi(Q)}{\widetilde{\Pi}_{\rm WI}(Q)}
\!+\!\frac{\Pi(Q)}{D_{\rm F}}
\biggl \{ \frac{1\!-\!\eta_1(Q)}{\eta_1(Q)}\frac{3i\omega_q}{v_{\rm F}^2{\bm q}^2}
\nonumber \\
& \hspace{0.05cm}  
\times \! \Bigl [ 
i\omega_q\!-\!(\varepsilon_{{\bm k}\!+\!{\bm q}}\!-\!\varepsilon_{\bm k})
\eta_1(Q)  \Bigr ] 
\!+\!{\tilde \eta}_{2}(K;Q)\!-\!\frac{\Pi_2(Q)}{\widetilde{\Pi}_{\rm WI}(Q)}
\biggr \}, 
\label{eq:29}
\end{align}
and $\widetilde{\Pi}_{\rm WI}(Q)$ is easily calculated by 
\begin{align}
\widetilde{\Pi}_{\rm WI}(Q)\!\equiv \! \widetilde{\Pi}_{\rm WI}({\bm q},i\omega_q)
=\frac{1}{\eta_1(Q)}\Pi_{\rm WI}\left ({\bm q},\frac{i\omega_q}{\eta_1(Q)}
\right ),
\label{eq:30}
\end{align}
with $\Pi_{\rm WI}({\bm q},i\omega_q)\!\equiv\!
\Pi_{\rm WI}(Q)$ in Eq.~(\ref{eq:16}).

%%%%%%%%%%%%%%%%%%%%< Paragraph 27: {\tilde \eta}_2(K;Q)  >%%%%%%%%%%%%%%%%%%%%%%
Since we may take ${\tilde \eta}_{2}(K;Q)$ at our disposal, we choose it 
so as to make numerical calculations as easy as possible. By carefully 
inspecting the structure of each term in Eq.~(\ref{eq:29}), we can think of a 
possible form for ${\tilde \eta}_{2}(K;Q)$ as
\begin{align}
&{\tilde \eta}_{2}(K;Q)=\eta_1(Q)\zeta_1(Q)\!+\!
\Bigl [ i\omega_q\!-\!(\varepsilon_{{\bm k}\!+\!{\bm q}}\!-\!\varepsilon_{\bm k})
\eta_1(Q) \Bigr ]
\nonumber \\
& \hspace{1.1cm}
\times \frac{3}{v_{\rm F}^2{\bm q}^2}
\Bigl [ i\omega_q \zeta_2(Q)-
(\varepsilon_{{\bm k}\!+\!{\bm q}}\!-\!\varepsilon_{\bm k})\zeta_3(Q) \Bigr ],
\label{eq:31}
\end{align}
with arbitrary functions $\zeta_i(Q)$ ($i\!=\!1,2$, and $3$). By substituting 
this ${\tilde \eta}_{2}(K;Q)$ into Eq.~(\ref{eq:29}), we immediately find the 
following: (i) $\zeta_1(Q)$ is irrelevant, because it is always cancelled by the 
corresponding term in $\Pi_2(Q)/\widetilde{\Pi}_{\rm WI}(Q)$. (ii) By choosing 
$\zeta_2(Q)$ as $[\eta_1(Q)-1]/\eta_1(Q)$, we can eliminate the first term in the 
curly brackets in Eq.~(\ref{eq:29}). (iii) Because the term containing $\zeta_3(Q)$ 
is rather difficult to treat, it would be better to approximate 
$(\varepsilon_{{\bm k}\!+\!{\bm q}}\!-\!\varepsilon_{\bm k})\zeta_3(Q)$ 
by $v_{\rm F}q\zeta_3(Q)$ with $q=|{\bm q}|$. As a result, we have arrived at the 
functional form for $\Gamma(K,K\!+\!Q)$, written as
\begin{align}
&\Gamma(K,K\!+\!Q) \!=\! 
\frac{G^{-1}(K\!+\!Q)\!-\! G^{-1}(K)} 
{i \omega_q  \!-\!  (\varepsilon_{{\bm k} \!+\! {\bm q}}
 \!-\! \varepsilon_{\bm k})\eta_1(Q)}\,
\frac{\Pi(Q)}{\widetilde{\Pi}_{\rm WI}(Q)}
\nonumber \\
& \hspace{1.0cm}  
\!-\! \bigl [ G^{-1}(K\!+\!Q)\! - \! G^{-1}(K) \bigr ]
\frac{\Pi(Q)}{D_{\rm F}}\frac{3\zeta_3(Q)}{v_{\rm F}q}.
\label{eq:32}
\end{align}
It should be noted that we can rigorously reproduce $\Pi(Q)$ irrespective of 
$\eta_1(Q)$ and $\zeta_3(Q)$ in substituting this functional form for 
$\Gamma(K,K\!+\!Q)$ into Eq.~(\ref{eq:14-b}), verifying the internal 
consistency of our formulation. 

%%%%%%%%%%%%%%%%%%< Paragraph 28: \eta_1(Q) & \zeta_3(K,K+Q)  >%%%%%%%%%%%%%%%%%%%%%
To determine $\eta_1(Q)$ and $\zeta_3(Q)$, we need to consider several 
constraints to be imposed on them to make $\Gamma(K,K\!+\!Q)$ behave 
correctly in various limits. In line with such constraints as discussed 
in Appendix C, we will use $\eta_1(Q)$ in Eq.~(C3) and $\zeta_3(Q)$ given by
\begin{align}
\zeta_3(Q)\!=\!-\frac{1}{3}\,\frac{\Pi_0(Q)\!-\!\Pi_{\rm WI}(Q)[1\!-\!\beta_3 G_s(Q)]}
                        {\Pi_{\rm WI}(Q)},
\label{eq:36}
\end{align}
in this paper.  

%%%%%%%%%%%%%%%%%%%%%%%%%%%%%%%%%%< Figure 3 >%%%%%%%%%%%%%%%%%%%%%%%%%%%%%%%%%%%%%%%
\begin{figure}[tbp]
\begin{center}
\includegraphics[scale=0.49,keepaspectratio]{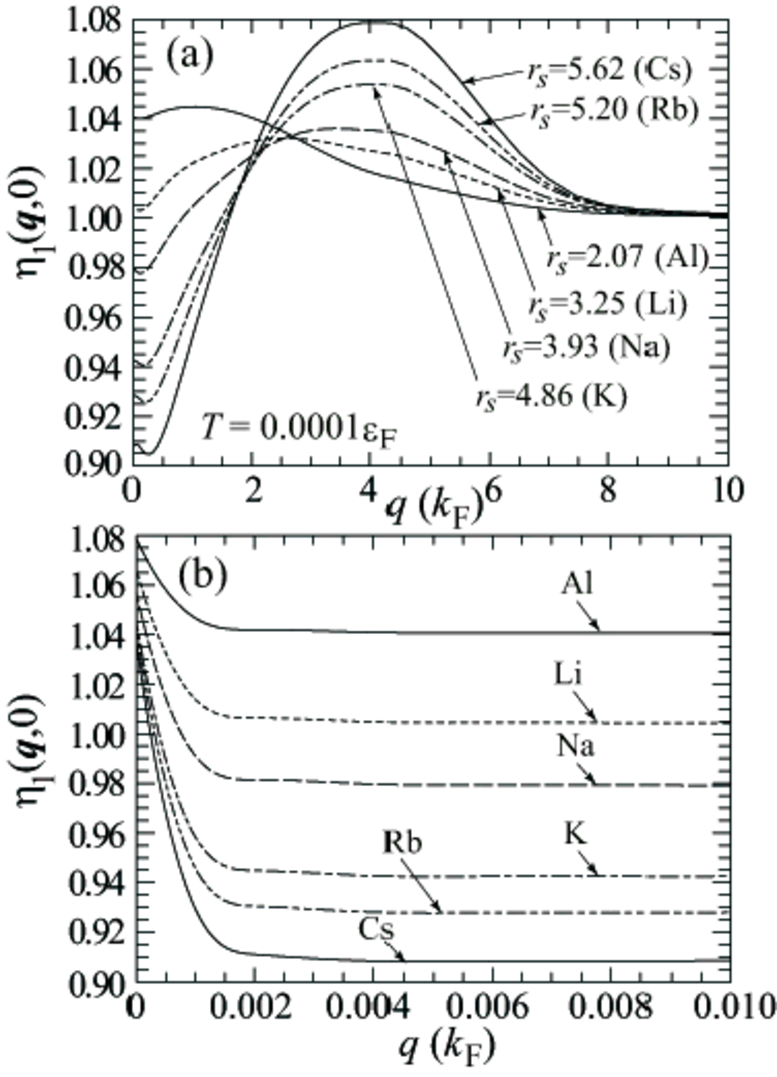}
\end{center}
\caption[Fig.03]{In panel (a), $\eta_1({\bm q},0)$ at $T\!=\!10^{-4}
\varepsilon_{\rm F}$ is plotted in the range of $0\!<\!q\!<\!10k_{\rm F}$ for the 
3D homogeneous electron gas with the density region corresponding to simple metals 
and in panel (b), the same function is drawn with $q$ in an enlarged scale near $q=0$.}
\label{fig:03}
\end{figure}
%%-------------------------------------------------------------====-----------------%

%%%%%%%%%%%%%%%%%%%%%%%%< Paragraph 29: \eta_1(q,0)  >%%%%%%%%%%%%%%%%%%%%%%%%% 
In Fig.~\ref{fig:03}, the self-consistently determined results of $\eta_1({\bm q},0)$ 
are plotted at $T\!=\!10^{-4}\varepsilon_{\rm F}$ for the 3D homogeneous electron gas 
with $2\!<\!r_s\!<\!6$. 
As seen from this figure, $\eta_1(Q)$ is smoothly converged to unity for $q\gtrsim 
10k_{\rm F}$, but it exhibits a rather rapid change near $q \approx 0$, indicating 
that $\Sigma(K)$ is not smooth enough for $K \approx K_{\rm F}$, a symptom of possible 
breakdown of FLT. Incidentally, in obtaining $\eta_1(Q)$, we need to know the 
static physical quantities, $E({\bm k},0)$ and $Z({\bm k},0)$, which can be 
obtained by an extrapolation procedure explained in Appendix D.

%%%%%%%%%%%%%%%%%%%%%%%%%%%%%%%%%%< Figure 4 >%%%%%%%%%%%%%%%%%%%%%%%%%%%%%%%%%%%%%%%
\begin{figure}[b]
\begin{center}
\includegraphics[scale=0.35,keepaspectratio]{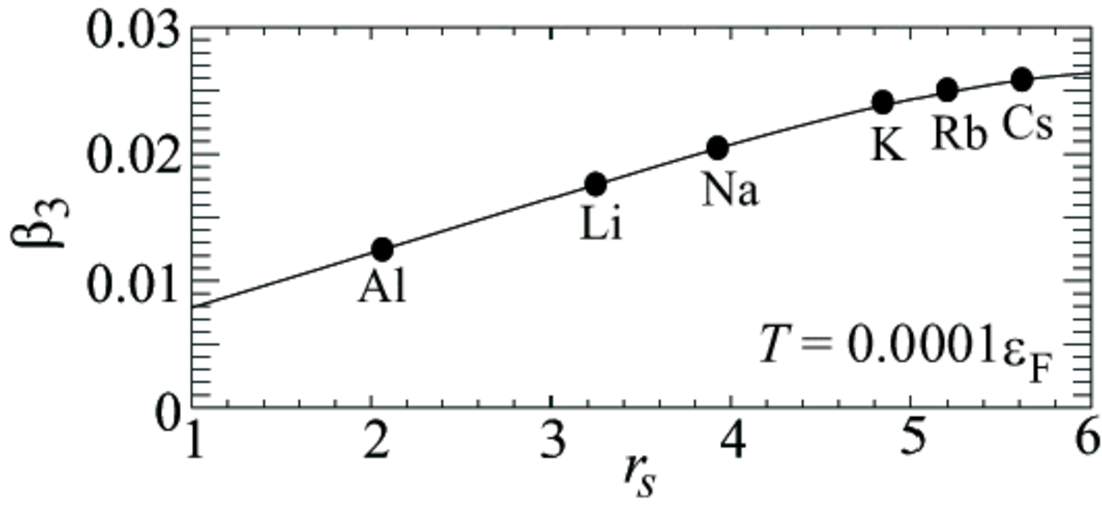}
\end{center}
\caption[Fig.04]{Results of $\beta_3$ in Eq.~(\ref{eq:36}) determined 
to reproduce the accurate $\mu_c$ for the 3D homogeneous electron gas with 
the density parameter $r_s$ in the range of $1\!<\!r_s\!<\!6$.}
\label{fig:04}
\end{figure}
%%----------------------------------------------------------------------------------%

%%%%%%%%%%%%%%%%%%%%%%%%< Paragraph 30: beta_3  >%%%%%%%%%%%%%%%%%%%%%%%%% 
In Eq.~(\ref{eq:36}), the term involving the parameter $\beta_3$ is introduced 
to rigorously reproduce $\mu_c$, the value accurately given in Eq.~(\ref{eq:06}). 
The appropriately determined values for $\beta_3$ providing the correct $\mu_c$ 
with $r_s$ in the region of $1<r_s<6$ are shown in Fig.~\ref{fig:04} in which 
we take $T$ as $10^{-4}\varepsilon_{\rm F}$. As seen from the figure, the magnitude 
of $\beta_3$ is small, i.e., of the order of 0.02, letting us know that 
the $\beta_3$ term exerts only limited effects on $\Sigma(K)$.

%%%%%%%%%%%%%%%%%%%%%%%%%%%%< Section 2-6 >%%%%%%%%%%%%%%%%%%%%%%%%%%%%%%%%%%%%%%%%%%
\subsection{Self-energy}
\label{sec:2F}
%%%%%%%%%%%%%%%%%%%%%%%< Paragraph 31: Self-energy formula  >%%%%%%%%%%%%%%%%%%%%%%%
By substituting Eq.~(\ref{eq:32}) with Eq.~(\ref{eq:36}) into Eq.~(\ref{eq:22}), 
we obtain an expression for the calculation of $\Sigma(K)$ as
\begin{align}
\Sigma(K) = \Sigma_a+\Sigma_b(K)+\Lambda(K)G^{-1}(K),
\label{eq:40}
\end{align}
with
\begin{subequations}
\label{eq:41abc}
\begin{align}
&\Sigma_a\ =-\sum_Q W_{\rm WI}(Q)R_{\rm WI}(Q)
\nonumber \\
& \hspace{0.6cm}  
=-T\sum_{\omega_q}\int_0^{\infty}\frac{q^2dq}{2\pi^2}
\,W_{\rm WI}(Q)R_{\rm WI}(Q),
\label{eq:41a}
\\
&\Sigma_b(K)\!=\!-\sum_{K'}\overline{W}_{\rm WI}(Q)G(K')
\overline{\Gamma}_{\rm WI}(K,K')
\nonumber \\
& \hspace{1.0cm}
=-T\sum_{\omega_{n'}}\frac{1}{k}\int_0^{\infty}\frac{qdq}{4\pi^2}\,
         \overline{W}_{\rm WI}({\bm q},i\omega_{n'}\!-\!i\omega_n)
\nonumber \\
& \hspace{1.8cm}
\times \int_{|k-q|}^{k+q}k'dk'G(K')\overline{\Gamma}_{\rm WI}(K,K'),
\label{eq:41b}
\\
& \Lambda(K)\,=\!\sum_{K'}W_{\rm WI}(Q)R_{\rm WI}(Q)G(K')
\nonumber \\
& \hspace{1.0cm}
=T\sum_{\omega_{n'}}\frac{1}{k}\int_0^{\infty}\frac{qdq}{4\pi^2}\,
W_{\rm WI}({\bm q},i\omega_{n'}\!-\!i\omega_n)
\nonumber \\
& \hspace{1.5cm}
\times \!R_{\rm WI}({\bm q},i\omega_{n'}\!-\!i\omega_n)\!
\int_{|k-q|}^{k+q}\!k'dk'G(K'),
\label{eq:41c}
\end{align}
\end{subequations}
where $W_{\rm WI}(Q)$, $\overline{W}_{\rm WI}(Q)$, $R_{\rm WI}(Q)$, and 
$\overline{\Gamma}_{\rm WI}(K',K)$ are, respectively, defined as
\begin{subequations}
\label{eq:42abcd}
\begin{align}
&W_{\rm WI}(Q)=W(Q)\frac{\Pi(Q)}{\Pi_{\rm WI}(Q)}
\nonumber \\
& \hspace{1.3cm}  
=\frac{V({\bm q})}{1+V({\bm q})\Pi_{\rm WI}(Q)[1-G_s(Q)]},
\label{eq:42a}
\\
&\overline{W}_{\rm WI}(Q)=W_{\rm WI}(Q)\,
\frac{\Pi_{\rm WI}(Q)}{\widetilde{\Pi}_{\rm WI}(Q)},
\label{eq:42b}
\\
&R_{\rm WI}(Q)\ =\,\frac{\Pi_0(Q)\!-\!\Pi_{\rm WI}(Q)[1-\beta_3G_s(Q)]}
{D_{\rm F}v_{\rm F}q},
\label{eq:42c}
\\
&\overline{\Gamma}_{\rm WI}(K,K')=
\frac{G^{-1}(K')\!-\! G^{-1}(K)} 
{i \omega_q  \!-\! (\varepsilon_{\bm k'}\!-\! \varepsilon_{\bm k})\eta_1(Q)}.
\label{eq:42d}
\end{align}
\end{subequations}
Here, $\Sigma_a$ is independent of $K$ and directly connected to the 
chemical potential shift, $\Sigma_b(K)$ is the main contribution to the 
self-energy, and $\Lambda(K)$ partially contributes to the renormalization 
factor. 

%%%%%%%%%%%%%%%%%%%%%%%%%%%%< Section 2-7 >%%%%%%%%%%%%%%%%%%%%%%%%%%%%%%%%%%%%%%%%%%
\subsection{Self-consistent iteration loop}
\label{sec:2G}

\begin{figure}[htbp]
\begin{center}
\includegraphics[scale=0.38,keepaspectratio]{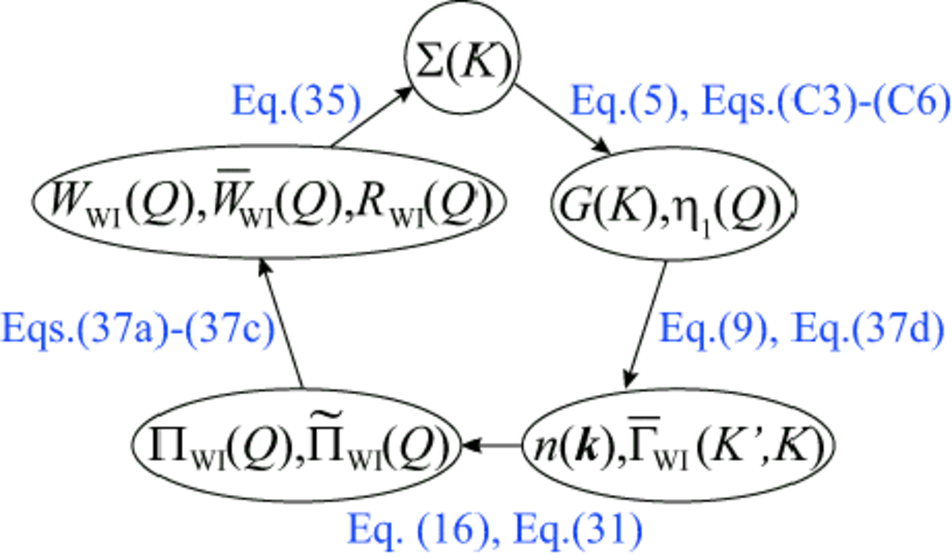}
\end{center}
\caption[Fig.05]{Self-consistent iteration loop to determine the self-energy 
$\Sigma(K)$ in the present calculation scheme.}
\label{fig:05}
\end{figure}

%%%%%%%%%%%%%%%%%%%%%%%%< Paragraph 32: iteration loop  >%%%%%%%%%%%%%%%%%%%%%%%%%
To summarize this section, Fig.~\ref{fig:05} schematically displays a self-consistent 
iteration loop to determine $\Sigma(K)$. 
In developing the actual code, we make it adaptable to calculations at unprecedentedly 
low temperatures, down to $T\!=\!10^{-4}\varepsilon_{\rm F}$, because no singularities 
in $\Sigma(K)$ leading to NFL have been obtained for $T\!=\! 10^{-2}\varepsilon_{\rm F}$ 
in the original GW$\Gamma$ scheme~\cite{YT_2001}, $T\! = \! 0.04 \varepsilon_{\rm F}$ 
in the recent variational diagrammatic Monte Carlo (VDMC) 
simulations~\cite{Haule_2019,Haule_2022}, and $T\! = \! 0.1 \varepsilon_{\rm F}$ 
in the algorithmic Matsubara-diagrammatic Monte Carlo (ADMC) 
technique~\cite{LeBlanc_2022}. In accordance with $T\! \approx \! 10^{-4}
\varepsilon_{\rm F}$, the mesh size $\Delta k$ in ${\bm k}$ space should 
also be small of the same order, i.e., $\Delta k \! \approx \! 10^{-4}k_{\rm F}$ 
near the Fermi surface, to detect any singularities in $\Sigma(K)$ 
appearing at such a low $T$, because $|\varepsilon_{\bm k}|$ 
at $|{\bm k}|\!=\!k_{\rm F} \pm  \Delta k$ is approximately 
equal to $k_{\rm F}\Delta k/m$ which must be comparable to $\pi T$. In this respect, 
no symptoms of NFL will be detected in the recent zero-temperature quantum 
Monte Carlo calculations~\cite{Holzmann_2023} in which $\Delta k (\approx \! 
k_{\rm F}/32)$ is not small enough.

%%%%%%%%%%%%%%%%%%%%%%%%%%%%%%%%%%< Figure 5 >%%%%%%%%%%%%%%%%%%%%%%%%%%%%%%%%%%%%%%%
%%----------------------------------------------------------------------------------%

%%%%%%%%%%%%%%%%%%%%< Paragraph 33: convergence criterion  >%%%%%%%%%%%%%%%%%%%%%%%%
The implementation of this iteration loop starts with $\Sigma_0(K)$ 
the self-energy in the random-phase approximation (RPA) (or the $G_0W_0$ 
approximation~\cite{YT_2016b}), given by 
\begin{align}
\Sigma_0(K)=-\sum_Q \frac{V({\bm q})}{1+V({\bm q})\Pi_0(Q)}\, G_0(K+Q),
\label{eq:43}
\end{align}
and ends up when the relative difference in $\Sigma(K)$ between input and output 
at each mesh point becomes less than $10^{-5}$. In revising the input $\Sigma(K)$ 
at each step during the iteration loop, we employ the second Broyden's 
method~\cite{Broyden_1965,Marks_2008,Fang_2009}. We need $10$-$100$ iteration steps 
depending on $r_s$ and $T$ to obtain converged results for $\Sigma(K)$. The 
calculated $\Sigma(K)$ is converted into the retarded self-energy 
$\Sigma^{R}({\bm k},\omega\!+\!i\gamma)$ with $\gamma\!=\!\pi T$ through numerical 
analytic continuation with the use of Pad\'{e} approximants~\cite{Vidberg_1977}.

%%%%%%%%%%%%%%%%%%%%< Paragraph 34: developing codes >%%%%%%%%%%%%%%%%%%%%%%%%%%%
In the actual numerical implementation, instead of $\Sigma(K)$, we calculate 
a couple of real functions in Eq.~(\ref{eq:11}), $Z(K)$ and $E(K)$, both of which 
are even in $\omega_n$ and depend on ${\bm k}$ only through $k\ (\equiv |{\bm k}|)$. 
Those functions will be evaluated only at a finite number of points in 
$(k,\omega_n)$-space, $\{k_i,\omega_j\}$, and if we need the values of those 
functions at $K$ other than those selected points, then we will employ the two-dimensional 
cubic spline interpolation method. In Appendix E, we make a more detailed explanation 
of $\{k_i,\omega_j\}$, together with some remarks on the numerical integration in 
Eqs~(\ref{eq:41a})-(\ref{eq:41c}) and the numerical analytic continuation. 

%%%%%%%%%%%%%%%%%%%%%%%%%%%%%%%%%%%%%%%%%%%%%%%%%%%%%%%%%%%%%%%%%%%%%%%%%%%%%%%%%%%%%
%%%%%%%%%%%%%%%%%%%%%%%%< Section 3 >%%%%%%%%%%%%%%%%%%%%%%%%%%%%%%%%%%%%%%%%%%%%%%%%
%%%%%%%%%%%%%%%%%%%%%%%%%%%%%%%%%%%%%%%%%%%%%%%%%%%%%%%%%%%%%%%%%%%%%%%%%%%%%%%%%%%%%
\section{Calculated Results}
\label{sec:3}
%%%%%%%%%%%%%%%%%%%%%%%%< Section 3-1 >%%%%%%%%%%%%%%%%%%%%%%%%%%%%%%%%%%%%%%%%%%%%%%
\subsection{One-particle spectral function}
\label{sec:3A}

%%%%%%%%%%%%%%%%%%%%%%%%%%%%%%%%%%< Figure 6 >%%%%%%%%%%%%%%%%%%%%%%%%%%%%%%%%%%%%%%%
\begin{figure}[b]
\begin{center}
\includegraphics[scale=0.45,keepaspectratio]{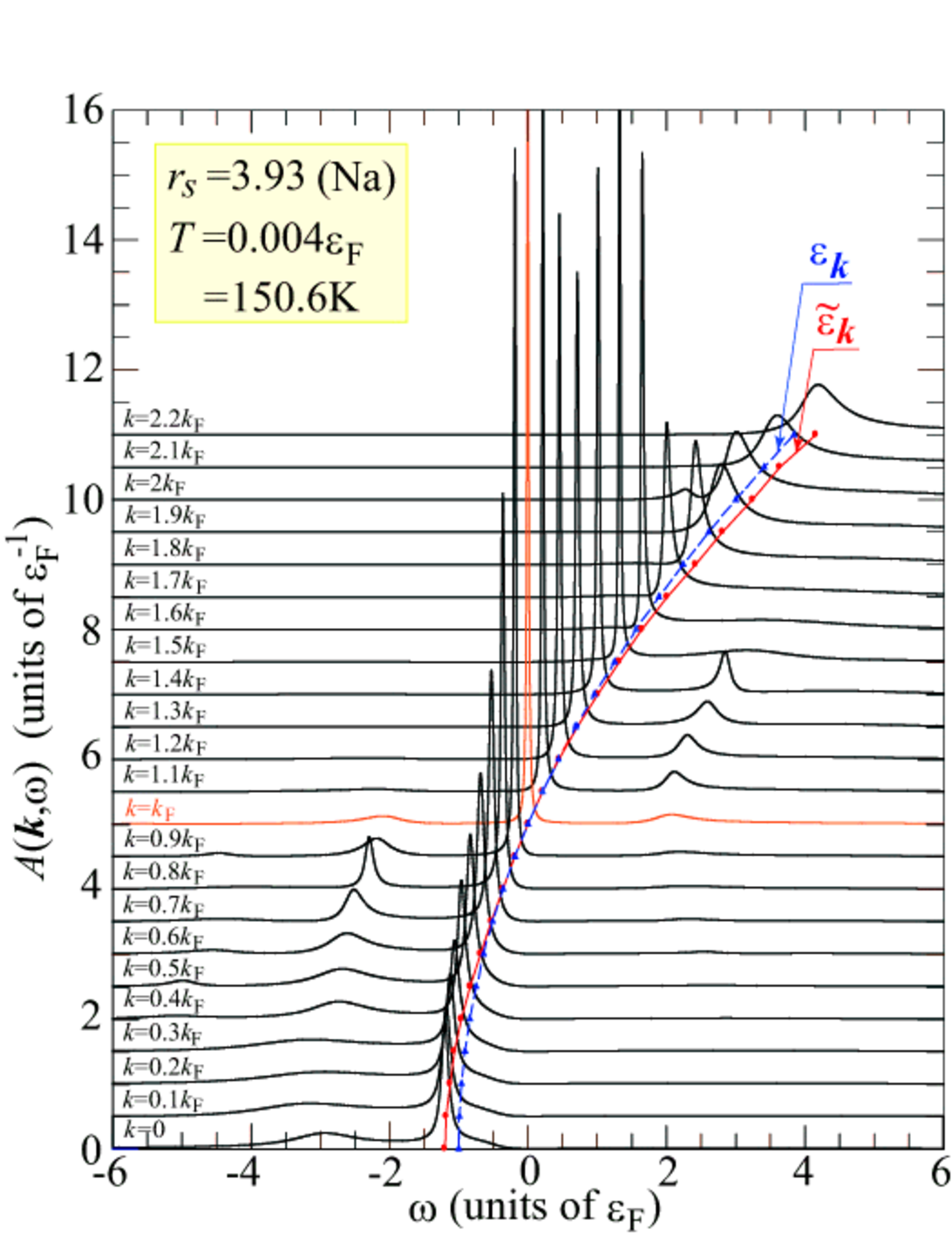}
\end{center}
\caption[Fig.06]{Overall structure of the one-particle spectral function 
$A({\bm k},\omega)$ in the 3D homogeneous electron gas at $r_s\!=\!3.93$ 
corresponding to sodium with $T\!=\!4\!\times \!10^{-3}\varepsilon_{\rm F}$ and 
$k\ (\equiv |{\bm k}|)=0.0,\, 0.1,\, 0.2,\, \cdots, \,2.0, \,2.1, \, 2.2$ 
in units of $k_{\rm F}$.}
\label{fig:06}
\end{figure}
%%----------------------------------------------------------------------------------%
%%%%%%%%%%%%%%%%%%%%%< Paragraph 35: formulation for A(p,w)  >%%%%%%%%%%%%%%%%%%%%%%%
In Fig.~\ref{fig:06}, the one-particle spectral function $A({\bm k},\omega)\ [\equiv 
\!-{\rm Im}G^R({\bm k},\omega\!+\!i\pi T)/\pi]$ is plotted as a function of $\omega$ 
for $r_s\!=\!3.93$, corresponding to sodium, at $T\!=\!4\!\times \!10^{-3}
\varepsilon_{\rm F}$ with $k\ (\equiv \! |{\bm k}|)\!=\!0.0-2.2$ in units of $k_{\rm F}$. 
This shows a well-known typical behavior, characterized by the dominant quasiparticle 
peak and the associated one-plasmon satellites. In fact, this result 
is essentially the same, even quantitatively, as the one given at $r_s=4$ in 
Fig.~3(a) in Ref.~\cite{Maebashi_2011} in which both ${\tilde \eta}_{1}$ and 
${\tilde \eta}_{2}$ were taken as unity. Therefore, we cannot find any indication 
of the breakdown of FLT at least for sodium at $T=4\!\times \!10^{-3}\varepsilon_{\rm F}$, 
as is the case in all previous studies on alkali metals.

\begin{figure}[htbp]
\begin{center}
\includegraphics[scale=0.36,keepaspectratio]{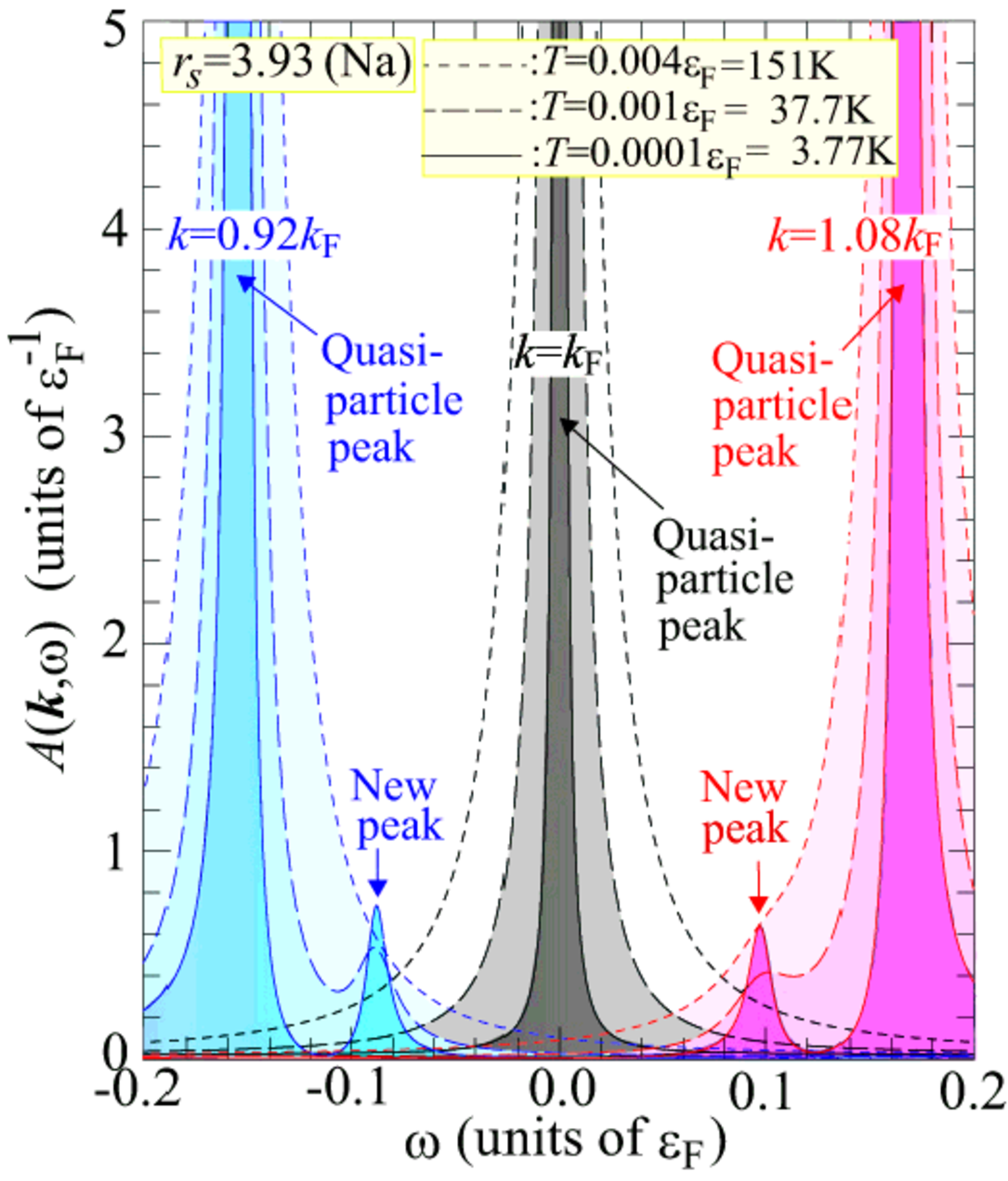}
\end{center}
\caption[Fig.07]{Change of $A({\bm k},\omega)$ with the decrease of $T$ from 
$4\!\times \!10^{-3}$ down to $10^{-4}$ in units of $\varepsilon_{\rm F}$ 
for $r_s=3.93$ with $k=0.92k_{\rm F}$ (in blue), $k_{\rm F}$ (in gray), and 
$1.08k_{\rm F}$ (in red).
}
\label{fig:07}
\end{figure}

%%%%%%%%%%%%%%%%%%%%%%%%%< Paragraph 36: T dependence  >%%%%%%%%%%%%%%%%%%%%%%%%%%%
By decreasing $T$ down to $10^{-4}\varepsilon_{\rm F}$, however, we find a totally 
new situation in the low-energy region of $\omega$ for $k \sim k_{\rm F}$ at $r_s=3.93$, 
as shown in Fig.~\ref{fig:07}. Although nothing special is seen at $k\!=\!k_{\rm F}$ 
at all temperatures, a shoulder or bump structure begins to develop at $T\!=\!10^{-3}
\varepsilon_{\rm F}$ and a clear peak structure emerges at $T\!=\!10^{-4}
\varepsilon_{\rm F}$ for $k$ not equal to $k_{\rm F}$ but close to it. The energy 
of this new peak is about a half of that of the corresponding quasiparticle peak 
at the same $k$ and thus a new mode, if any, associated with this peak (to be dubbed 
``excitron'') is characterized by the very low excitation energy of the order of 
$0.1\varepsilon_{\rm F}$ or less. 

%%%%%%%%%%%%%%%%%%%%%%%%%%%%%%%%%%< Figure 7 >%%%%%%%%%%%%%%%%%%%%%%%%%%%%%%%%%%%%%%%
%%----------------------------------------------------------------------------------%

\begin{figure}[htbp]
\begin{center}
\includegraphics[scale=0.44,keepaspectratio]{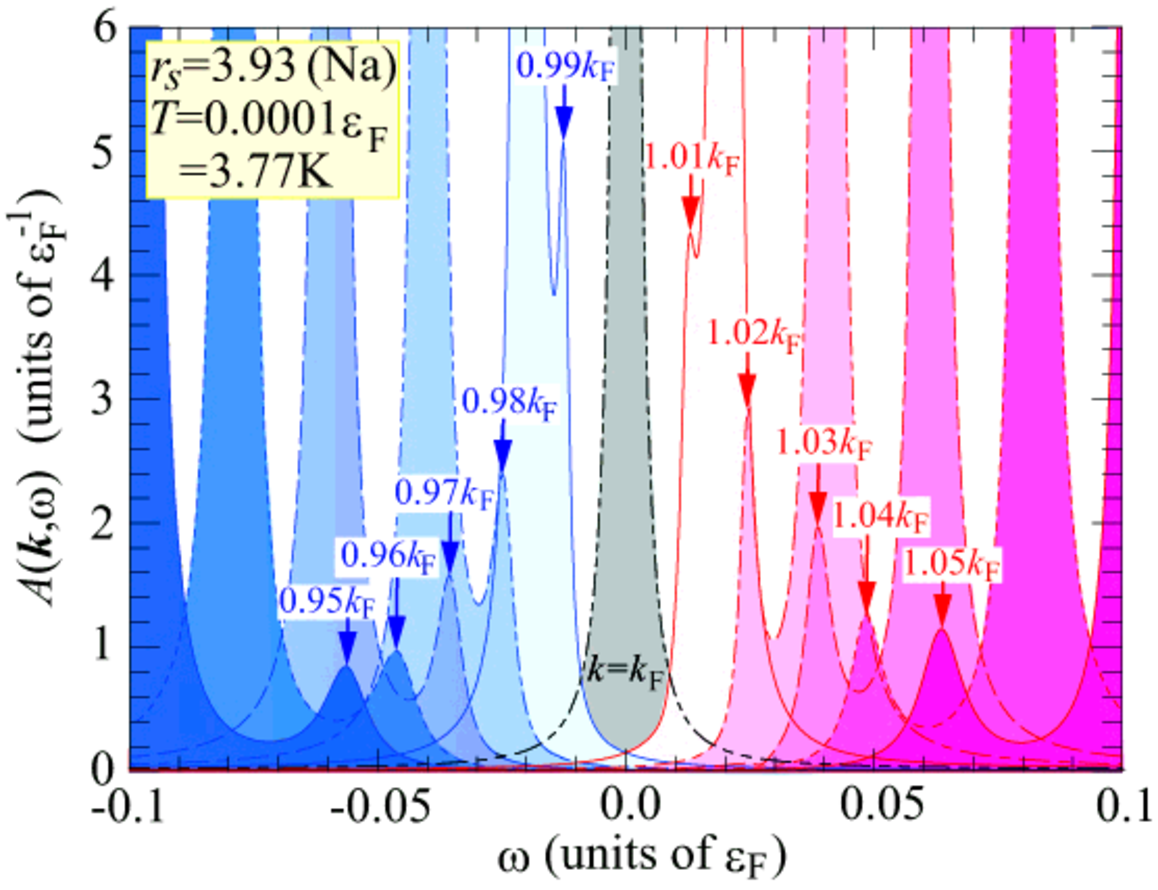}
\end{center}
\caption[Fig.08]{Change in shape and position of the new peak in $A({\bm k},\omega)$ 
with the increase of $k$ from $0.95k_{\rm F}$ to $1.05k_{\rm F}$ for $r_s=3.93$ 
at $T=10^{-4}\varepsilon_{\rm F}$. The new peak position is indicated by an arrow 
at each $k$.
}
\label{fig:08}
\end{figure}

%%%%%%%%%%%%%%%%%%%%%%%%%< Paragraph 37: k-dependence  >%%%%%%%%%%%%%%%%%%%%%%%%%%%
In Fig.~\ref{fig:08}, we show the change in shape and position of this new peak with the 
increase of $k$ from $0.95k_{\rm F}$ to $1.05k_{\rm F}$ through $k_{\rm F}$ for 
$r_s\!=\!3.93$ at $T\!=\!10^{-4}\varepsilon_{\rm F}$, from which we see that this new peak
 is absorbed into (or perfectly overlapped with) the dominant quasiparticle peak 
 at $k=k_{\rm F}$. Thus, we need to keep $k$ away from $k_{\rm F}$ to detect this new 
peak, but at the same time, as $|k-k_{\rm F}|$ increases, the peak height decreases, 
making the detection rather difficult. To compromise between these competing factors, 
it would be good to search for this peak at $k$ in the range $|k-k_{\rm F}|\sim 
0.03k_{\rm F}-0.2 k_{\rm F}$. We have also calculated $A({\bm k},\omega)$ at $T\!=\!10^{-4}
\varepsilon_{\rm F}$ for other values of $r_s$ in the range of $1<r_s<6$ to find that 
this new peak always appears with qualitatively the same features as those described 
above, including the behavior with the change of $k$ and $T$, though quantitatively 
the peak strength (or height) depends rather strongly on $r_s$; as $r_s$ increases, 
the peak emerges more vividly and strongly. A more detailed analysis on the character 
of this new peak (or excitron) will be made in Sec.~\ref{sec:4}. 

%%%%%%%%%%%%%%%%%%%%%%%%%%%%%%%%%%< Figure 8 >%%%%%%%%%%%%%%%%%%%%%%%%%%%%%%%%%%%%%%%
%-----------------------------------------------------------------------------------%

\begin{figure}[bhtp]
\begin{center}
\includegraphics[scale=0.43,keepaspectratio]{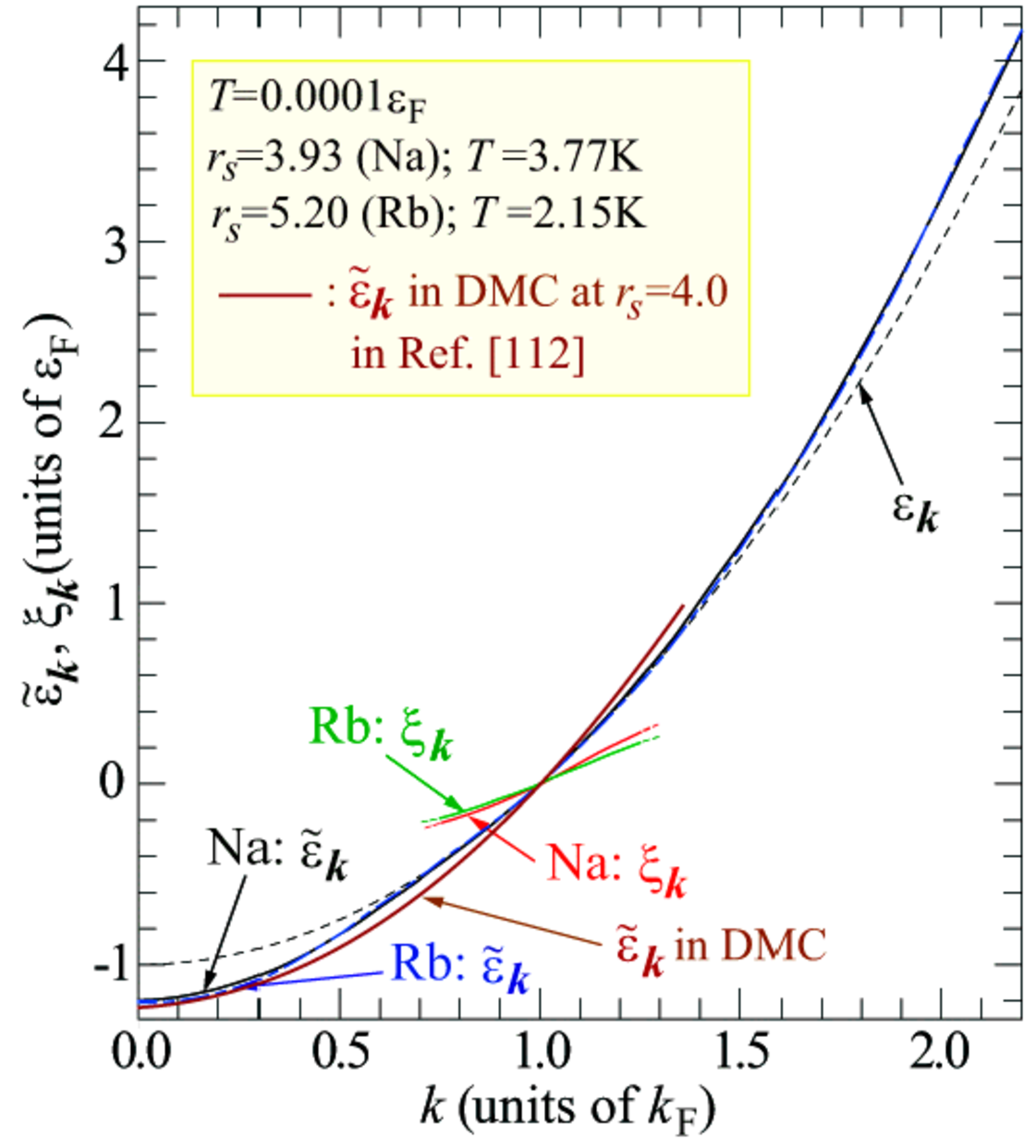}
\end{center}
\caption[Fig.09]{Dispersion relation $\xi_{\bm k}$ of the new peak (excitron) 
in comparison with the quasiparticle dispersion relation ${\tilde \varepsilon}_{\bm k}$ 
and the bare dispersion $\varepsilon_{\rm k}$ (the dotted curve) for $r_s\!=\!3.93$ and 
$5.20$ corresponding to Na and Rb, respectively, at $T\!=\!10^{-4}\varepsilon_{\rm F}$. 
For comparison, ${\tilde \varepsilon}_{\bm k}$ at $r_s\!=\!4$ in diffusion Monte Carlo 
(DMC) simulations is given by the brown solid curve.}
\label{fig:09}
\end{figure}

%%%%%%%%%%%%%%%%%%%%%< Paragraph 38: dispersion relation  >%%%%%%%%%%%%%%%%%%%%%%%%
The dispersion relation of excitron (or the peak position of this new mode), 
$\xi_{\bm k}$, at $T\!=\!10^{-4}\varepsilon_{\rm F}$ for $r_s=3.93$ and $5.20$ 
corresponding to Na and Rb, respectively, is drawn in Fig.~\ref{fig:09}, together 
with the quasiparticle dispersion relation ${\tilde \varepsilon}_{\bm k}$ determined 
by the quasiparticle peak position and the bare dispersion $\varepsilon_{\rm k}$. 
Although the results for $r_s=4.86$ corresponding to K are not shown here, they enter 
just between those of Na and Rb. For $|k-k_{\rm F}|\gtrsim 0.25 k_{\rm F}$, $\xi_{\bm k}$ 
is not shown, because the new peak in either $k\lesssim 0.75k_{\rm F}$ or 
$k \gtrsim 1.25k_{\rm F}$ becomes broad and its peak height is very low, making us 
very difficult to identify the peak position or the peak itself. For $k \sim k_{\rm F}$, 
the excitron dispersion $\xi_{\bm k}$ is linear and can be written as 
$\xi_{\bm k} = v_{\rm excitron}(k-k_{\rm F})$, but for $k$ not in the vicinity of 
$k_{\rm F}$, $\xi_{\bm k}$ deviates from this linear relation. 
The ratio of $v_{\rm excitron}/v_{\rm F}$ is plotted as a function of 
$r_s$ in Fig.~\ref{fig:13} in Sec.~\ref{sec:3D}. 

%%%%%%%%%%%%%%%%%%%%%%%%%%%%%%%%%%< Figure 9 >%%%%%%%%%%%%%%%%%%%%%%%%%%%%%%%%%%%%%%%
%-----------------------------------------------------------------------------------%

%%%%%%%%%%%%%%%%%%%%%%%%%< Paragraph 39: physics on m*  >%%%%%%%%%%%%%%%%%%%%%%%%%%%
The quasiparticle effective mass $m^*$ estimated by ${\tilde \varepsilon}_{\bm k}$ 
with $k$ in the range of $|k-k_{\rm F}|\lesssim 0.1 k_{\rm F}$ is about the same 
as $m$, but it becomes smaller than $m$ for $k$ outside of this range. This change of 
$m^*$ with $k$ can easily be understood by the fact that $m^*$ is determined by the 
competition of exchange and correlation effects; the former makes $m^*$ small as understood 
by the fact that $m^* \!\to \! 0$ at $k\!=\!k_{\rm F}$ in the exchange-only 
(or Hartree-Fock) approximation, while the latter makes $m^*$ large as easily guessed 
by just considering the heavy-fermion physics. Because the exchange effect is evaluated 
in first-order perturbation theory (and thus without energy denominators), 
it persists even for $k$ far away from $k_{\rm F}$. This is not the case for 
the correlation effect to which the second- and higher-order perturbation terms 
contribute. Thus, as $k$ goes away from the Fermi level, the correlation effect 
becomes weaker than that of exchange, making $m^*$ smaller than $m$. In this way, 
the dispersion relation ${\tilde \varepsilon}_{\bm k}$ can never be parabolic 
in the whole region of $k$ from $0$ to $k_{\rm F}$, making the occupied bandwidth 
wider than that of the free-electron band. This widened bandwidth is also seen 
in the diffusion Monte Carlo (DMC) simulations~\cite{Maezono_2003,Azadi_2021} 
as shown by the brown curve in Fig.~\ref{fig:09}; the magnitude of bandwidth in DMC 
is about the same as that in the present calculation, though its behavior of 
${\tilde \varepsilon}_{\bm k}$ for $k\! \sim \!k_{\rm F}$ is much different, providing 
$m^*$ much smaller than $m$ which is not correct as we shall see in Sec.~\ref{sec:3D}. 
Note that DMC was feared to be an unreliable method in determining $m^*$~\cite{Eich_2017}.

%%%%%%%%%%%%%%%%%%%< Paragraph 40: bandwidth in ARPES  >%%%%%%%%%%%%%%%%%%%%%%%%%%%
In contrast to this widening of the computed occupied bandwidth, much narrower bandwidths 
have been observed by ARPES experiments~\cite{Lyo_1988,Potorochin_2022}. 
To account for this discrepancy, various explanations were proposed; in particular, 
the effect of final states in the optical measurements attracted 
attention~\cite{Yasuhara_1999,YT_2001}, but very recently yet another plausible 
explanation was proposed by emphasizing the importance of local 
dynamical correlations associated with an atom based on a more realistic model 
beyond the electron gas~\cite{Mandal_2022}.  

%%%%%%%%%%%%%%%%%%%%%%%%%%%%%%%%%%< Figure 10 >%%%%%%%%%%%%%%%%%%%%%%%%%%%%%%%%%%%%%%%
\begin{figure}[hbtp]
\begin{center}
\includegraphics[scale=0.48,keepaspectratio]{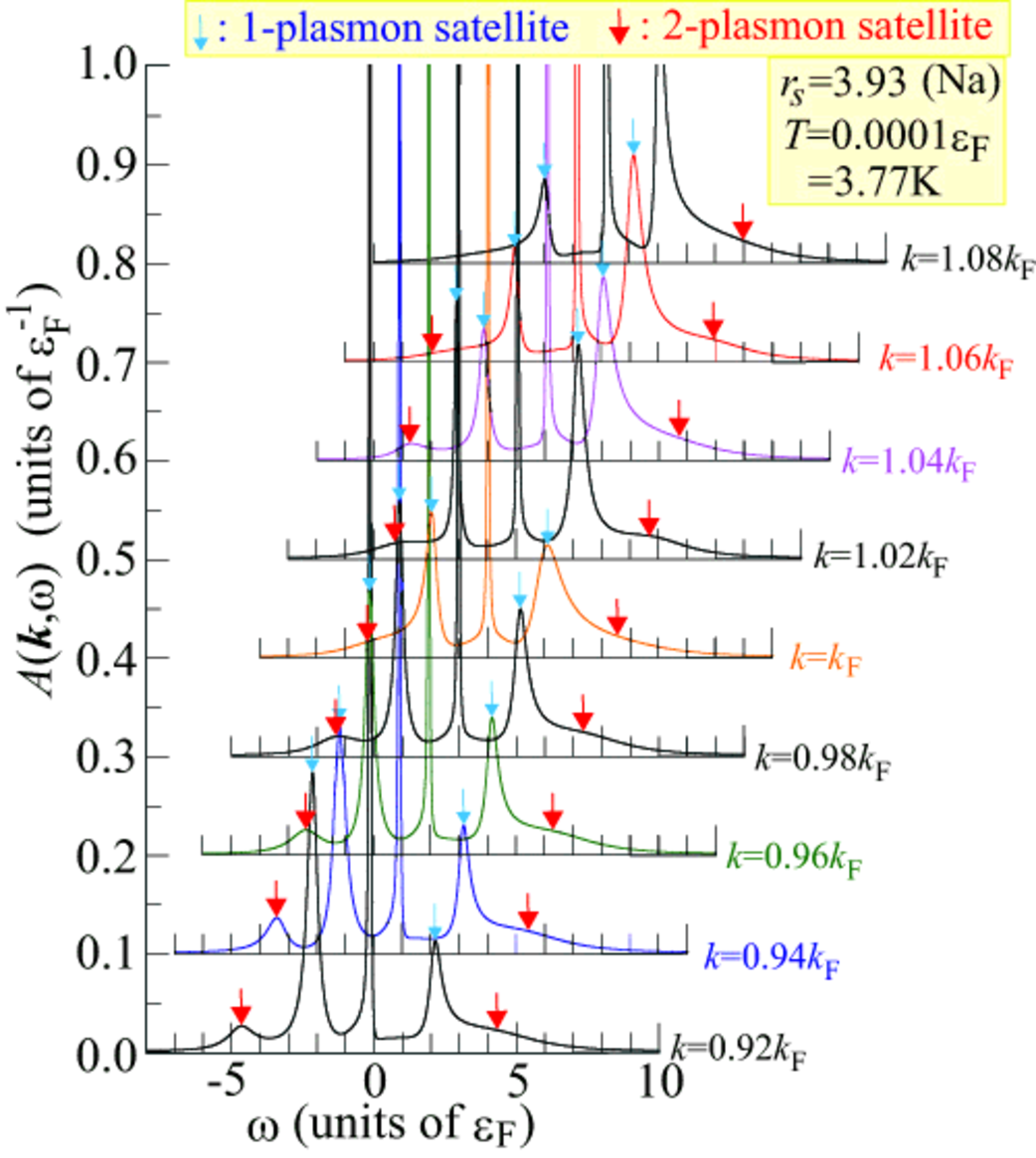}
\end{center}
\caption[Fig.10]{Change of $A({\bm k},\omega)$ for $k$ in the range from 
$0.92k_{\rm F}$ to $1.08k_{\rm F}$ at $r_s\!=\!3.93$ and $T\!=\!10^{-4}
\varepsilon_{\rm F}$ with paying special attention to one- and two-plasmon 
satellites, indicated by blue smaller and red larger arrows, respectively.}
\label{fig:10}
\end{figure}
%-----------------------------------------------------------------------------------%

%%%%%%%%%%%%%%%%%%%%%< Paragraph 41: two-plasmon satellites  >%%%%%%%%%%%%%%%%%%%%%%%%
In Fig.~\ref{fig:10}, $A({\bm k},\omega)$ is plotted on the hundred times wider scale 
of $\omega$ for $k$ in the range of $0.92-1.08$ in units of $k_{\rm F}$ at $r_s\!=\!3.93$ 
and $T\!=\!10^{-4}\varepsilon_{\rm F}$. On this large (not logarithmic but linear) scale, 
the excitron peak is not seen well as a separate structure from the dominant 
quasiparticle peak even at this very low $T$, but we can detect the structures 
associated with the energies of the order of $\varepsilon_{\rm F}$ instead. In fact, 
we can easily find the one-plasmon satellites (shown by blue smaller arrows) and also even 
the two-plasmon ones (shown by red larger arrows). 

%%%%%%%%%%%%%%%%%%%%< Paragraph 42: two-plasmon satellites >%%%%%%%%%%%%%%%%%%%%%%%
The two-plasmon satellite is a challenging issue in the theoretical studies 
of photoemission in the electron gas in connection with experiments in simple metals. 
In the usual $GW$ and related schemes, it is known to be very difficult to provide 
this two-plasmon satellite structure, which  urged Aryasetiawan {\it et al.} 
to invent a $GW$ plus cumulant-expansion approach~\cite{Aryasetiawan_1996}, 
a method manually including the multiplasmon satellites. Afterwards, many other 
works followed in that direction~\cite{Kas_2014,McClain_2016,Carusoa_2016,
Mayers_2016,Kas_2017}, whereas Pavlyukh {\it et al.} succeeded in obtaining 
the structure without resort to the cumulant expansion for the first 
time~\cite{Pavlyukh_2016}. As seen in Figs.~\ref{fig:02} and ~\ref{fig:10}, 
our present method is the second one to accomplish the goal of obtaining 
the two-plasmon satellites without manually including the multiplasmon satellites, 
as far as the author knows. More details on this important achievement 
may be published elsewhere in the future, along with data for 
wider ranges of $k$ and comparisons with other related works.

%%%%%%%%%%%%%%%%%%%%%%%%< Section 3-2 >%%%%%%%%%%%%%%%%%%%%%%%%%%%%%%%%%%%%%%%%%%%%%%
\subsection{Momentum distribution function}
\label{sec:3B}

\begin{figure}[t]
\begin{center}
\includegraphics[scale=0.51,keepaspectratio]{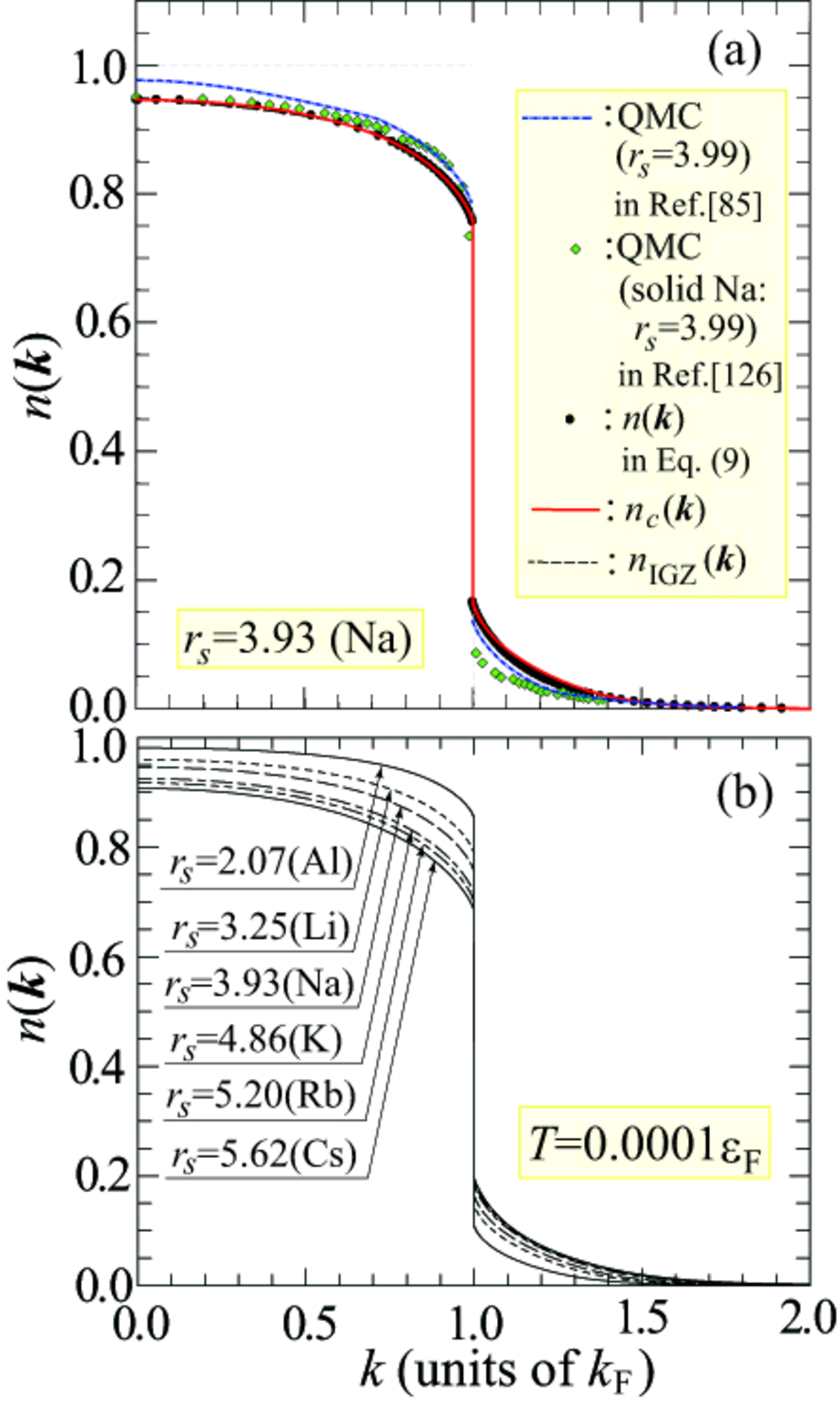}
\end{center}
\caption[Fig.11]{Momentum distribution function $n({\bm k})$ at 
$T\!=\!10^{-4}\varepsilon_{\rm F}$ for (a) $r_s\!=\!3.93$ corresponding to sodium 
and (b) $r_s$ in the range of $2.07-5.62$. In panel (a), data at $r_s\!=\!3.99$ in QMC 
are also included for comparison.}
\label{fig:11}
\end{figure}

%%%%%%%%%%%%%%%%%%%%%< Paragraph 43: n(k)  >%%%%%%%%%%%%%%%%%%%%%%%
The momentum distribution function $n({\bm k})$ is calculated for $r_s$ in the range 
of $2.07-5.62$ at $T=10^{-4}\varepsilon_{\rm F}$ in accordance with the prescription 
described in Sec.~\ref{sec:2C}, together with Appendix B, in which three functions, 
$n({\bm k})$ defined in Eq.~(\ref{eq:12}), $n_{\rm IGZ}(k/k_{\rm F})$, and 
$n_c(k/k_{\rm F})$, are introduced. As shown in Fig.~\ref{fig:11}(a), we cannot see 
the difference among those three functions at $r_s=3.93$ on the scale of this figure. 
In fact, as long as $r_s\lesssim 5.0$, those three functions provide virtually the same 
result. Even for $r_s>5$, a small difference between $n({\bm k})$ and $n_c(k/k_{\rm F})$ 
appears only for $k \gtrsim 1.5k_{\rm F}$. Therefore, in Fig.~\ref{fig:11}(b), we give 
only the results of $n_c(k/k_{\rm F})$ which satisfies all the three sum rules 
associated with $I_2$, $I_4$, and $I_6$ very accurately up to seven digits. We have 
also calculated $n({\bm k})$ at several other temperatures up to $4\times 10^{-3}
\varepsilon_{\rm F}$ to find that the obtained results are virtually independent of $T$.
Incidentally, the present results of $n_c(k/k_{\rm F})$ are essentially the same 
as those of the momentum distribution function given in Sec.~II\,F 
in Ref.~\cite{Takada_2016}.

%%%%%%%%%%%%%%%%%%%%%%%%%%%%%%%%%%< Figure 11 >%%%%%%%%%%%%%%%%%%%%%%%%%%%%%%%%%%%%%%%
%-----------------------------------------------------------------------------------%

%%%%%%%%%%%%%%%%%%%%%%< Paragraph 44: n(k) in QMC  >%%%%%%%%%%%%%%%%%%%%%%%%%%
For comparison, the results of $n({\bm k})$ at $r_s=3.99$ in quantum Monte Carlo (QMC) 
simulations for the 3D homogeneous electron gas~\cite{Holzmann_2011}
(blue dotted-dashed curve) and the solid sodium~\cite{Olevano_2012} (green diamonds) 
are depicted in Fig.~\ref{fig:11}(a). The data in QMC are seen to be in reasonably 
good agreement with our present ones, but we do not regard the QMC data as sufficiently 
accurate ones for the following reasons: (i) The QMC data are not verified to satisfy 
the three sum rules. In fact, by just looking at the difference between $n({\bm k})$ 
in QMC and $n_c(k/k_{\rm F})$ perfectly satisfying the three sum rules, we would consider 
that the QMC data might satisfy the $I_2$ sum rule, but they never satisfy other 
sum rules. (ii) The size extrapolation, an inevitable process in QMC to obtain physical 
quantities in the bulk system, is not reliable enough to produce a definite and 
well-converged result, as mentioned in Ref.~\cite{Maebashi_2011}. 
(iii) If we compare the results given by the blue dotted-dashed curve with those by 
the green diamonds, the difference might be ascribed to the band effect. 
This effect, however, should not be large for $|{\bm k}|\! \ll \! k_{\rm F}$, while 
the actual difference between them is unphysically large at ${\bm k}\! \approx 
\! {\bm 0}$, indicating that the magnitude of errors in the QMC evaluation is 
of the order of this size. 

%%%%%%%%%%%%%%%%%%%%%%%%< Section 3-3 >%%%%%%%%%%%%%%%%%%%%%%%%%%%%%%%%%%%%%%%%%%%%%%
\subsection{Quasiparticle renormalization factor}
\label{sec:3C}

%%%%%%%%%%%%%%%%%%%%%< Paragraph 45: z and \delta z  >%%%%%%%%%%%%%%%%%%%%%%%
In all preceding works in which FLT was assumed to be valid, the quasiparticle 
renormalization factor $z^*$ is nothing but $Z({\bm k}_{\rm F},0)^{-1}$. 
In the present study, however, the quasiparticle peak is overlapped with 
that of excitron at $k=k_{\rm F}$ as seen in Fig.~\ref{fig:08}, implying a 
possibility that $z^*$ differs from $Z({\bm k}_{\rm F},0)^{-1}$. Because 
we will come to know in Sec.~\ref{sec:4} that the singularity associated with 
the excitron is well described by a branch cut, we do not expect any contribution 
from the excitron to the jump of $n({\bm k})$ at the Fermi level, 
suggesting us to consider this jump as $z^*$. At the same time, we may 
regard the difference between $Z({\bm k}_{\rm F},0)^{-1}$ and the jump as the 
strength of the excitron peak at the Fermi level, $\delta z$, namely, 
\begin{align}
z^* &\equiv n_c(1\!-\!0^+)-n_c(1\!+\!0^+),
\label{eq:44}
\\
\delta z &\equiv \frac{1}{Z({\bm k}_{\rm F},0)}-z^*.
\label{eq:45}
\end{align}
In actual calculations, we find that $\delta z$ defined in Eq.~(\ref{eq:45}) is 
always positive, which is consistent with our interpretation of $\delta z$ from 
a physical point of view. 

%%%%%%%%%%%%%%%%%%%%%%%%%%%%%%%%%%< Figure 12 >%%%%%%%%%%%%%%%%%%%%%%%%%%%%%%%%%%%%%%%
\begin{figure}[t]
\begin{center}
\includegraphics[scale=0.44,keepaspectratio]{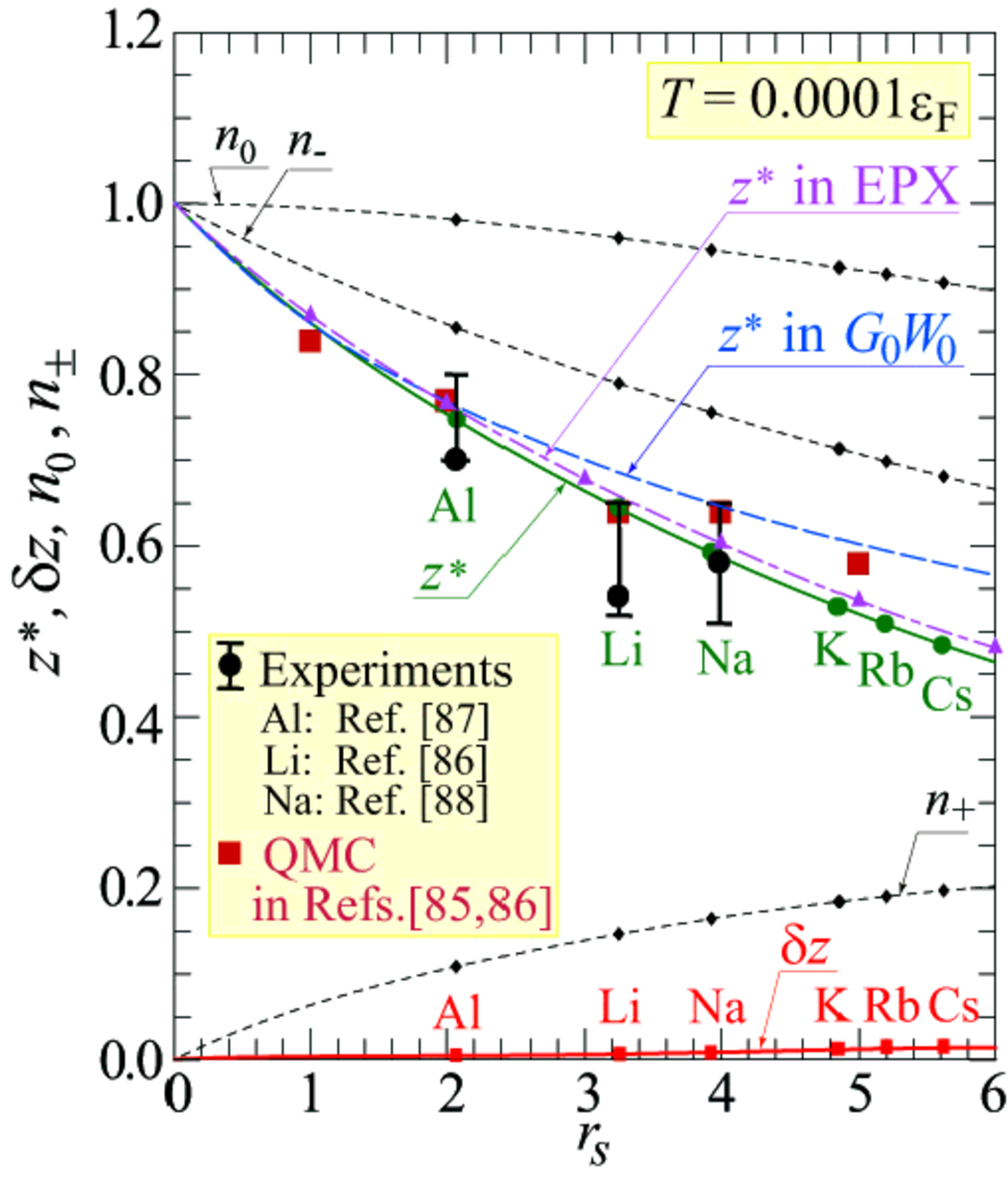}
\end{center}
\caption[Fig.12]{Quasiparticle renormalization factor $z^*$ as a function of 
$r_s$ given by the green solid curve with circles at $T\!=\!10^{-4}\varepsilon_{\rm F}$. 
For comparison, the results in $G_0W_0$ (blue dashed curve) and in EPX (purple 
dotted-dashed curve) are shown, together with the data in QMC simulations 
(big brown squares) and the experimental data (solid circles with error bars) for Al, 
Li, and Na. The data for $\delta z$ representing the strength of the excitron 
peak (red solid curve with squares) are also given, together with those of 
$n_0\equiv n_c(0)$ and $n_{\pm} \equiv n_c(1\!\pm \!0^+)$ by dotted curves.}
\label{fig:12}
\end{figure}
%-----------------------------------------------------------------------------------%

%%%%%%%%%%%%%%%%%< Paragraph 46: Results for z* and \delta z  >%%%%%%%%%%%%%%%%%%%%%%
In Fig.~\ref{fig:12}, we plot our results of $z^*$ and $\delta z$ as a function 
of $r_s$ by green solid curve with circles and red solid curve with squares, 
respectively. For reference, the data for $n_0\ [=\!n_c(0)]$ and $n_{\pm}\ 
[=\!n_c(1\!\pm \!0^+)]$ are also given by the black dotted curves. This figure 
clearly shows that $z^*$ is larger than $\delta z$ by $50\!-\!100$ times, implying 
that the excitron will exert its effect, if any, on bulk physical 
quantities by only a very small amount for $r_s<6$. 

%%%%%%%%%%%%%%%%%%%%< Paragraph 47: comparison  >%%%%%%%%%%%%%%%%%%%%%%%%
For comparison, the preceding results of $z^*$ in both experiments and 
theories are added to the figure: Compton-scattering studies were done 
on Al~\cite{Suortti_2000}, Li~\cite{Schulke_1996,Hiraoka_2020}, and 
Na~\cite{Huotari_2010} and the obtained results are indicated by the big black solid 
circles with error bars, while the data in QMC~\cite{Holzmann_2011,Hiraoka_2020} 
are by the big brown squares. For the sake of reference, the results of $z^*$ 
in $G_0W_0$ and EPX methods are, respectively, shown by the blue 
dashed and purple dotted-dashed curves. We find that (i) our results of 
$z^*$ perfectly reproduce those experimental ones, (ii) they are also 
in very good agreement with the QMC data, and (iii) they are virtually the same 
as the old data in EPX, confirming the accuracy of the results in 
Ref.~\cite{YT_1991c}. 
 
%%%%%%%%%%%%%%%%%%%%< Paragraph 48: comments on z*  >%%%%%%%%%%%%%%%%%%%%%%%%
In the literature, it is often the case that the results of $z^*$ are given 
to be much higher than our present results or even those in $G_0W_0$, but those 
results are not correct, simply because such results originate from an inappropriate 
treatment of the correlation effect near the Fermi level where the energy 
denominators diverge. As discussed in rather details in Ref.~\cite{YT_1991c}, 
any theoretical frameworks without correctly taming the divergent energy 
denominators will fail to produce correct values of $z^*$. 
To be more concrete, a theory with the use of Jastrow-type variational trial 
functions is, in general, not a good choice. Even in QMC or DMC, if those 
simulations start with Jastrow-based variational Monte Carlo, then final results 
may inherit the demerits of Jastrow functions. This might be one of the reasons why 
DMC does not provide correct ${\tilde \varepsilon}_{\bm k}$ near the Fermi 
level in Fig.~\ref{fig:09}. 

%%%%%%%%%%%%%%%%%%%%%%%%< Section 3-4 >%%%%%%%%%%%%%%%%%%%%%%%%%%%%%%%%%%%%%%%%%%%%%%
\subsection{Quasiparticle effective mass}
\label{sec:3D}

\begin{figure}[hbtp]
\begin{center}
\includegraphics[scale=0.46,keepaspectratio]{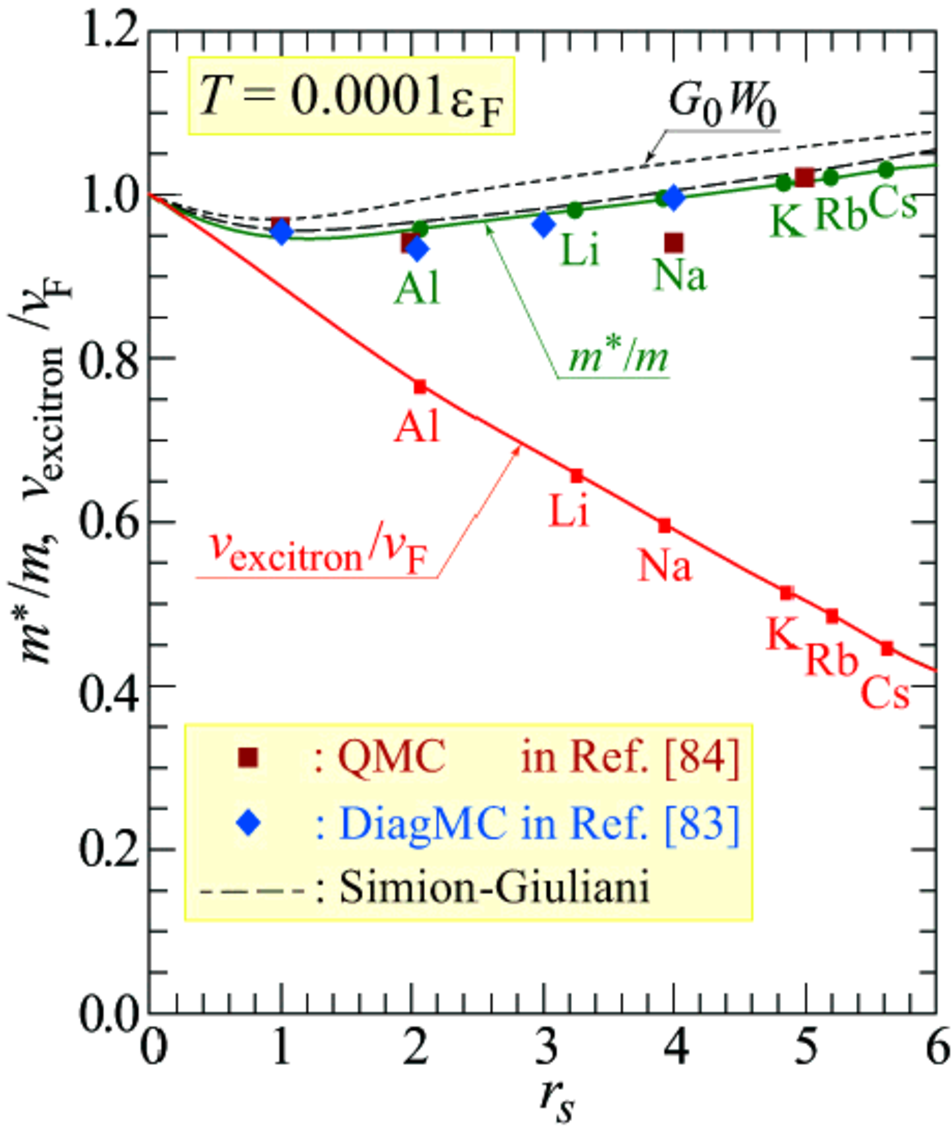}
\end{center}
\caption[Fig.13]{Quasiparticle effective mass in units of the free-electron mass 
$m^*/m$ as a function of $r_s$, together with the excitron velocity in units 
of $v_{\rm F}$. The results in $G_0W_0$, the method by Simion-Giuliani, diagrammatic 
Monte Carlo (DiagMC), and QMC are also shown for comparison.}
\label{fig:13}
\end{figure}

%%%%%%%%%%%%%%%%%%%%< Paragraph 49: Results for m*  >%%%%%%%%%%%%%%%%%%%%%%%%
The quasiparticle effective mass $m^*$ can be determined through the derivative of 
${\tilde \varepsilon}_{\bm k}$ at $k\!=\!k_{\rm F}$ and the obtained result is drawn 
by the green solid curve with circles in Fig.~\ref{fig:13}. Our present result is very 
close to that of Simion and Giuliani~\cite{Simion_2008} given by the black dashed curve 
and it is also in excellent agreement with the very recent data provided by 
diagrammatic Monte Carlo calculation~\cite{Haule_2022} (blue diamonds) and 
QMC~\cite{Holzmann_2023} (brown squares). The result in $G_0W_0$ shown by the black 
dotted curve is not quite the same as ours, but it may still be regarded as a 
semiquantitatively good result. This success of $G_0W_0$ probably reflects 
the fact that ${\tilde \varepsilon}_{\bm k}$ itself is well reproduced by 
$G_0W_0$ due to the strong mutual cancellation between the self-energy effect and 
the vertex correction as a consequence of the Ward identity~\cite{YT_2016b}.
Thus, if we are not concerned with the physics of highly correlated phenomena 
such as excitron, then $G_0W_0$ is a good choice for many purposes, because 
it is computationally very cheap yet provides qualitatively correct results.  

%%%%%%%%%%%%%%%%%%%%%%%%%%%%%%%%%%< Figure 13 >%%%%%%%%%%%%%%%%%%%%%%%%%%%%%%%%%%%%%%%
%-----------------------------------------------------------------------------------%

%%%%%%%%%%%%%%%%%%%%< Paragraph 50: v_excitron  >%%%%%%%%%%%%%%%%%%%%%%%%
Compared with the quasiparticle velocity at the Fermi level $v^*_{\rm F}
\! \equiv \!(m/m^*)v_{\rm F}\!\approx\!v_{\rm F}$, the velocity of excitron 
at the Fermi level $v_{\rm excitron}$, which can be obtained through the derivative 
of $\xi_{\bm k}$ at $k\!=\!k_{\rm F}$, is found to be typically about a half of 
$v^*_{\rm F}$, as seen in Fig.~\ref{fig:13} in which $v_{\rm excitron}$ is shown 
in units of $v_{\rm F}$ by the red solid curve with squares.

%%%%%%%%%%%%%%%%%%%%%%%%%%%%%%%%%%%%%%%%%%%%%%%%%%%%%%%%%%%%%%%%%%%%%%%%%%%%%%%%%%%%%
%%%%%%%%%%%%%%%%%%%%%%%%%%%%%< Section 4 >%%%%%%%%%%%%%%%%%%%%%%%%%%%%%%%%%%%%%%%%%%%
\section{Details of Excitron}
\label{sec:4}

%%%%%%%%%%%%%%%%%%%%< Paragraph 51: preface  >%%%%%%%%%%%%%%%%%%%%%%%%
So far, the excitron is introduced only as a new low-energy peak in 
$A({\bm k},\omega)$, but here we try to understand its features in terms of the 
self-energy, either the retarded one $\Sigma^R({\bm k},\omega\!+\!i\gamma)$ or the thermal 
one $\Sigma({\bm k},i\omega_n)$. For this purpose, we focus exclusively on the 
case of $r_s\!=\!3.93$ at $T\!=\!2\times\! 10^{-4}\varepsilon_{\rm F}$ 
in Secs.~\ref{sec:4A}$-$\ref{sec:4D}, partly because this is a typical example 
exhibiting a clear excitron peak in $A({\bm k},\omega)$ and partly because we can 
obtain a completely convergent result of $\Sigma(K)$ much more easily in this system 
than in those at larger $r_s$ and/or lower $T$. In Sec.~\ref{sec:4E}, we examine 
the $T$-dependence of $\Sigma^R({\bm k},\omega\!+\!i\gamma)$ by changing 
$T$ in the range of $(1$-$8)\times 10^{-4}\varepsilon_{\rm F}$. 

%%%%%%%%%%%%%%%%%%%%%%%%< Section 4-1 >%%%%%%%%%%%%%%%%%%%%%%%%%%%%%%%%%%%%%%%%%%%%%%
\subsection{Characteristics of excitron 
in $\mbox{\boldmath$\Sigma$}^{R}({\bm k},\omega\!+\!i\gamma)$}
\label{sec:4A}

%%%%%%%%%%%%%%%%%%%%%%%%%%%%%%%%%%< Figure 14 >%%%%%%%%%%%%%%%%%%%%%%%%%%%%%%%%%%%%%%%
\begin{figure}[b]
\begin{center}
\includegraphics[scale=0.45,keepaspectratio]{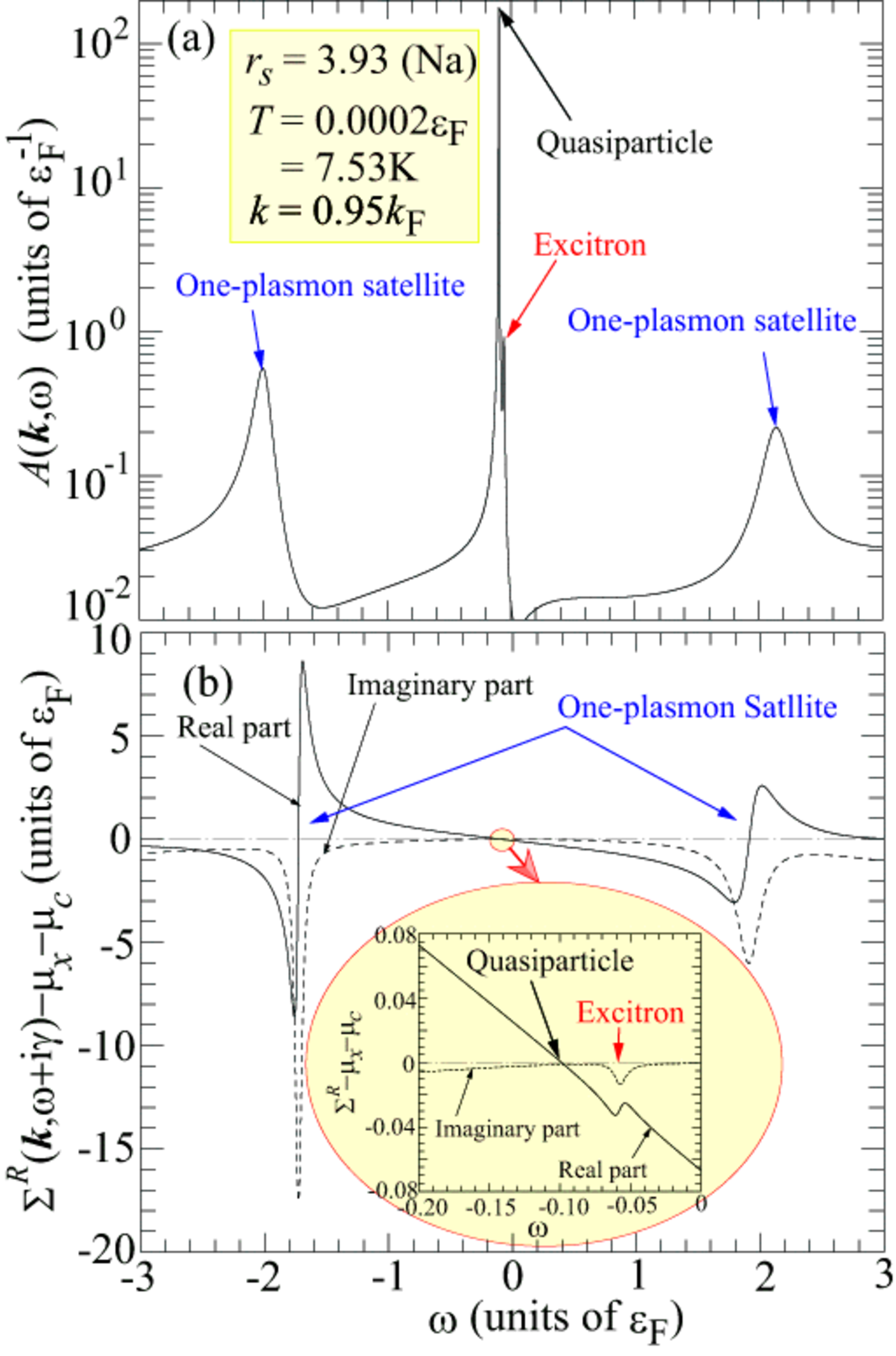}
\end{center}
\caption[Fig.14]{Peaks in the one-particle spectral function $A({\bm k},\omega)$ 
in panel (a) and the corresponding structure in the retarded self-energy 
$\Sigma^R({\bm k},\omega\!+\!i\gamma)$ in panel (b) at $k \!\equiv \!|{\bm k}|\!=
\!0.95k_{\rm F}$ for the 3D homogeneous electron gas at $r_s\!=\!3.93$ and 
$T\!=\!2\times \!10^{-4}\varepsilon_{\rm F}$. The region of $\omega \sim 0$ is 
enlarged to show the behavior leading to the quasiparticle and excitron peaks 
in the inset in panel (b).}
\label{fig:14}
\end{figure}
%-----------------------------------------------------------------------------------%

%%%%%%%%%%%%%%%%%%< Paragraph 52: Retarded self-energy  >%%%%%%%%%%%%%%%%%%%%%%%%
In Fig.~\ref{fig:14}(a), $A({\bm k},\omega)$ in this system is drawn in a logarithmic 
scale as a function of $\omega$ at $k\!\equiv \!|{\bm k}|\!=\!0.95k_{\rm F}$. 
For $\omega$ in the range $(-3\varepsilon_{\rm F},3\varepsilon_{\rm F})$, 
there exist four peaks in $A({\bm k},\omega)$ and the corresponding structure in 
$\Sigma^R({\bm k},\omega\!+\!i\gamma)\!-\!\mu_x\!-\!\mu_c$ is given in Fig.~\ref{fig:14}(b); 
at the quasiparticle peak position, both real and imaginary parts in 
$\Sigma^R({\bm k},\omega\!+\!i\gamma)\!-\!\mu_x\!-\!\mu_c$ vary very smoothly with vanishingly 
small magnitudes, as shown in the inset, in accordance with the assumption in FLT. 
On the other hand, the one-plasmon satellites are associated with large variations 
in both real and imaginary parts in the shape of functions as can be found 
in the Lorentz oscillator model, reflecting an electron motion in the electric field 
induced by the plasmon. At the excitron peak position, although the 
magnitudes of the variations are by far small, about a one-hundredth, the real and 
imaginary parts behave in a way very similar to those at the one-plasmon satellites, 
implying that the excitron peak must also be connected with the motion of an electron 
in the field induced by some kind of low-energy (of the order of $0.1\varepsilon_{\rm F}$) 
excitations. 

%%%%%%%%%%%%%%%%%%%%%%%%%%%%%%%%%%< Figure 15 >%%%%%%%%%%%%%%%%%%%%%%%%%%%%%%%%%%%%%%%
\begin{figure}[b]
\begin{center}
\includegraphics[scale=0.44,keepaspectratio]{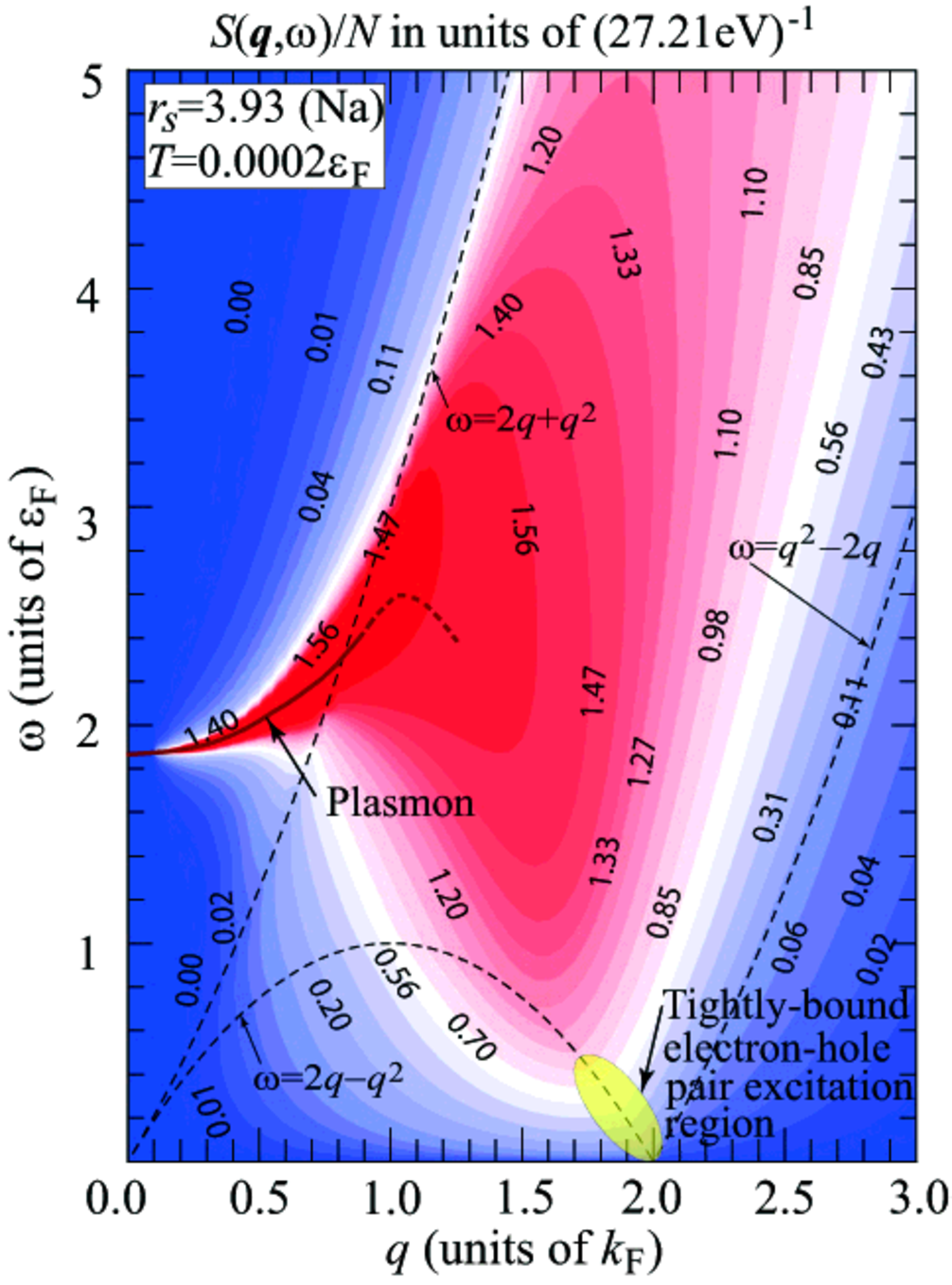}
\end{center}
\caption[Fig.15]{Two-dimensional contour map of the dynamical structure factor 
$S(q,\omega)$ for the electron gas characterized by the same parameters 
as those in Fig.~\ref{fig:14}.}
\label{fig:15}
\end{figure}
%-----------------------------------------------------------------------------------%

%%%%%%%%%%%%%%%%%%%%< Paragraph 53: S(q,w)  >%%%%%%%%%%%%%%%%%%%%%%%%
To pinpoint the relevant low-energy excitations to bring about 
the excitron, we show the calculated result of the structure factor 
$S({\bm q},\omega)$, defined by
\begin{align}
S({\bm q},\omega)=-\frac{1}{\pi}\frac{1}{1-e^{-\omega/T}}\,
{\rm Im}\, Q_c^R({\bm q},\omega),
\label{eq:46}
\end{align}
in Fig.~\ref{fig:15} to demonstrate that, though the plasmon contribution in the range 
of $\omega \approx 1.9-2.4\varepsilon_{\rm F}$ with $q \lesssim k_{\rm F}$ overwhelmingly 
dominates, the most important contribution in the range of $\omega \approx 0-0.2
\varepsilon_{\rm F}$ comes from the tightly bound electron-hole pair excitations with 
$q\! \approx \!2k_{\rm F}$, the region indicated by the yellow shaded area in the figure. 
Incidentally, the result for $S({\bm q},\omega)$ 
in Fig.~\ref{fig:15} is virtually the same as that given for $r_s=4$ in Fig.~1(a) in 
Ref.~\cite{Takada_2016}, basically because the present result for $Q_c^R({\bm q},\omega)$ 
is essentially the same as those in the previous publications, not only in 
Ref.~\cite{Takada_2016} but also in Refs.~\cite{YT_2002,YT_2005}.

\begin{figure}[h]
\begin{center}
\includegraphics[scale=0.40
,keepaspectratio]{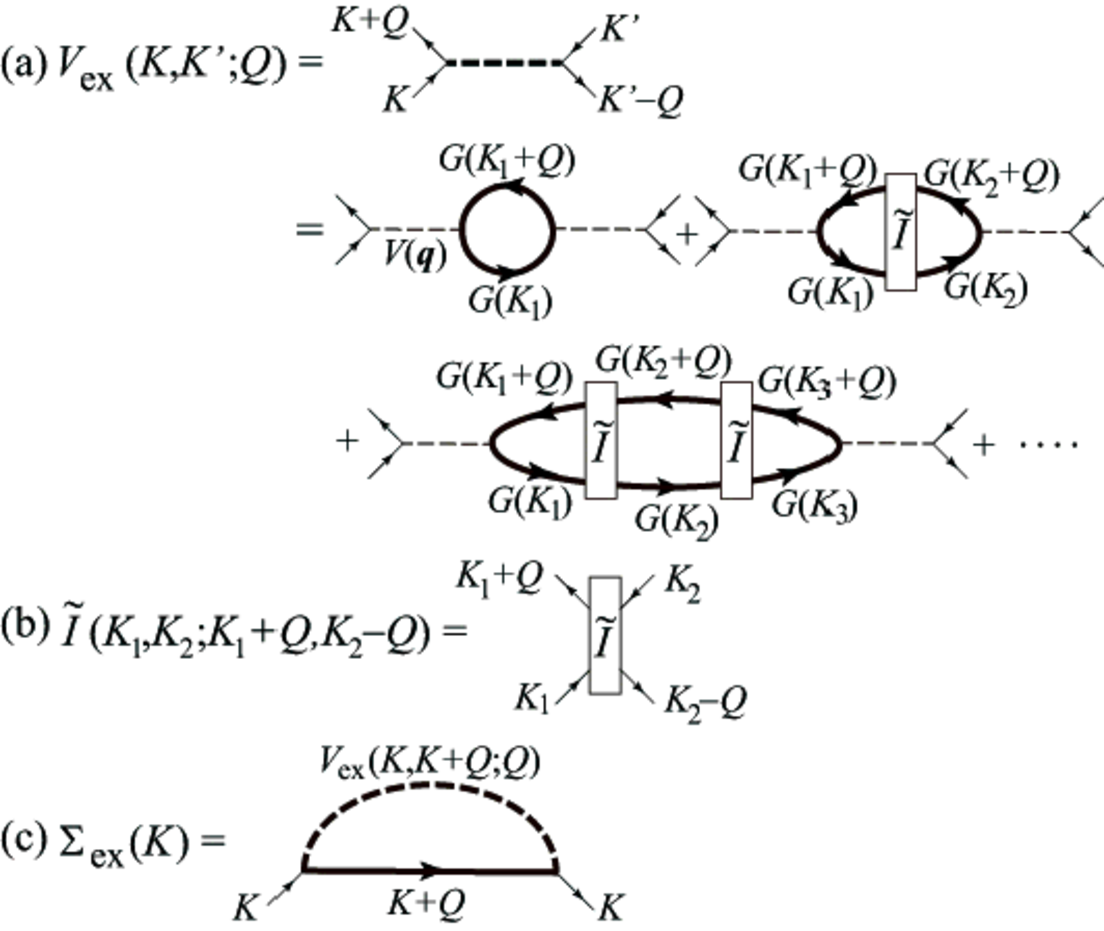}
\end{center}
\caption[Fig.01]{Diagrammatic representation of $V_{\rm ex}(K,K';Q)$ 
the electron-electron attractive interaction induced by multiple excitations of 
tightly bound electron-hole pairs in panel (a), with the electron-hole irreducible 
four-point interaction $\tilde{I}$ in panel (b), and its contribution 
to the self-energy $\Sigma_{\rm ex}(K)$ in panel (c).}
\label{fig:01}
\end{figure}

%%%%%%%%%%%%%%%%%%%%< Paragraph 54: V_ex(K,K+Q)  >%%%%%%%%%%%%%%%%%%%%%%%% 
Now, our task is to study the electron-electron interaction mediated by the 
tightly bound electron-hole pair excitations $V_{\rm ex}(K,K';Q)$ 
and its contribution to the self-energy $\Sigma_{\rm ex}(K)$; diagrammatically, 
they are given in Figs.~\ref{fig:01}(a) and \ref{fig:01}(c), respectively, 
with the electron-hole irreducible (defined in the horizontal view) four-point interaction 
$\tilde{I}(K_1,K_2;K_1\!+\!Q,K_2\!-\!Q)$ in Fig.~\ref{fig:01}(b). 
For $Q \approx 2K_{\rm F}$ and an electron on the Fermi sphere, i.e., 
$|{\bm k}| \! \approx k_{\rm F}$, the dominant contribution to $\Sigma_{\rm ex}(K)$ 
comes from the scattered states $K\!+\!Q$ also on the Fermi sphere. 
Since $|{\bm q}|\! \approx \! 2k_{\rm F}$, this is only possible for ${\bm k}\!+\!{\bm q}\!
\approx \!-{\bm k}$ and ${\bm q}\approx -2{\bm k}$. A similar restriction also applies 
to each pair polarization process in $V_{\rm ex}(K,K+Q;Q)$ in Fig.~\ref{fig:01}(a); 
the important contribution arises only for ${\bm k}_n\! \approx \! {\bm k}$ in $G(K_n)$ 
and ${\bm k}_n\!+\!{\bm q}\! \approx \! -{\bm k}$ in $G(K_n\!+\! Q)$. The above restrictions 
clearly indicate that once we choose an electron characterized by ${\bm k}$, all the 
processes of its scatterings and the associated electron-hole pair excitations occur 
predominantly in either parallel or antiparallel to the ${\bm k}$-direction. 

%%%%%%%%%%%%%%%%%%%%%%%%%%%%%< Figure 15a; formerly Figure 01 >%%%%%%%%%%%%%%%%%%%%%%%%%%%%%%%
%%----------------------------------------------------------------------------------%

%%%%%%%%%%%%%%%%%%%%%%%%%%%%%%%%%%< Figure 16 >%%%%%%%%%%%%%%%%%%%%%%%%%%%%%%%%%%%%%%%
\begin{figure}[b]
\begin{center}
\includegraphics[scale=0.44,keepaspectratio]{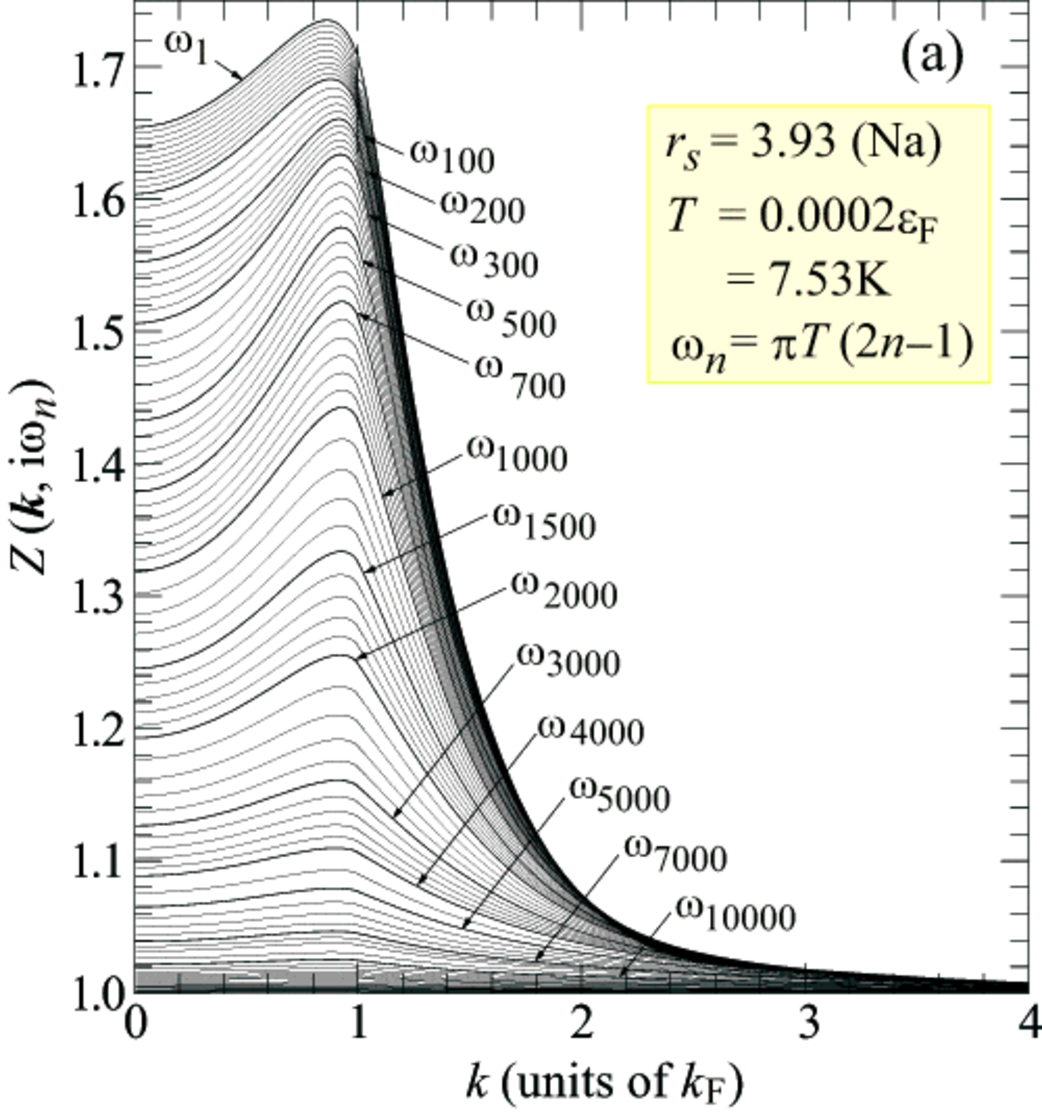}
\includegraphics[scale=0.44,keepaspectratio]{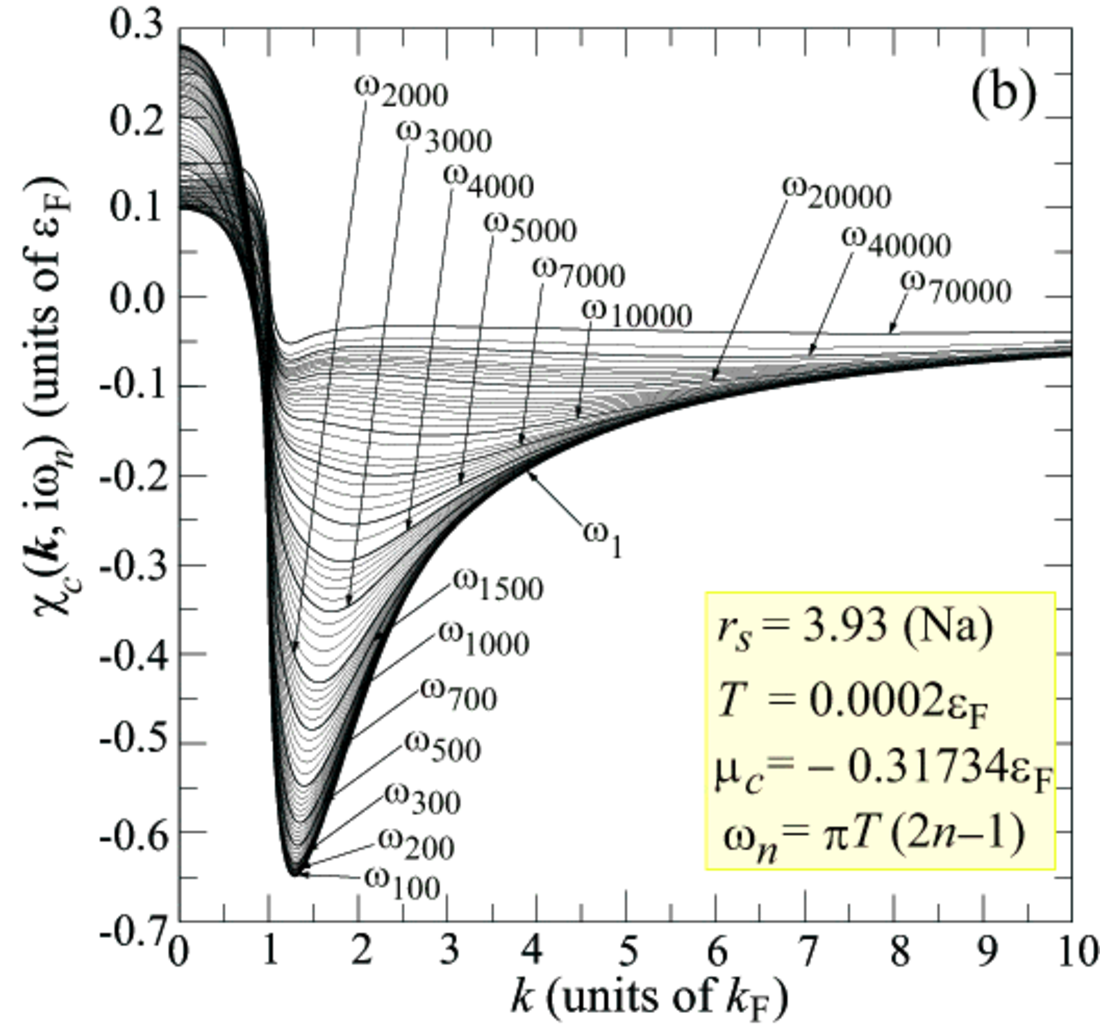}
\end{center}
\caption[Fig.16]{Overall structure of the renormalization function $Z(K)$ in panel (a) and 
that of the correlation part in the self-energy $\chi_c(K)$ in panel (b) for the electron 
gas characterized by the same parameters as those in Fig.~\ref{fig:14}.
}
\label{fig:16}
\end{figure}
%-----------------------------------------------------------------------------------%

%%%%%%%%%%%%%%%%%%%%< Paragraph 55: approximation toV_ex(K,K+Q)  >%%%%%%%%%%%%%%%%%%% 
If we approximate $G(K_n)$ and $G(K_n\!+\! Q)$ by $G_0(K_n)$ and $G_0(K_n\!+\! Q)$, 
respectively, in Fig.~\ref{fig:01}(a), then we can easily obtain an approximate expression for 
$V_{\rm ex}$ as 
\begin{align}
\label{eq:03}
V_{\rm ex} \approx 
-\frac{V({2\bm k}_{\rm F})^2\Pi_0(2K_{\rm F})}
{1\!\!-\tilde{I}(K_{\rm F},-K_{\rm F};-K_{\rm F},K_{\rm F})\Pi_0(2K_{\rm F})/2}.
\end{align}
This is an attractive interaction and its importance was well appreciated long ago 
by the systematic and unbiased survey of the electron-electron interaction in the 
problem of superconductivity in the electron gas~\cite{YT_1988}.
Because the relevant interaction $V_{\rm ex}$ is attractive, we can understand 
why the excitron energy $\xi_{\bm k}$ is lower than ${\tilde \varepsilon}_{\bm k}$.

%%%%%%%%%%%%%%%%%%%%%%%%< Section 4-2 >%%%%%%%%%%%%%%%%%%%%%%%%%%%%%%%%%%%%%%%%%%%%%%
\subsection{Extraction of the singular part in $\mbox{\boldmath$Z$}(K)$}
\label{sec:4B}

%%%%%%%%%%%%%%%%%%%%%%%%%%%%%%%%%%< Figure 17 >%%%%%%%%%%%%%%%%%%%%%%%%%%%%%%%%%%%%%%%
\begin{figure}[b]
\begin{center}
\includegraphics[scale=0.48,keepaspectratio]{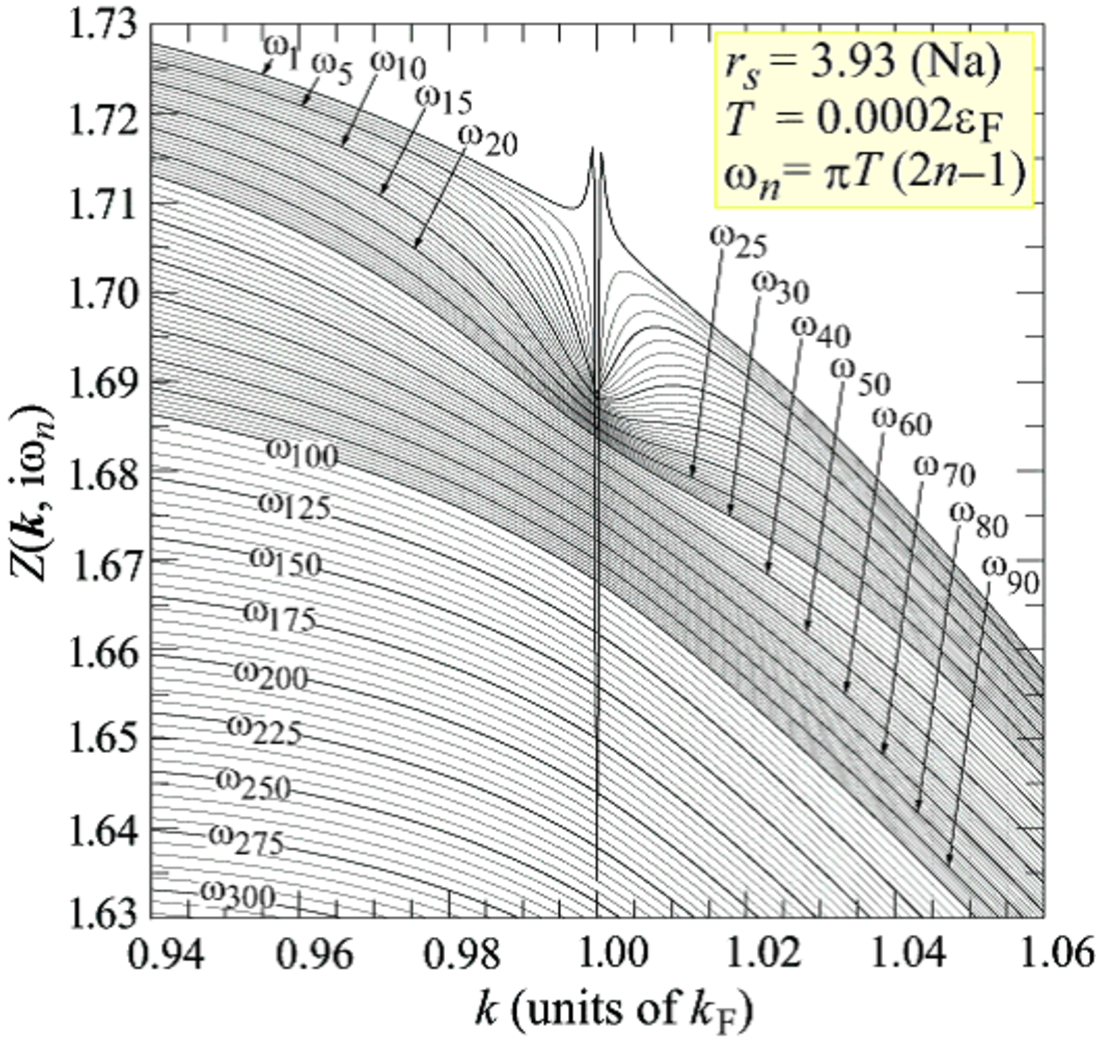}
\end{center}
\caption[Fig.17]{Enlarged view of $Z({\bm k},i\omega_n)$ given in Fig.~\ref{fig:16}(a) 
for $k/k_{\rm F}$ in the range of $0.94$-$1.06$ and $\omega_n$ with $n  \lesssim  300$ 
(or $\omega_n  \lesssim 0.38\varepsilon_{\rm F})$.}
\label{fig:17}
\end{figure}
%-----------------------------------------------------------------------------------%

%%%%%%%%%%%%%%< Paragraph 56: self-energy on imaginary axis  >%%%%%%%%%%%%%%%%%%%%%
After much trial and error, we come to realize that the mathematical feature is better 
seized in terms of $\Sigma({\bm k},i\omega_n)$ rather than $\Sigma^R({\bm k},\omega\!+\!i\gamma)$, 
as long as only numerically obtained data are available to us at the present stage. 
Therefore, let us draw $Z(K)$ the renormalization function and $\chi_c(K)$ 
the correlation contribution to $\chi(K)$ in Figs.~\ref{fig:16}(a) and \ref{fig:16}(b), 
respectively, as a function of $k$ in a wide range from $0$ to $4k_{\rm F}$ [and 
even up to $10k_{\rm F}$ for $\chi_c(K)$] with changing Matsubara frequency $\omega_n$ 
also in a very wide range, i.e., for $n$ from $1$ ($\omega_1\approx 0.0006
\varepsilon_{\rm F}$) to $7\times 10^4$ ($\omega_{70000}\approx 80\varepsilon_{\rm F}$). 
where $\chi_c(K)$ is defined by 
\begin{align}
\chi_c(K) \equiv \chi(K)-\chi_x(k),
\label{eq:47}
\end{align}
with the exchange part of the self-energy $\chi_x(k)$ which is independent of 
$\omega_n$ and calculated as 
\begin{align}
\frac{\chi_x(k)}{\varepsilon_{\rm F}}
= -\frac{2\alpha r_s}{\pi}\left [1+\frac{k_{\rm F}^2-k^2}{2k_{\rm F}k}
\ln \left | \frac{k_{\rm F}+k}{k_{\rm F}-k} \right | \,\right ].
\label{eq:48}
\end{align}
As we see, in the scale of these figures, both $Z(K)$ and $\chi_c(K)$ seem to 
behave quite normally in the whole $\{k,\omega_n\}$ space, in accordance with FLT. 
In fact, even in a much-enlarged scale with $\omega_n$ in the limited range of $n$ 
from $1$ to $100$, we hardly see noticeable variations, much less anomaly, 
in $\chi_c(K)$ even for $k$ in the vicinity of $k_{\rm F}$.

%%%%%%%%%%%%%%%%%%%%< Paragraph 57: anomaly in Z(K)  >%%%%%%%%%%%%%%%%%%%%%%%%
In $Z(K)$, however, we find a conspicuous spike structure for $k$ in the vicinity 
of $k_{\rm F}$ (or $0.99 \lesssim k/k_{\rm F} \lesssim 1.01$) and small $\omega_n$ 
(or $\omega_n/\varepsilon_{\rm F}\lesssim 0.03$). Actually, this anomalous behavior 
can also be faintly seen in Fig.~\ref{fig:16}(a), but it is very clearly found 
through an enlarged view of $Z(K)$, as given in Fig.~\ref{fig:17}. 
This kind of anomaly in $Z(K)$ is never expected in FLT, but here it emerges 
as a strong symptom of possible breakdown of FLT. At the same time, it is found 
to be very localized in $\{k,\omega_n\}$ space as a characteristic feature.

%%%%%%%%%%%%%%%%%%%%%%%%%%%%%%%%%%< Figure 18 >%%%%%%%%%%%%%%%%%%%%%%%%%%%%%%%%%%%%%%%
\begin{figure}[b]
\begin{center}
\includegraphics[scale=0.50,keepaspectratio]{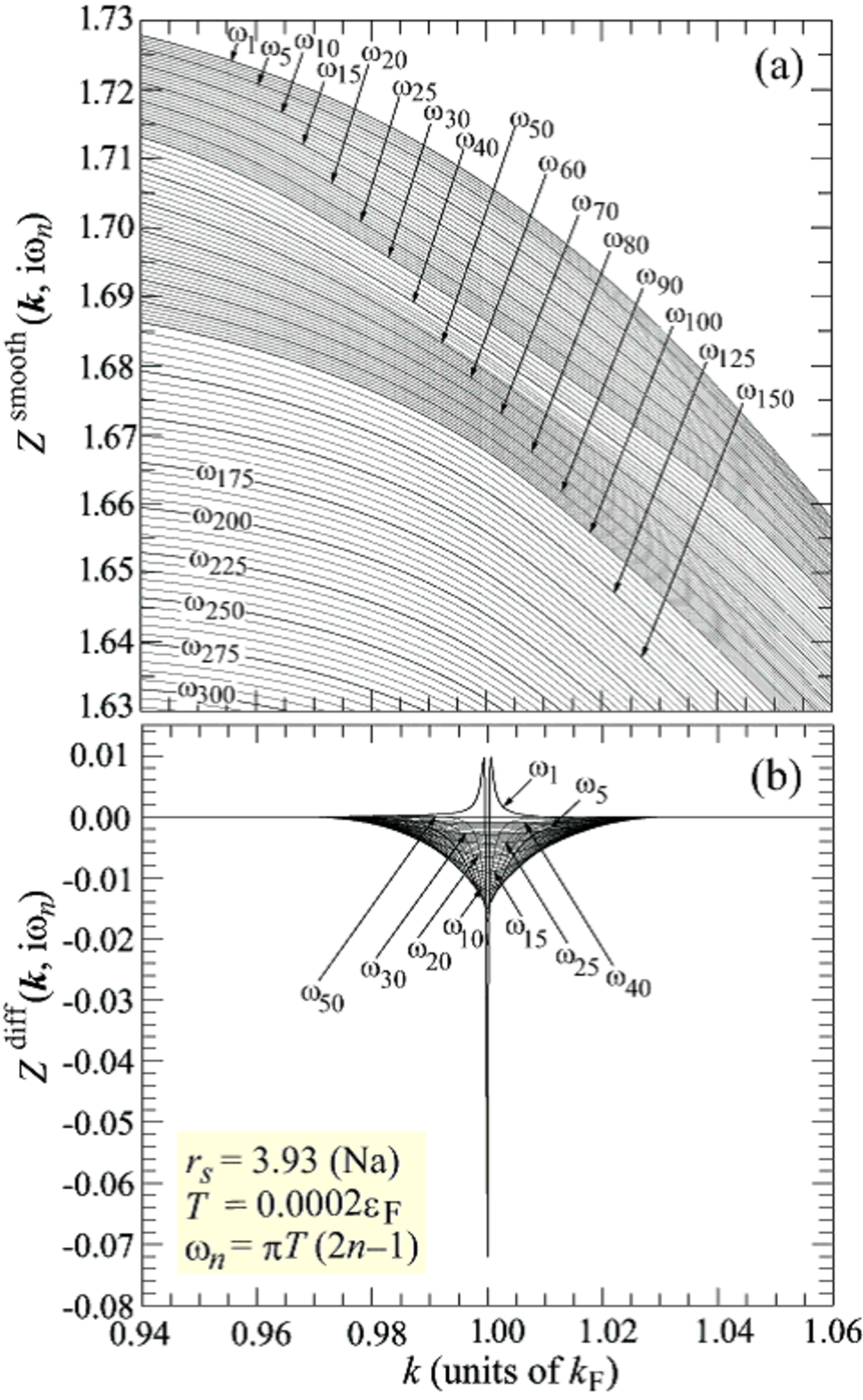}
\end{center}
\caption[Fig.18]{Division of $Z(K)$ given in Fig.~\ref{fig:17} into (a) the smoothed 
part $Z^{\rm smooth}(K)$ and (b) the singular part $Z^{\rm diff}(K)$.}
\label{fig:18}
\end{figure}
%-----------------------------------------------------------------------------------%

%%%%%%%%%%%%%%%%%%%%< Paragraph 58: smoothed self-energy  >%%%%%%%%%%%%%%%%%%%%%%%%
Because of this locality, we are tempted to extract this anomalous structure from 
$Z(K)$ by taking the difference between $Z(K)$ and $Z^{\rm smooth}(K)$, 
the latter of which is a smoothed part of $Z(K)$ obtained by the cubic-spline 
interpolation along $k$-axis with the exclusion of the mesh points in the interval 
$(0.97k_{\rm F},1.03k_{\rm F})$ at each $\omega_n$ with $n \leq 50$. For $n > 50$, 
$Z^{\rm smooth}(K)$ is tentatively taken as $Z(K)$ itself. In Figs.~\ref{fig:18}(a) and (b), 
both $Z^{\rm smooth}(K)$ and $Z^{\rm diff}(K) \ [\equiv \! Z(K)-Z^{\rm smooth}(K)]$ 
are drawn, respectively. As we see, $Z^{\rm smooth}(K)$ is rather smooth and has no 
abnormal structure in the whole $\{k,\omega_n\}$ space. On the other hand, 
$Z^{\rm diff}(K)$ has a distinctive particle-hole symmetric structure and its magnitude 
is not small only in the very limited region in $\{k,\omega_n\}$ space, or more concretely, 
for $|k-k_{\rm F}| \! \lesssim \! 0.025k_{\rm F}$ and $\omega_n\! \lesssim \! 
0.05\varepsilon_{\rm F}$ (or $n \lesssim 40$). 

%%%%%%%%%%%%%%%%%%%%< Paragraph 59: comments on \chi(K)  >%%%%%%%%%%%%%%%%%%%%%%%%
Incidentally, a similar extraction procedure cannot be adopted to produce 
$\chi_c^{\rm smooth}(K)$ and $\chi_c^{\rm diff}(K)$ at this stage, mainly because 
$\chi_c(K)$ does not change much with $\omega_n$ in the small-$\omega_n$ region, 
making it difficult to clearly identify the anomalous structure in $\chi_c(K)$. 
In Sec.~\ref{sec:4C}, however, we shall explain an alternative procedure to 
unambiguously define both $\tilde{\chi}_c^{\rm smooth}(K)$ and 
$\tilde{\chi}_c^{\rm diff}(K)$ from the division of $\chi_c(K)$ in perfectly 
consistent with the redefined division of $Z(K)$ into $\widetilde{Z}^{\rm smooth}(K)$ 
plus $\widetilde{Z}^{\rm diff}(Z)$.

%%%%%%%%%%%%%%%%%%%%%%%%< Section 4-3 >%%%%%%%%%%%%%%%%%%%%%%%%%%%%%%%%%%%%%%%%%%%%%%
\subsection{Branch-cut singularity}  % in $\mbox{\boldmath$G$}(K)$}
\label{sec:4C}

%%%%%%%%%%%%%%%%%%%%< Paragraph 60: 1D Luttinger liquid  >%%%%%%%%%%%%%%%%%%%%%%%%
Fascinated by the mathematically beautiful mirror-symmetry in $Z^{\rm diff}(K)$ with 
respect to $k$ in reference to $k_{\rm F}$ at each $\omega_n$, we proceed to express 
it in an analytically closed form. The discussion on $\Sigma_{\rm ex}$ and $V_{\rm ex}$ 
in Sec.~\ref{sec:4A} indicates that we might be able to treat our problem in reference 
to 1D physics, because the virtual scattering processes of an electron specified 
by ${\bm k}$ on the Fermi surface, along with the multiple electron-hole pair 
excitations to bring about the excitron, occur predominantly in either parallel 
or antiparallel to the ${\bm k}$-direction. Now, in some particularly simple models 
in the 1D Tomonaga-Luttinger liquids~\cite{Tomonaga_1950,Luttinger_1963}, an analytic 
form for $G(K) \! \equiv \! G_{\rm TL}(K)$ is exactly known as~\cite{Maslov_2005} 
\begin{align}
G_{\rm TL}(K)=\frac{1}{\sqrt{i\omega_n-\varepsilon_h(k)}\sqrt{i\omega_n-\varepsilon_s(k)}},
\label{eq:49}
\end{align}
where $\varepsilon_h(k)$ and $\varepsilon_s(k)$ are, respectively, ``holon'' and 
``spinon'' dispersion relations in 1D physics.
 
%%%%%%%%%%%%%%%%%%< Paragraph 61: division of the self-energy  >%%%%%%%%%%%%%%%%%%%%%
Inspired by this simple expression for $G_{\rm TL}(K)$ with possessing branch-cut 
singularities, we shall take a heuristic approach to developing an analytic 
expression for $Z^{\rm diff}(K)$ by starting with the redefined division of $\Sigma(K)$ as
\begin{align}
\Sigma(K)=\widetilde{\Sigma}_{\rm smooth}(K)+\widetilde{\Sigma}_{\rm diff}(K),
\label{eq:50}
\end{align}
with
\begin{subequations}
\label{eq:51ab}
\begin{align}
\widetilde{\Sigma}_{\rm smooth}(K)=& 
\left [1-\widetilde{Z}^{\rm smooth}(K) \right ]i\omega_n
\nonumber \\
&+\chi_x(k)+\tilde{\chi}_c^{\rm smooth}(K),
\label{eq:51a}
\\
\widetilde{\Sigma}_{\rm diff}(K)=& -\widetilde{Z}^{\rm diff}(K)i\omega_n
+\tilde{\chi}_c^{\rm diff}(K).
\label{eq:51b}
\end{align}
\end{subequations}
Here, $\widetilde{\Sigma}_{\rm smooth}(K)$ is regarded as such a self-energy 
in the electron gas as has been considered in FLT and 
$\widetilde{\Sigma}_{\rm diff}(K)$ is supposed to accurately take account of 
the anomalous structure in $\Sigma(K)$ by the assumption of the following 
analytic form: 
\begin{align}
\widetilde{\Sigma}_{\rm diff}(K)=&-\sqrt{i\omega_n-\xi_{\bm k}}
\sqrt{i\omega_n-\xi^*_{\bm k}(i\omega_n)}
\nonumber \\
 &+\frac{\sqrt{\alpha_{\infty}(i\omega_n)}}{2}(i\omega_n-\xi_{\bm k})
 \nonumber \\
 &+\frac{1}{2\sqrt{\alpha_{\infty}(i\omega_n)}}
  \left [i\omega_n-\xi^*_{\bm k}(i\omega_n)\right ],   
\label{eq:52}
\end{align}
where $\xi_{\bm k}$ is chosen as the excitron dispersion relation given 
in Fig.~\ref{fig:09} and $\xi^*_{\bm k}(i\omega_n)$ is defined by 
$\xi^*_{\bm k}(i\omega_n) \equiv \alpha_{\bm k}(i\omega_n)\xi_{\bm k}$ 
with the introduction of a parameter $\alpha_{\bm k}(i\omega_n)$. As we shall see, 
in the limit of $|\xi_{\bm k}/\omega_n| \gg 1$, $\alpha_{\bm k}(i\omega_n)$ 
becomes independent of ${\bm k}$ and the value in this limit is written as 
$\alpha_{\infty}(i\omega_n)$, a parameter involved in Eq.~(\ref{eq:52}).
 
%%%%%%%%%%%%%%%%%< Paragraph 62: condition for \alpha(iw)  >%%%%%%%%%%%%%%%%%%%%%
Physically, $\xi_{\bm k}$ is supposed to represent a ``collective-charge'' 
excitation (or a holon-like excitation in the terminology of 1D physics) and 
thus it is considered as the excitron dispersion. On the other hand, 
we presume that $\xi^*_{\bm k}(i\omega_n)$ corresponds to a spinon-like excitation 
whose dispersion relation is not much different from the quasiparticle dispersion 
${\tilde \varepsilon}_{\bm k}$ in the electron gas, leading us to the condition 
that $\alpha_{\bm k}(i\omega_n)$ should not be less than unity due to the fact that 
$\xi_{\bm k}/\tilde{\varepsilon}_{\bm k}<1$ in the present case.   

%%%%%%%%%%%%%%%%%%%%< Paragraph 63: more on \alpha(iw)  >%%%%%%%%%%%%%%%%%%%%%%%%
In this theoretical framework, $\alpha_{\bm k}(i\omega_n)$ is a single free parameter 
and plays an important role in $\widetilde{\Sigma}_{\rm diff}(K)$; if 
$\alpha_{\bm k}(i\omega_n)$ is taken to be unity, $\widetilde{\Sigma}_{\rm diff}(K)$ 
vanishes completely, implying that the strength of the anomaly is determined only 
by the degree of deviation of $\alpha_{\bm k}(i\omega_n)$ from unity. We also note 
that the $\omega_n$-dependence in $\alpha_{\bm k}(i\omega_n)$ seems to be indispensable 
in describing a branch-cut singularity in not exactly 1D systems, as opposed to 
purely 1D systems in which such dependence is absent as seen in Eq.~(\ref{eq:49}). 
Effects of the motions tangential to the 1D axis on the branch-cut singularity 
are assumed, to a large extent, to be effectively included by this $\omega_n$-dependence.

%%%%%%%%%%%%%%%%%%%%< Paragraph 64: Z_diff  >%%%%%%%%%%%%%%%%%%%%%%%%
In view of the importance of $\alpha_{\bm k}(i\omega_n)$, we make a rather detailed 
explanation of the procedure to determine it by using the numerically-obtained 
data of $Z^{\rm diff}(K)$. From Eq.~(\ref{eq:52}), we can write 
$\widetilde{Z}^{\rm diff}(K)$ and $\tilde{\chi}_c^{\rm diff}(K)$ as
\begin{subequations}
\label{eq:53ab}
\begin{align}
\widetilde{Z}^{\rm diff}(K) =& [S_{-}(K)S^*_{+}(K)+S_{+}(K)S^*_{-}(K)]/\omega_n
\nonumber \\
&-\frac{\sqrt{\alpha_{\infty}(i\omega_n)}}{2}
\left [ 1+\frac{1}{\alpha_{\infty}(i\omega_n)}\right],
\label{eq:53a}
\\
\tilde{\chi}_c^{\rm diff}(K) =& S_{+}(K)S^*_{+}(K)-S_{-}(K)S^*_{-}(K)
\nonumber \\
&-\frac{\sqrt{\alpha_{\infty}(i\omega_n)}}{2}
\left [\xi_{\bm k}+\frac{\xi^*_{\bm k}(i\omega_n)}{\alpha_{\infty}(i\omega_n)}\right ],
\label{eq:53b}
\end{align}
\end{subequations}
respectively, where $S_{\pm}(K)$ is defined as
\begin{align}
S_{\pm}(K) \equiv \sqrt{\frac{\sqrt{\omega_n^2+\xi_{\bm k}^2}\pm \xi_{\bm k}}{2}},
\label{eq:54}
\end{align}
and $S^*_{\pm}(K)$ is defined analogously with the replacement of $\xi_{\bm k}$ with 
$\xi^*_{\bm k}(i\omega_n)$. From Eqs.~(\ref{eq:53a}) and (\ref{eq:53b}), together 
with Eq.~(\ref{eq:54}), it is easy to see that both $\widetilde{Z}^{\rm diff}(K)$ 
and $\tilde{\chi}_c^{\rm diff}(K)$ vanish in the limit of $|\xi_{\bm k}/\omega_n| \gg 1$. 

%%%%%%%%%%%%%%%%%%%%%%%%%%%%%%%%%%< Figure 19>%%%%%%%%%%%%%%%%%%%%%%%%%%%%%%%%%%%%%%%
\begin{figure}[b]
\begin{center}
\includegraphics[scale=0.46,keepaspectratio]{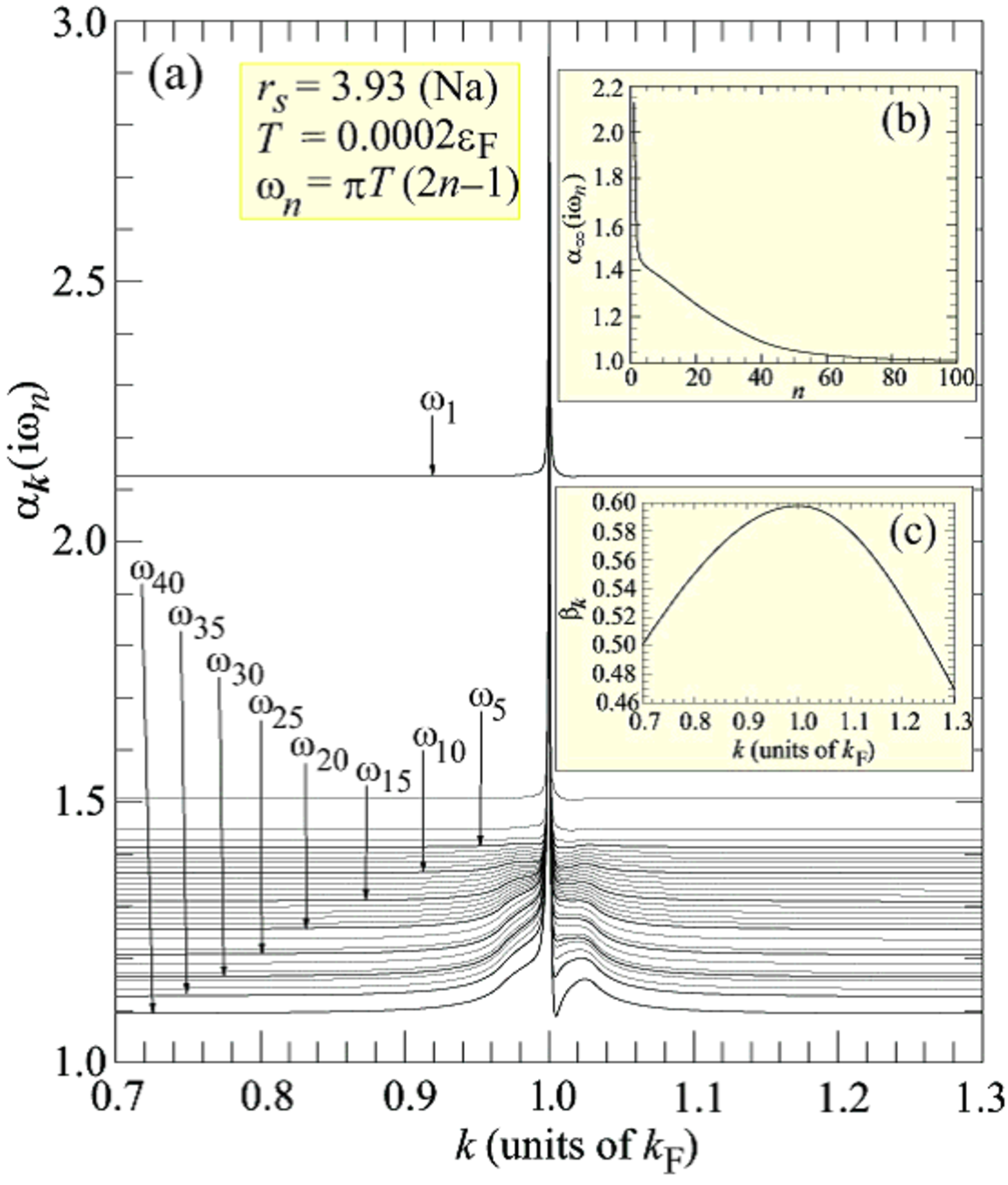}
\end{center}
\caption[Fig.19]{Overall structure of the parameter $\alpha_{\bm k}(i\omega_n)$ 
as a function of $k\!=\!|{\bm k}|$ for $\omega_n$ in the range of $n\!=\!1\!-\!50$ 
in panel (a), its asymptotic form in the limit $|\xi_{\bm k}| \!\gg \!|\omega_n|$, 
$\alpha_{\infty}(i\omega_n)$, with the increase of $n$ in panel (b), and 
the parameter $\beta_{\bm k}$ related to the excitron dispersion relation 
as a function of $k\!=\!|{\bm k}|$ in panel (c).}
\label{fig:19}
\end{figure}
%-----------------------------------------------------------------------------------%

%%%%%%%%%%%%%%%%%%%%< Paragraph 65: Results for \alpha_0(iw)  >%%%%%%%%%%%%%%%%%%%%%%%%
Now, because of $\xi_{\bm k}\!=\!\xi^*_{\bm k}(i\omega_n)\!=\!0$ at the fermi level, 
we obtain $\widetilde{Z}^{\rm diff}({\bm k}_{\rm F},i\omega_n)\!=\!1\!-\!
[\sqrt{\alpha_{\infty}(i\omega_n)}\!+\!1/\sqrt{\alpha_{\infty}(i\omega_n)}]/2$. Then, 
by equating this value of $\widetilde{Z}^{\rm diff}({\bm k}_{\rm F},i\omega_n)$ 
to the numerically obtained $Z^{\rm diff}({\bm k}_{\rm F},i\omega_n)$, we can determine 
$\alpha_{\infty}(i\omega_n)$ as
\begin{align}
\alpha_{\infty}(i\omega_n)=\left (Z_0+\sqrt{Z_0^2-1}\right )^2,
\label{eq:55}
\end{align}
with $Z_0\!\equiv\! 1\!-\!Z^{\rm diff}({\bm k}_{\rm F},i\omega_n)$. This value of 
$\alpha_{\infty}(i\omega_n)$ is chosen under the condition of $\alpha_{\infty}(i\omega_n) 
\geq 1$ and the obtained result is shown in Fig.~\ref{fig:19}(b) as a function 
of $n$. Actually, because $Z^{\rm diff}({\bm k}_{\rm F},i\omega_n)$ is available 
only for $n \leq 50$, $\alpha_{\infty}(i\omega_n)$ for $n>50$ is not determined 
by Eq.~(\ref{eq:55}) but by $\alpha_{\infty}(i\omega_n)\!=\!1\!+\!\delta 
\alpha_{\infty}(i\omega_n)$ with the extrapolation of the data 
$\{ \delta \alpha_{\infty}(i\omega_n)\}_{n=1,\cdots,50}$ under the assumption of 
the power-law decay of $\delta \alpha_{\infty}(i\omega_n)\ [\equiv 
\alpha_{\infty}(i\omega_n)-1]$ with the increase of $n$ from $50$.  

%%%%%%%%%%%%%%%%%%%%< Paragraph 66: Results for \alpha(iw)  >%%%%%%%%%%%%%%%%%%%%%%%%
Once the data of $\{ \alpha_{\infty}(i\omega_n) \}_{n=1,2,3,\cdots}$ are known, 
we can determine $\alpha_{\bm k}(i\omega_n)$ at each ${\bm k}$ by accurately 
solving the equation of $\widetilde{Z}^{\rm diff}(K)=Z^{\rm diff}(K)$ with 
the use of the Newton-Raphson method~\cite{Ypma_1995} in which we employ 
the partial derivative of $\widetilde{Z}^{\rm diff}(K)$ with respect to 
$\alpha_{\bm k}(i\omega_n)$ as
\begin{align}
\frac{\partial \widetilde{Z}^{\rm diff}(K)}{\partial \alpha_{\bm k}(i\omega_n)}
=&\frac{\xi_{\bm k}}{2\omega_n\sqrt{\omega_n^2+{\xi^*_{\bm k}(i\omega_n)}^2}}
\nonumber \\
&\times [S_{-}(K)S^*_{+}(K)-S_{+}(K)S^*_{-}(K)],
\label{eq:56}
\end{align}
and choose $\alpha_{\infty}(i\omega_n)$ as an initial input in the iterative 
solution for $n \leq 50$. For $n>50$, $\alpha_{\bm k}(i\omega_n)$ is determined 
by an extrapolation method similar to that for $\alpha_{\infty}(i\omega_n)$.
The obtained result of $\alpha_{\bm k}(i\omega_n)$ for $n \leq 50$ 
is given in Fig.~\ref{fig:19}(a) from which we see that for $|k\!-\!k_{\rm F}| 
\gtrsim 0.02k_{\rm F}$ in the important range of $\omega_n$ with $n \lesssim 30$, 
$\alpha_{\bm k}(i\omega_n)$ is independent of ${\bm k}$ and essentially 
the same as $\alpha_{\infty}(i\omega_n)$. Even for $n \gtrsim 30$, we see that 
$\alpha_{\bm k}(i\omega_n) \approx \alpha_{\infty}(i\omega_n)$ for 
$|k\!-\!k_{\rm F}| \gtrsim 0.05k_{\rm F}$, revealing that $\alpha_{\infty}(i\omega_n)$ 
is the most important parameter to describe $\widetilde{\Sigma}_{\rm diff}(K)$. 
In Fig.~\ref{fig:19}(c), the parameter $\beta_{\bm k}$ defined by $\beta_{\bm k} 
\equiv \xi_{\bm k}/\varepsilon_{\bm k}$ is also shown. Note that 
$\beta_{\bm k}$ at $|{\bm k}|\!=\!k_{\rm F}$ is nothing but $v_{\rm excitron}/v_{\rm F}$. 
The deviation of $\beta_{\bm k}$ from $\beta_{{\bm k}_{\rm F}}$ represents 
the degree of the departure from the linear dispersion in $\xi_{\bm k}$.

%%%%%%%%%%%%%%%%%%%%%%%%%%%%%%%%%%< Figure 20 >%%%%%%%%%%%%%%%%%%%%%%%%%%%%%%%%%%%%%%%
\begin{figure}[b]
\begin{center}
\includegraphics[scale=0.46,keepaspectratio]{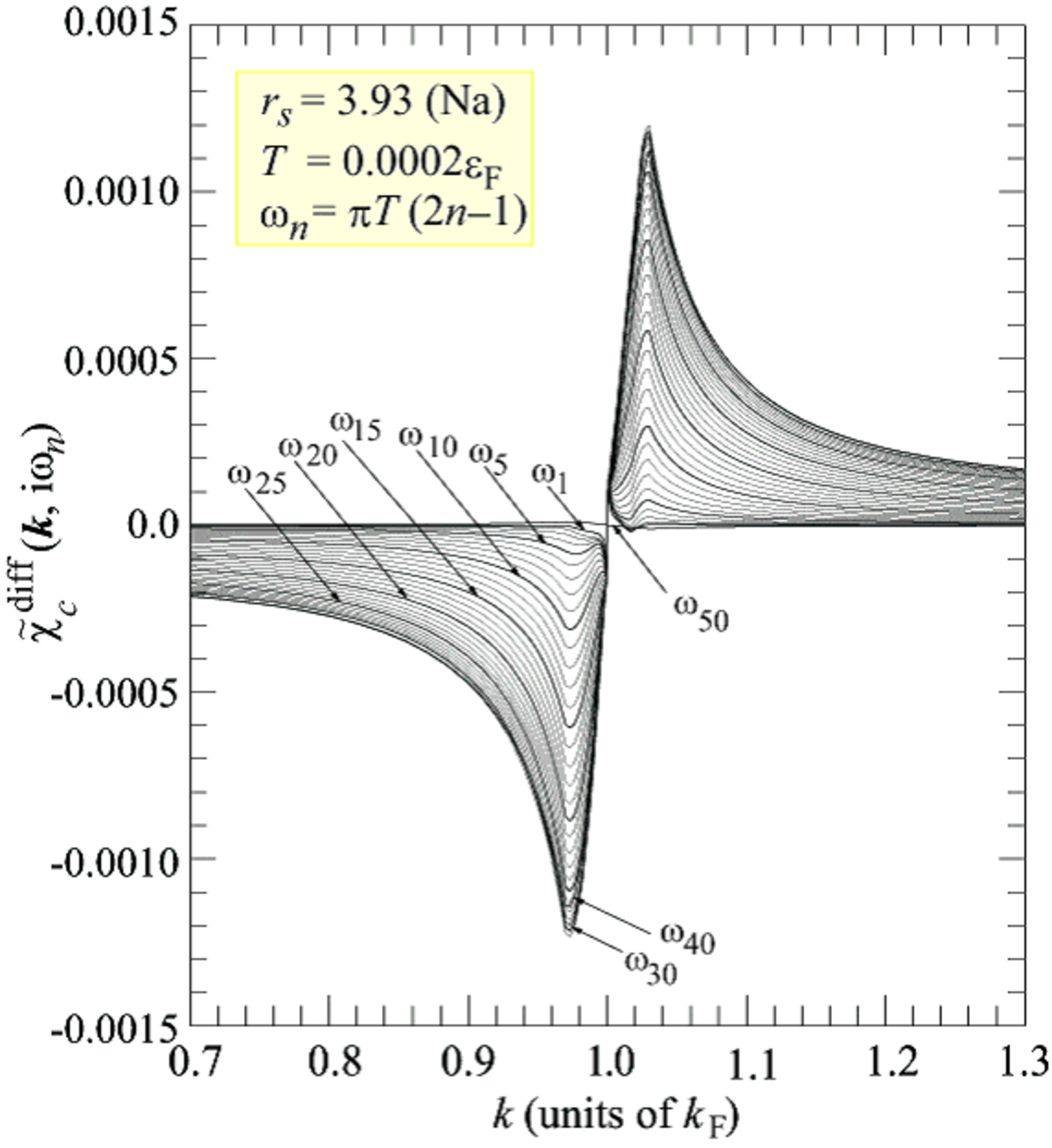}
\end{center}
\caption[Fig.20]{Singular part $\tilde{\chi}_c^{\rm diff}(K)$ in $\chi_c(K)$ 
as a function of $k\!=\!|{\bm k}|$ for small $\omega_n$ with $n\!=\!1\!-\!50$.}
\label{fig:20}
\end{figure}
%-----------------------------------------------------------------------------------%

%%%%%%%%%%%%%%%%%%%%< Paragraph 67: redefined Z_diff  >%%%%%%%%%%%%%%%%%%%%%%%%
Having completely specified the parameter $\alpha_{\bm k}(i\omega_n)$ in the whole 
$\{{\bm k},\omega_n\}$ space, we can employ Eqs.~(\ref{eq:53a}) and (\ref{eq:53b})
to calculate both $\widetilde{Z}^{\rm diff}(K)$ and $\tilde{\chi}_c^{\rm diff}(K)$, 
with which we can also determine $\widetilde{Z}^{\rm smooth}(K)\ [\equiv \! Z(K)
\!-\!\widetilde{Z}^{\rm diff}(K)]$ and $\tilde{\chi}_c^{\rm smooth}(K)\ [\equiv \! 
\chi_c(K)\!-\!\tilde{\chi}_c^{\rm diff}(K)]$ unambiguously. The obtained 
$\widetilde{Z}^{\rm diff}(K)$ and $\widetilde{Z}^{\rm smooth}(K)$ are, 
respectively, found to be virtually the same as $Z^{\rm diff}(K)$ and 
$Z^{\rm smooth}(K)$ given in Fig.~\ref{fig:18} and thus we suppress to show those 
redefined functions here. 

%%%%%%%%%%%%%%%%%%%%< Paragraph 68: redefined \chi(K)  >%%%%%%%%%%%%%%%%%%%%%%%%
In Fig.~\ref{fig:20}, $\tilde{\chi}_c^{\rm diff}(K)$ is drawn to show that its behavior 
is qualitatively different from that of $\widetilde{Z}^{\rm diff}(K)$; as opposed to 
the locality of $Z^{\rm diff}(K)$ or $\widetilde{Z}^{\rm diff}(K)$, 
$\tilde{\chi}_c^{\rm diff}(K)$ is considerably extended in the $k$-axis. We also note 
that its magnitude is quite small, of the order of $0.001\varepsilon_{\rm F}$, 
compared with that of $\chi_c(K)$ of the order of $\varepsilon_{\rm F}$, 
making $\tilde{\chi}_c^{\rm smooth}(K)$ virtually the same as $\chi_c(K)$. 
Those features specific to $\tilde{\chi}_c^{\rm diff}(K)$ are the reasons 
why we could not find any anomalous behavior in $\chi_c(K)$ in the first place. 
Asymmetry with respect to $k$ in reference to $k_{\rm F}$ in 
$\tilde{\chi}_c^{\rm diff}(K)$ is another interesting point to note.
 
%%%%%%%%%%%%%%%%%%%%%%%%< Section 4-4 >%%%%%%%%%%%%%%%%%%%%%%%%%%%%%%%%%%%%%%%%%%%%%%
\subsection{Analysis of 
$\mbox{\boldmath$\widetilde{\Sigma}$}_{\rm diff}^{R}({\bm k},\omega\!+\!i\gamma)$}
\label{sec:4D}

%%%%%%%%%%%%%%%%%%%%%%%%%%%%%%%%%%< Figure 21 >%%%%%%%%%%%%%%%%%%%%%%%%%%%%%%%%%%%%%%%
\begin{figure}[b]
\begin{center}
\includegraphics[scale=0.45,keepaspectratio]{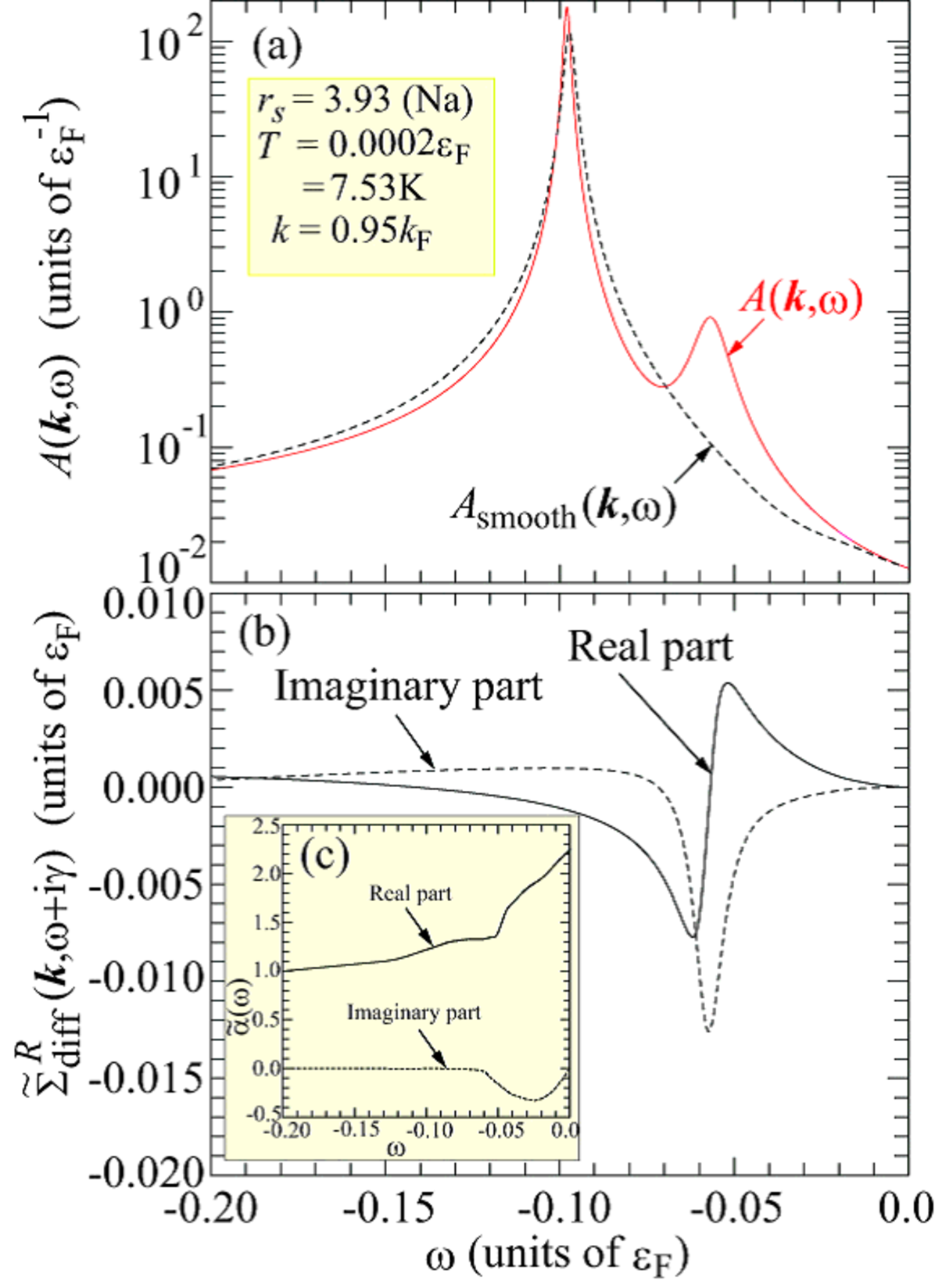}
\end{center}
\caption[Fig.21]{Comparison of $A({\bm k},\omega)$ with $A_{\rm smooth}({\bm k},\omega)$ 
the one-particle spectral function corresponding to the smoothed self-energy is made 
in panel (a) for the case of $k\!=\!|{\bm k}|\!=\!0.95k_{\rm F}$. The singular part of 
the retarded self-energy and the analytically continued ${\tilde \alpha}(\omega)$ from 
$\alpha_{\infty}(i\omega_n)$ are, respectively, shown in panels (b,c).}
\label{fig:21}
\end{figure}
%-----------------------------------------------------------------------------------%

%%%%%%%%%%%%%%%%%%%%< Paragraph 69: retarded \Sigma_diff  >%%%%%%%%%%%%%%%%%%%%%%%%
At first glance, one might think that we can easily make an analytic continuation of 
$\widetilde{\Sigma}_{\rm diff}(K)$ to 
$\widetilde{\Sigma}_{\rm diff}^R({\bm k},\omega\!+\!i\gamma)$ 
by just changing $i\omega_n$ into $\omega+i\gamma$ in Eq.~(\ref{eq:52}), but actually 
it is not so simple due to the presence of $\alpha_{\bm k}(i\omega_n)$; 
the analyticity property of $\alpha_{\bm k}(i\omega_n)$ is not precisely known. 
Thus, we employ the usual analytic continuation method (or Pad\'{e} approximants) 
to obtain $\widetilde{\Sigma}_{\rm smooth}^R({\bm k},\omega\!+\!i\gamma)$ from 
$\widetilde{\Sigma}_{\rm smooth}(K)$ defined in Eq.~(\ref{eq:51a}) and then we 
determine $\widetilde{\Sigma}_{\rm diff}^R({\bm k},\omega\!+\!i\gamma)$ as the difference 
between $\Sigma^R({\bm k},\omega\!+\!i\gamma)$ and 
$\widetilde{\Sigma}_{\rm smooth}^R({\bm k},\omega\!+\!i\gamma)$, 
the former of which is already obtained from $\Sigma(K)$. The result of 
$\widetilde{\Sigma}_{\rm diff}^R({\bm k},\omega\!+\!i\gamma)$ 
in the low-energy region is drawn in Fig.~\ref{fig:21}(b) 
which is perfectly consistent with the anomalous structure of 
$\Sigma^R({\bm k},\omega\!+\!i\gamma)$ around the excitron position in the inset 
in Fig.~\ref{fig:14}. As related to 
$\widetilde{\Sigma}_{\rm smooth}^R({\bm k},\omega\!+\!i\gamma)$, 
the one-particle spectral function $A_{\rm smooth}({\bm k},\omega)$ is given 
in Fig.~\ref{fig:21}(a) by the black dotted curve in comparison with $A({\bm k},\omega)$ 
indicated by the red solid curve, where $A_{\rm smooth}({\bm k},\omega)$ is defined by
\begin{align}
A_{\rm smooth}({\bm k},\omega) = -\frac{1}{\pi} {\rm Im}\,
G_{\rm smooth}^R({\bm k},\omega\!+\!i\gamma)
\label{eq:57}
\end{align}
with
\begin{align}
{G_{\rm smooth}^R({\bm k},\omega\!+\!i\gamma)}^{-1}
=&\,\omega\!+\!i\gamma\!+\!\mu_x\!+\!\mu_c^{\rm smooth}\!-\!\varepsilon_{\bm k}
\nonumber \\
&-\widetilde{\Sigma}_{\rm smooth}^R({\bm k},\omega\!+\!i\gamma),
\label{eq:58}
\end{align}
where $\mu_c^{\rm smooth}$ is the correlation contribution to the chemical 
potential as calculated through $\widetilde{\Sigma}_{\rm smooth}(K)$. Actually, 
its difference from $\mu_c$ is negligibly small. There exists no signature 
of the excitron in $A_{\rm smooth}({\bm k},\omega)$ as it should be 
in the case of FLT. 

%%%%%%%%%%%%%%%%%%< Paragraph 70: analytic continuation  >%%%%%%%%%%%%%%%%%%%%%%
To investigate the function analytically continued from 
$\alpha_{\bm k}(i\omega_n)$, we write down $\Sigma^R({\bm k},\omega\!+\!i\gamma)$ 
in reference to Eq.~(\ref{eq:52}) as 
\begin{align}
\widetilde{\Sigma}_{\rm diff}^R({\bm k},\omega\!+\!i\gamma)=&
-\sqrt{\omega\!+\!i\gamma-\xi_{\bm k}}
\sqrt{\omega\!+\!i\gamma-{\tilde \alpha}(\omega)\xi_{\bm k}}
\nonumber \\
 &+\frac{\sqrt{{\tilde \alpha}(\omega)}}{2}(\omega\!+\!i\gamma-\xi_{\bm k})
 \nonumber \\
 &+\frac{1}{2\sqrt{{\tilde \alpha}(\omega)}}
  \left [\omega\!+\!i\gamma-{\tilde \alpha}(\omega)\xi_{\bm k}\right ],   
\label{eq:59}
\end{align}
where ${\tilde \alpha}(\omega)$ is introduced as the analytically continued 
function from $\alpha_{\infty}(i\omega_n)$, but because $\alpha_{\bm k}(i\omega_n)$ 
is well approximated by $\alpha_{\infty}(i\omega_n)$ at $|{\bm k}|\!=\!
0.95k_{\rm F}$, it also represents the analytically continued one 
from $\alpha_{\bm k}(i\omega_n)$. The branch cut in the square root $\sqrt{z}$ 
in Eq.~(\ref{eq:59}) is taken along the negative real axis in complex-$z$ plane.

%%%%%%%%%%%%%%%%%%%%< Paragraph 71: \alpha^R(w)  >%%%%%%%%%%%%%%%%%%%%%%%%
By comparing the result of $\widetilde{\Sigma}_{\rm diff}^R({\bm k},\omega\!+\!i\gamma)$ 
in Fig.~\ref{fig:21}(b) with that in Eq.~(\ref{eq:59}), we can determine not rigorously 
correct but reasonably accurate values of ${\tilde \alpha}(\omega)$ for $\omega$ 
in the range from $-0.2\varepsilon_{\rm F}$ to $0$ and the obtained results are 
given in Fig.~\ref{fig:21}(c), from which we can raise a couple of points to 
appreciate the importance of $\omega$-dependence in ${\tilde \alpha}(\omega)$:  
(i) Although ${\tilde \alpha}(\omega)$ is real at $\omega=0$ (more concretely, 
${\tilde \alpha}(0)\!=\!2.247$ as given by an extrapolation of the data 
$\{\alpha_{\infty}(i\omega_n)\}_{n=1,2,3,\cdots}$), it is generally complex 
with a negative imaginary part. Due to the existence of this imaginary part in 
${\tilde \alpha}(\omega)$, $A({\bm k},\omega)$ is not characterized by such a 
square-root singularity, $1/\sqrt{\xi_{\rm k}-\omega}$, as typically seen in purely 
1D Luttinger liquids~\cite{Schonhammer_1993a,Schonhammer_1993b,Maebashi_2014,Maebashi_2015}, 
but is well approximated by a Lorentzian-type function~\cite{Khodas_2007}. 
(ii) One might expect to see another anomaly at $\omega={\tilde \alpha}(\omega)\xi_{\bm k}$ 
originating from the second square root $\sqrt{\omega\!+\!i\gamma-{\tilde \alpha}(\omega)
\xi_{\bm k}}$ in Eq.~(\ref{eq:59}), but it does not seem to be the case, primarily because 
the deviation of ${\tilde \alpha}(\omega)$ from unity at the relevant $\omega$ is not 
large enough to provide a noticeable structure in $A({\bm k},\omega)$. The only effect 
from this contribution is found to make the quasiparticle peak position in 
$A({\bm k},\omega)$ shift slightly from that in $A_{\rm smooth}({\bm k},\omega)$ 
as seen in Fig.~\ref{fig:21}(a). 

%%%%%%%%%%%%%%%%%%%%%%%%< Section 4-5 >%%%%%%%%%%%%%%%%%%%%%%%%%%%%%%%%%%%%%%%%%%%%%%
\subsection{$T$-dependence of the excitron liftime}
\label{sec:4E}

%%%%%%%%%%%%%%%%%%%%< Paragraph 72: \tau_*)  >%%%%%%%%%%%%%%%%%%%%%%%%
In order to better understand the excitron, it is useful to determine its lifetime 
$\tau_{\rm excitron}$, especially, its dependence on $T$ in connection with 
the coherence nature of the excitron. In the conventional Fermi liquids in which 
$G(K)$ is expressed in the form of Eq.~(\ref{eq:01}), the quasiparticle lifetime 
$\tau_*$ is obtained by
\begin{align}
\tau_*^{-1}=2z^*|{\rm Im} \Sigma^{R}({\bm k}_{\rm F},0)|,
\label{eq:60}
\end{align}
and scales as $T^2$. Thus, on general grounds, one expects that 
${\rm Im} \Sigma^{R}({\bm k_{\rm F}},\omega)$ also contains some relevant information 
on $\tau_{\rm excitron}^{-1}$, but because the excitron peak is hidden behind 
the quasiparticle peak at $k=k_{\rm F}$, it is not easy to extract information on 
the excitron from ${\rm Im}\Sigma^R({\bm k}_{\rm F},\omega)$ without removing 
the dominant quasiparticle contribution in the first place. According to the 
first-Matsubara-frequency rule~\cite{Chubukov_2012}, the quasiparticle 
part vanishes in the $T$-dependent ${\rm Im}\Sigma({\bm k}_{\rm F},i\pi T)$, 
implying the possibility that we may directly connect 
${\rm Im}\Sigma({\bm k}_{\rm F},i\pi T)$ with $\tau_{\rm excitron}^{-1}$.  

%%%%%%%%%%%%%%%%%%%%%%%%%%%%%%%%%%< Figure 22 >%%%%%%%%%%%%%%%%%%%%%%%%%%%%%%%%%%%%%%%
\begin{figure}[bht]
\begin{center}
\includegraphics[scale=0.45,keepaspectratio]{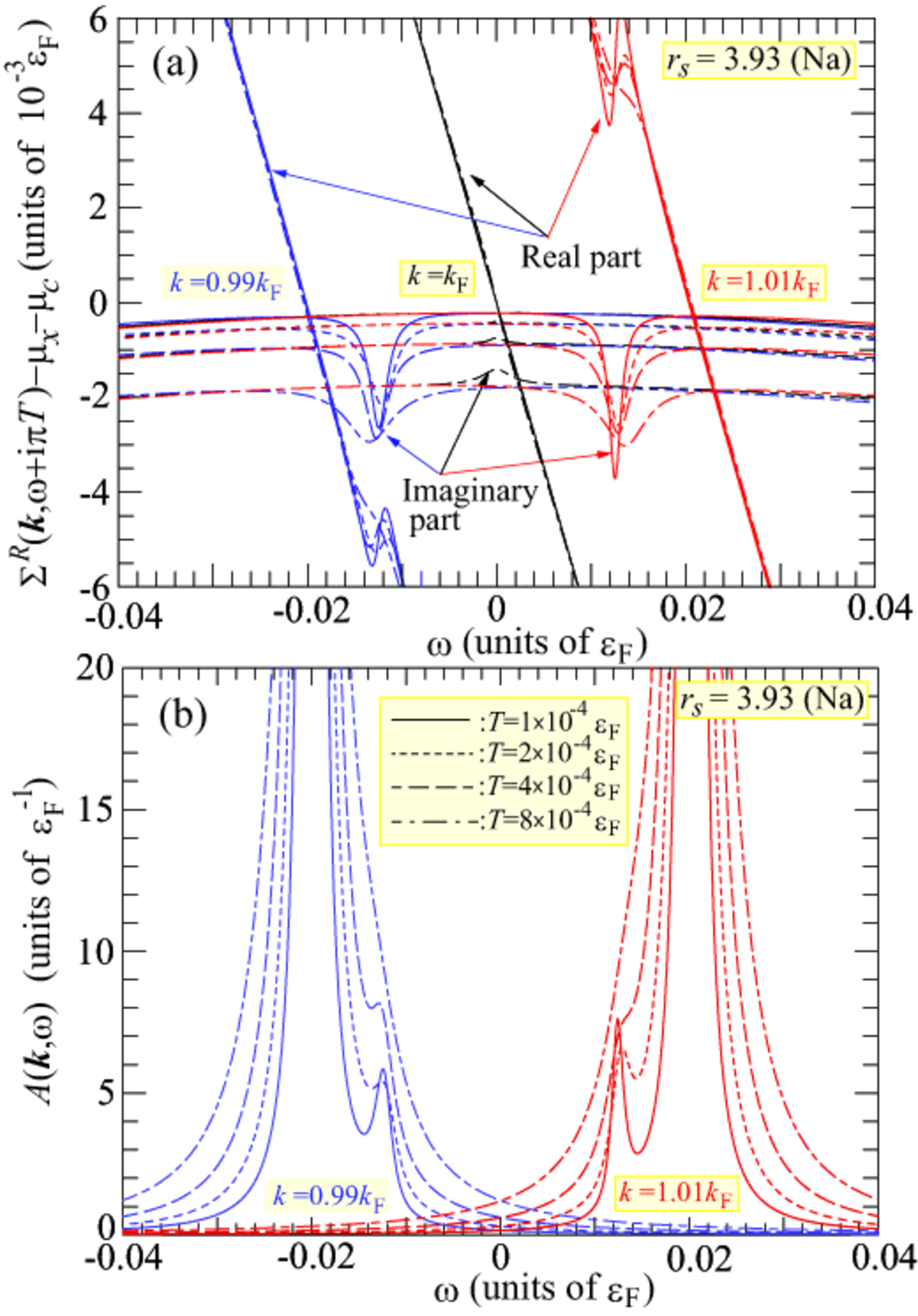}
\end{center}
\caption[Fig.22]{(a) Change of $\Sigma^R({\bm k},\omega\!+\!i\pi T)$ with $T$ in the 
range from $10^{-4}\varepsilon_{\rm F}$ to $8 \!\times\! 10^{-4}\varepsilon_{\rm F}$ 
at $k=0.99k_{\rm F}$, $k=k_{\rm F}$, and $1.01k_{\rm F}$ for $r_s=3.93$. (b) Temperature 
dependence of $A({\bm k},i\pi T)$ at $k=0.99k_{\rm F}$ and $1.01k_{\rm F}$ for $T$ 
in the same range as that in panel (a).
}
\label{fig:22}
\end{figure}
%-----------------------------------------------------------------------------------%

%%%%%%%%%%%%%%%%%%%%< Paragraph 73: \tau_*)  >%%%%%%%%%%%%%%%%%%%%%%%%
In these circumstances, we have examined the $T$-dependence of 
$\Sigma^{R}({\bm k},\omega\!+\!i\pi T)$ not only at $k\!=\!k_{\rm F}$ but also in the 
very vicinity of $k_{\rm F}$, namely, at $k\!=\!0.99k_{\rm F}$ and $k\!=\!1.01k_{\rm F}$ 
and the obtained results are shown in Fig.~\ref{fig:22}(a). The corresponding ones for 
$A({\rm k},\omega)$ are also given in Fig.~\ref{fig:22}(b) from which we estimate 
$\tau_{\rm excitron}^{-1}$ by measuring a full width at a half maximum (FWHM). Note that 
we cannot obtain this FWHM unambiguously, mainly because the excitron peak is located 
at the foot of the dominant quasiparticle peak. Thus, a rather large error bar is associated 
with this estimation, particularly for $T\!\gtrsim \!4\!\times\! 10^{-4}\!\varepsilon_{\rm F}$. 
Those data for $\tau_{\rm excitron}^{-1}$ can be summarized in the form of 
$\tau_{\rm excitron}^{-1} \!\propto \!T^{\alpha}$ with $\alpha \!\approx \!1.0\!\pm\!0.2$. 
Similar analysis is also made in reference to $-{\rm Im}\Sigma^R({\bm k},i\pi T)$ 
to find $\alpha \!\approx \!0.6\!-\!0.9$. Incidentally, we find that 
the numerical data for ${\rm Im}\Sigma^R({\bm k},i\pi T)$ are very accurately 
expressed by 
\begin{align}
{\rm Im}\Sigma^R((1\pm0.01){\bm k}_{\rm F},i\pi T)=-(2.22\mp 0.02)T^{1.00}, 
\label{eq:61}
\end{align}
at $k\!=\!(1 \!\pm \!0.01)k_{\rm F}$, while at $k\!=\!k_{\rm F}$, 
\begin{align}
{\rm Im}\Sigma^R({\bm k}_{\rm F},i\pi T)=-0.924\varepsilon_{\rm F}
(T/\varepsilon_{\rm F})^{0.91}, 
\label{eq:62}
\end{align}
for $T$ less than $8\!\times\! 10^{-4}\!\varepsilon_{\rm F}$. If we assume that 
$\tau_{\rm excitron}^{-1}$ is directly proportional to 
${\rm Im}\Sigma^R({\bm k}_{\rm F},i\pi T)$, then we obtain another estimate of $\alpha$ 
as $\alpha\!=\!0.91$. In this way, $\alpha$ seems to be close to unity which is 
exactly the value evaluated in 1D Luttinger liquids~\cite{Hur_2006}. The result 
of $\alpha=1$ is also expected in the marginal Fermi liquids. In the present case, 
however, it is likely that $\alpha$ is less than unity, though a definite value 
of $\alpha$ should be determined by some other analytic method in the future.

%%%%%%%%%%%%%%%%%%%%< Paragraph 74: \tau_*)  >%%%%%%%%%%%%%%%%%%%%%%%%
Because of $\alpha < 1$, the excitron is not a coherent but an incoherent excitation. 
Thus, it should formally be included in the incoherent background in Eq.~(\ref{eq:01}), 
but it still provides a clear peak structure in $A({\rm k},\omega)$ contrary to the 
general belief that the incoherent background will be represented by a smooth 
function in the Fermi liquids. 

%%%%%%%%%%%%%%%%%%%%< Paragraph 75: \tau_*)  >%%%%%%%%%%%%%%%%%%%%%%%%
Intuitively, we can think of the following: The diagram in Fig.~\ref{fig:01}(c) 
indicates that the excitron is an electron surrounded by multiple electron-hole 
pairs excited mostly in the longitudinal direction. In this sense, the excitron 
(or an electron-exciton-cloud composite in a shape elongated along the electron motion) 
may be regarded as an entity akin to a polaron (or an electron-phonon-cloud composite).  
In the case of polarons, however, the associated 
mediating modes are phonons which are coherent bosons in the whole crystal. 
On the other hand, the excitron is associated with the incipient excitonic mode, 
an incoherent boson mode which is damped in a short distance. Thus, it is very 
reasonable to reach the conclusion that the excitron is an incoherent excitation.  

%%%%%%%%%%%%%%%%%%%%%%%%%%%%%%%%%%%%%%%%%%%%%%%%%%%%%%%%%%%%%%%%%%%%%%%%%%%%%%%%%%%%%
%%%%%%%%%%%%%%%%%%%%%%%%%%%%%< Section 5 >%%%%%%%%%%%%%%%%%%%%%%%%%%%%%%%%%%%%%%%%%%%
\section{Conclusion and Discussion}
\label{sec:5}

%%%%%%%%%%%%%%%%%%%%%%%%%< Paragraph 76: summary I >%%%%%%%%%%%%%%%%%%%%%%%%%%%%
In this paper, we have developed a feasible nonperturbative scheme to accurately 
determine $\Sigma(K)$ through a fully self-consistent iterative calculation 
with rigorously satisfying the Ward identity and the total-momentum conservation law, 
while fulfilling all other known conservation laws, sum rules, and 
correct asymptotic behaviors in $G(K)$, $\Gamma(K,K\!+\!Q)$, $\Pi(Q)$, and $n({\bm k})$. 
The scheme has been successfully implemented in the 3D homogeneous electron gas 
for the range of $r_s$ corresponding to all simple metals at $T$ down to 
$10^{-4}\varepsilon_{\rm F}$ with tiny mesh as small as $10^{-4}k_{\rm F}$ 
near the Fermi surface in ${\bm k}$ space. Our results on $n({\bm k})$, the quasiparticle 
renormalization factor $z^*$, and the quasiparticle effective mass $m^*$, 
all of which are the long-standing challenges in the electron gas, are in very good 
agreement with the recent data given by quantum Monte Carlo simulations and available 
experiments, confirming that our present scheme actually provides sufficiently accurate 
results of $\Sigma(K)$.
 
%%%%%%%%%%%%%%%%%%%%%%%%%< Paragraph 77: summary II >%%%%%%%%%%%%%%%%%%%%%%%%%%%%
By analytic continuation onto the real $\omega$ axis through Pad\'{e} approximants, 
$G(K)$ is transformed into $G^R({\bm k},\omega)$, from which we obtain 
$A({\bm k},\omega)$ exhibiting a new sharp low-energy peak for $|{\bm k}|$ 
not just at $k_{\rm F}$ but in its vicinity, in addition to the dominant quasiparticle 
peak as well as high-energy one- and two-plasmon satellites. The appearance of 
two-plasmon satellites without resort to the ad hoc combination of the $GW$ approximation 
with a cumulant expansion is a notable theoretical achievement, but the most important 
issue is the discovery of the new low-energy peak that emerges 
for all simple-metal densities at $T \lesssim 10^{-3}\varepsilon_{\rm F}$. 
Its origin is attributed to the excitonic attraction arising from the multiple 
excitations of tightly bound electron-hole pairs in $\Pi({\bm q},\Omega)$ for 
$|{\bm q}| \approx 2k_{\rm F}$ and $|\Omega| \ll \varepsilon_{\rm F}$, 
suggesting that it should be dubbed ``excitron''. This excitonic 
scattering process occurs only in very restricted angles along the longitudinal 
direction, which motivates us to characterize the excitron as a branch-cut singularity 
in analogy with 1D physics. From a viewpoint of QCP physics, the excitron is 
also regarded as an anomaly induced by quantum fluctuations of the incipient excitonic 
mode around the quantum-critical CDW transition. In either way, this anomalous 
low-energy phenomenon poses an interesting question as to the validity of the Landau's 
hypothesis on the one-to-one correspondence of low-energy excitations 
between a free Fermi gas and an interacting normal Fermi liquid. 
Taken together, our results indicate that 
non-Fermi liquid physics may already play a role in the description 
of simple metals at sufficiently low temperatures.

%%%%%%%%%%%%%%%%%%%%%%%%%< Paragraph 78: Discussion 1 >%%%%%%%%%%%%%%%%%%%%%%%%%%%%
Four comments are in order: 

(i) Since we start with the rigorous equation to determine $\Sigma(K)$ 
in Eq.~(\ref{eq:22}) in which $W(Q)$ is taken as an accurately known quantity, 
the vertex function $\Gamma(K,K\!+\!Q)$ is the only unknown quantity. Thus, we have examined 
various forms for $\Gamma(K,K')$ in our theoretical framework, looking for a necessary 
and sufficient condition for the appearance of excitron. As a result, we come 
to know that the excitron appears, if and only if we include either 
$\overline{\Gamma}_{\rm WI}(K,K')$ in Eq.~(\ref{eq:42d}) or $\Gamma_{\rm WI}(K,K')$ 
without $\eta_1(Q)$ in Eq.~(\ref{eq:28}) in the definition of $\Gamma(K,K')$. 
We can easily understand the necessity of this kind of the Ward-identity-related vertex part, 
because this is the crux to make our scheme nonperturbative; remember that we need to 
go beyond simple perturbation expansion from $G_0(K)$ for describing a situation 
intimately connected with the breakdown of the Fermi-liquid theory. Incidentally, 
the quantitative details of excitron, such as the peak position $\xi_{\bm k}$, 
the peak height, and the peak width, depend on the choice of $\Gamma(K,K')$ by not 
negligible amounts. Therefore, more useful information on excitron is needed 
in the future to further improve on $\Gamma(K,K')$. 

%%%%%%%%%%%%%%%%%%%%%%%%%< Paragraph 79: Discussion 2 >%%%%%%%%%%%%%%%%%%%%%%%%%%%%
(ii) In the simple-metal density region, the strength of the excitron peak is found 
to be so weak that its existence will not be detected by bulk measurements 
such as electric conductivity and specific heat. In ARPES experiments, however, 
it will be detected, if the energy resolution is much smaller, of the order 
of 1 meV or less, than those in preceding experiments. In fact, in the previous 
ARPES studies, the resolution was about 0.2-0.4eV in 
1980s~\cite{Jensen_1885,Plummer_1887,Lyo_1988,Itchkawitz_1990}, 80-200meV in 
1990s~\cite{Wertheim_1995}, and still 30meV in 2020s~\cite{Potorochin_2022}. 
This is probably the reason why the excitron peak has not been detected, 
though some interesting unresolved features were observed near the Fermi 
level in the past~\cite{Jensen_1885,Plummer_1887,Itchkawitz_1990}. 
At the present time, it is encouraging to know that experimental equipments 
with the energy resolution of the order of 1 meV or less~\cite{Zhang_2022} do exist, 
but they have not been applied to simple metals so far. If they are actually 
applied with due attention to the possible appearance of excitron, then it will be very 
exciting to see the experimental results from a perspective of fundamental physics. 
Detection of excitron by ARPES is also very important from a viewpoint of further 
developments of our theoretical framework, as mentioned in the previous paragraph.

%%%%%%%%%%%%%%%%%%%%%%%%%< Paragraph 80: Discussion 3 >%%%%%%%%%%%%%%%%%%%%%%%%%%%%
(iii) By making the electron density lower than those of simple metals to approach 
the CDW transition, we can expect a more interesting situation in which the effects 
of excitron become so strong that FLT apparently breaks down, leading to the 
emergence of NFL. With this expectation in mind, we are now trying to obtain 
a fully self-consistent solution for such low densities (or $r_s>6$), but 
at present it takes too many iterative steps to obtain a completely convergent 
result of $\Sigma(K)$ for $r_s>8$. We shall report our efforts in this 
direction in the near future.

%%%%%%%%%%%%%%%%%%%%%%%%%< Paragraph 81: Discussion 4 >%%%%%%%%%%%%%%%%%%%%%%%%%%%%
(iv) In Sec.~\ref{sec:4}, we have not taken a mathematically rigorous but a heuristic 
approach to the analysis of the excitron. Admittedly, it would be better to 
derive the branch-cut singularity in $\Sigma(K)$ analytically by explicitly including 
the effect of $V_{\rm ex}(K,K';Q)$ defined in Fig.~\ref{fig:01}(a), but we have to 
understand that this is a very difficult task. In fact, this problem of accurately 
treating the local charge fluctuations induced by correlated multiple electron-hole 
pair excitations is as difficult as that of the local spin fluctuations in the 
heavy-Fermion superconductors~\cite{Hewson_1993,Stewart_2001,Lohneysen_2007,Gegenwart_2008,
Steglich_2013,Takada_2015,Coleman_2015} and high-$T_c$ cuprate 
superconductors~\cite{Moriya_1990,Millis_1990,Monthoux_1993,Chubukov_1994,Anderson_2004,
Lee_2006,Anderson_2009,Varma_2020}, suggesting that we should leave this problem 
for future analysis. To put it the other way around, our present approach to treating 
$\Sigma(K)$ as a whole by imposing various conservation laws and sum rules may 
provide a new route to the solution of local spin fluctuation problems in those 
strongly-correlated materials. We would expect a new development from this perspective 
in those hot fields, including high-$T_c$ superconductivity.  

%-----------------------------------------------------------------------------------%

\acknowledgments

The author thanks Kazuhiro Matsuda for useful discussions at the very early 
stage of this work. He is also grateful to Hiroyuki Yata and Naoki Kawashima 
at Institute for Solid State Physics, The University of Tokyo 
for maintaining the cluster machines to efficiently perform the computations 
reported in this paper. 

\appendix

%%%%%%%%%%%%%%%%%%%%%%%%%< Appendix A >%%%%%%%%%%%%%%%%%%%%%%%%%%%%
\section{Matsubara sum}

The Matsubara sum of a given function $f(i\omega_n)$ with $\omega_n=\pi T(2n-1)$ 
for an integer $n$ is calculated numerically in the following way:
\begin{align}
T\sum_{\omega_n}f(i\omega_n)&=T\sum_{n=1}^{\infty}F(\omega_n)
\nonumber \\
&=T\sum_{n=1}^{N}F(\omega_n)+\frac{T}{12}[F(\omega_N)
+5F(\omega_{N+1})]
\nonumber \\
& \hspace{2.4cm}
+\int_{\omega_{N+1}}^{\infty}\frac{dx}{2\pi}\,F(x),
\label{eq:A1}
\end{align} 
where $F(\omega_n) \equiv f(i\omega_n)+f(-i\omega_n)$ and we increase $N$ 
from $N \approx 10$ until a convergent result is obtained; in most cases, $N$ as 
small as $10$ is already good enough, but it is safe to choose $N=100$. We can 
derive Eq.~(\ref{eq:A1}) from the Euler-Maclaurin formula~\cite{Mathews_1970}:
\begin{align}
\int_a^b dx\,F(x)&=h\Bigl[\frac{1}{2}F(a)+F(a+h)+\cdots +F(b-h)
\nonumber \\
&\ +\frac{1}{2}F(b) \Bigr ]-\frac{B_2}{2!}h^2F'(x)\Bigr|_a^b- \cdots,
\label{eq:A2}
\end{align}
with the Bernoulli number $B_2=1/6$. By taking $h\!=\!2\pi T$, 
$a\!=\!\omega_{N+1}$, $b \to \infty$, and $F'(a)\!=\![F(a)\!-\!F(a\!-\!h)]/h$, 
we can easily arrive at Eq.~(\ref{eq:A1}). The relative error incurred 
in cutting off the series in Eq.~(\ref{eq:A2}) at the level of $F'(x)$ is 
of the order of $T^4$, negligibly small for sufficiently low $T$.

%%%%%%%%%%%%%%%%%%%%%%%%%< Appendix B >%%%%%%%%%%%%%%%%%%%%%%%%%%%%
\section{Modification of the momentum distribution function}

We take the following strategy to improve on $n({\bm k})$: 

(i) Obtain $n({\bm k})\ [=\!n(x)$ with $x\!=\!|{\bm k}|/k_{\rm F}]$ through 
Eq.~(\ref{eq:12}). (ii) Determine $n_0\!\equiv\!n(0)$ and $n_{\pm}\!\equiv\!
n(1\!\pm \!0^{+})$. (iii) With the use of these $n_0$ and $n_{\pm}$, obtain 
$n_{\rm IGZ}(x)$ in the parametrization scheme described in Ref.~\cite{Takada_2016}. 
Note that ``IGZ'' stands for ``improved Gori-Giorgi and Ziesche''~\cite{Ziesche_2002}. 
(iv) Because $n_{\rm IGZ}(x)$ is almost the same as $n(x)$, construct 
a corrected function $n_c(x)$ which changes smoothly from $n(x)$ 
for $x\!\lesssim \!1.1$ to $n_{\rm IGZ}(x)$ for $x\! \gtrsim \! 2.0$. 
Concretely, we define $n_c(x)$ as $n_c(x)\!\equiv\!n(x)$ for $x \le x_c$ and $n_c(x)
\!\equiv \!n_{\rm IGZ}(x)\!+\!\Delta n(x)$ for $x\! >\! x_c$ with choosing 
an appropriate $x_c$ in the region of $1.1\!<\!x_c \!<\!2.0$. The small additional 
term $\Delta n(x)$ decreases exponentially as $x$ increases. 
At $x\!=\!x_c$, $\Delta n(x)$ is so determined as to make $n_c(x)$ smoothly 
connected to $n(x)$ up to second derivative. It is also tuned to satisfy the three 
sum rules for the momentum distribution function as accurately as possible. 

%%%%%%%%%%%%%%%%%%%%%%%%%< Appendix C >%%%%%%%%%%%%%%%%%%%%%%%%%%%%
\section{Consideration on $\mbox{\boldmath$\eta$}_1(Q)$ and 
$\mbox{\boldmath$\zeta$}_3(Q)$}

The behavior of $\Gamma(K,K\!+\!Q)$ at $K\!=\!K_{\rm F}$ in the limit of 
$Q \to Q_0$ is exactly known; in the $\omega$-limit, it approaches  
$\Gamma^{\omega}(K_{\rm F},K_{\rm F})$, given by
\begin{align}
\Gamma^{\omega}(K_{\rm F},K_{\rm F})
=\left . \frac{\partial G^{-1}(K)}{\partial (i\omega_n)}\right |_{K=K_{\rm F}}
=Z(K_{\rm F}),
\label{eq:C1}
\end{align}
as a direct consequence of WI, while in the $q$-limit, it approaches  
$\Gamma^{q}(K_{\rm F},K_{\rm F})$, given by 
\begin{align}
\Gamma^{q}(K_{\rm F},K_{\rm F})
=\left . \frac{\partial G^{-1}(K)}{\partial \mu}\right |_{K=K_{\rm F}}
=\frac{\kappa}{\kappa_{\rm F}}\frac{\partial E(K_{\rm F})}
{\partial \varepsilon_{{\bm k}_{\rm F}}},
\label{eq:C2}
\end{align}
as one can convince oneself by considering the one-to-one correspondence of 
each Feynman diagram representing $\Gamma(K_{\rm F},K_{\rm F})$ with the one 
obtained by the differentiation of an arbitrary $G$ line in each Feynman diagram 
for $\Sigma$ with respect to $\mu$~\cite{Takada_1995,Higuchi_1995,Maebashi_2011}.
If we use the expression in Eq.~(\ref{eq:24}) as it is, then the above limiting behavior 
is automatically satisfied, but in arriving at Eq.~(\ref{eq:32}), we have 
introduced a few approximations and simplifications which may deteriorate this 
favorable feature. In fact, $\Gamma(K,K\!+\!Q)$ in 
Eq.~(\ref{eq:32}) reduces to $\Gamma^{\omega}(K_{\rm F},K_{\rm F})/\eta_1(Q_0)$ 
in the $\omega$-limit and to $\Gamma^{q}(K_{\rm F},K_{\rm F})[1/I_0+3\zeta_3(Q_0)]$ 
in the $q$-limit, compelling us to impose the following constraints; $\eta_1(Q_0)$ 
in the $\omega$-limit, $\eta_1^{\omega}\!=\!1$ and $\zeta_3(Q_0)$ in the $q$-limit, 
$\zeta_3^q\!=\!(I_0\!-\!1)/(3I_0)$. Note that $\eta_1(Q_0)$ in the $q$-limit should be 
equal to $\eta_1^q \!\equiv \! {\tilde \eta}_{1}(K_{\rm F})\!=\!
(\partial E(K_{\rm F})/\partial \varepsilon_{{\bm k}_{\rm F}})
/Z(K_{\rm F})$ from the very definition of $\eta_1(Q)$.

Taking account of those constraints as well as the basic feature that 
$\Gamma(K,K\!+\!Q)$ should rapidly approach unity for either $K$ or $K\!+\!Q$ 
far away from $K_{\rm F}$, we have chosen $\zeta_3(Q)$ in Eq.~(\ref{eq:36}) 
and $\eta_1(Q)$ in the following form: 
\begin{align}
\eta_1(Q)=\frac{1}{[\eta_a^{-1}(q)-1]\eta_b(i\omega_q/2)\eta_c(Q)+1},
\label{eq:C3}
\end{align}
where, $\eta_a(q)$ is the angular average of 
${\tilde \eta}_1({\bm k},0)$ with respect to the angle $\theta$ between 
${\bm k}_{\rm F}$ and ${\bm q}$ in the definition of ${\bm k}\! \equiv 
\!{\bm k}_{\rm F}\!+\!{\bm q}/2$ (and thus $k^2\!=\!k_{\rm F}^2\!+\!
k_{\rm F}q\cos \theta \!+\!q^2/4)$, given by
\begin{align}
\eta_a(q)&=\langle {\tilde \eta}_1({\bm k},0) \rangle 
\equiv \frac{\int_0^{\pi}\sin \theta d\theta\, {\tilde \eta}_1({\bm k},0)}
            {\int_0^{\pi}\sin \theta d\theta}
\nonumber \\
&=\int_{|k_{\rm F}\!-\!q/2|}^{k_{\rm F}\!+\!q/2} \frac{kdk}{k_{\rm F}q}\,
\frac{\partial E({\bm k},0)/\partial \varepsilon_{\bm k}}
{Z({\bm k},0)}.         
\label{eq:C4}
\end{align}
The overall $\omega_q$-dependence is described by $\eta_b(i\omega_n)$ with 
$i\omega_n \to i\omega_q/2$ and the functional $\eta_b(i\omega_n)$ is defined as 
\begin{align}
\eta_b(i\omega_n)=\frac{Z({\bm k}_{\rm F},i\omega_n)-1}
{Z({\bm k}_{\rm F},0)-1},
\label{eq:C5}
\end{align}
by considering the fact that the dominant $\omega_n$ dependence comes from 
$Z(K)$ in $\Sigma(K)$ for $K$ near $K_{\rm F}$. The conversion function 
from $\omega$ to $q$ limits, $\eta_c(Q)$, is taken as
\begin{align}
\eta_c(Q)=\frac{v_{\rm F}^2q^2}{v_{\rm F}^2q^2+3\omega_q^2},
\label{eq:C6}
\end{align}
in reference to the conversion in $\Pi_0(Q)$ at $Q \to Q_0$, as shown 
in Eqs.~(\ref{eq:18}) and (\ref{eq:19}).

%%%%%%%%%%%%%%%%%%%%%%%%%< Appendix D >%%%%%%%%%%%%%%%%%%%%%%%%%%%%
\section{Extrapolation to static quantity}

From a set of data $\{f_1,f_2,\cdots,f_N\}$ for an even function $f(i\omega_n)$ 
at $n=1,\cdots,N$ with $\omega_n\!=\!\pi T(2n\!-\!1)$, we can estimate the static 
value $f(0)$ by the following extrapolation: First, we regard the data set as 
that of the size twice as large by considering $\{f_N,f_{N-1}, \cdots, 
f_2,f_1,f_1,f_2,\cdots,f_N\}$ given at $\{-\omega_N,-\omega_{N-1},\cdots,-\omega_2,
-\omega_1,\omega_1,\omega_2,\cdots,\omega_N \}$. Then, we apply a Lagrange's 
polynomial interpolation formula to this enlarged data set to obtain the 
interpolation function $\tilde{f}(\omega)$ as
\begin{align}
\tilde{f}(\omega)=\sum_{n=1}^{N}f_n\prod_{j\neq n}^N\frac{\omega^2-\omega_j^2}
{\omega_n^2-\omega_j^2}.
\label{eq:D1}
\end{align}
By substituting $\omega=0$ in Eq.~(\ref{eq:D1}), we obtain $f(0)$ as
\begin{align}
f(0) \approx \tilde{f}(0)=\sum_{n=1}^{N}f_n\prod_{j\neq n}^N\frac{\omega_j^2}
{\omega_j^2-\omega_n^2}.
\label{eq:D2}
\end{align}
We can check the convergence of the result $f(0)$ by increasing $N$ from $N \approx 10$ 
to find that in all cases $N=30$ is large enough.

%%%%%%%%%%%%%%%%%%%%%%%%%< Appendix E >%%%%%%%%%%%%%%%%%%%%%%%%%%%%
\section{Mesh points}

The selected mesh points in $(k,\omega_n)$ space, $\{k_i,\omega_j\}$, are chosen 
in the following way:
On $k$-axis, the first point $k_1$ is taken as $k_1\!=\!0.007k_{\rm F}$; 
for $0\!<\!k_i\!\leq \! 2k_{\rm F}$ and $|k_i\!-\!k_{\rm F}|\!\gtrsim \!
0.01k_{\rm F}$, $\Delta_{i} \ (\equiv \!k_{i+1}\!-\!k_i) \sim |k_i\!-\!k_{\rm F}|/10$;
for $k_i \sim k_{\rm F}$, $\Delta_{i} \sim 10^{-4}k_{\rm F}$; 
for $k_i \!> \!2k_{\rm F}$, $\Delta_{i} \sim k_i/10$; and 
the last point $k_M$ is taken as $120k_{\rm F}$ with $M=240$. 
Since the minimum value of $\{\Delta_i\}$ is as small as $10^{-4}k_{\rm F}$, 
the integrals in Eqs.~(\ref{eq:41a})-(\ref{eq:41c}) should be very accurately 
performed, i.e., up to at least six digits, to obtain significant difference 
between the results at adjacent points. To achieve this accuracy, we employ 
the double-exponential formula for numerical integration~\cite{Takahashi_1973}.
 
As for $\omega_n$-axis, we take account of only the positive side of the axis,  
because we consider only even functions; for $j=1-24$, $\omega_j=\pi T(2j-1)$;
for $j=25-35$, $\omega_j=\pi T(4j-49)$; for $j=36-50$, $\omega_j=\pi T(8j-189)$; 
$\cdots$ ; and the last point $\omega_N$ is about $10^7 \times \pi T$ with $N=310$. 
This set $\{ \omega_j \}$ is used in the Pad\'{e} approximants for the numerical 
analytic continuation~\cite{Vidberg_1977} in which it is quite useful to calculate  
in quadruple precision.

%%%%%%%%%%%%%%%%%%%%%%%%%%%%%%%%%%  References  %%%%%%%%%%%%%%%%%%%%%%%%%%%%%%%%%%%%%

\end{document}